\documentclass[trackchanges, twocolumn]{aastex7}
\usepackage{graphicx}
\usepackage{subfig}
\usepackage{subfloat}
\usepackage{rotating}
\usepackage{booktabs}
\usepackage{float}
\usepackage{color}
\usepackage{ulem}
\usepackage{amsmath}
\usepackage{soul}

\newcommand{\OIII}{{\ion{O}{3}}}

\newcommand{\FeII}{{\ion{Fe}{2}}}

\newcommand{\kms}{km s$^{-1}$}

\newcommand{\snu}{\affil{Department of Physics and Astronomy, Seoul National University, 1 Gwanak-ro, Gwanak-gu, Seoul 08826, Republic of Korea}}

\shorttitle{Modified 3D Biconical Outflow Model}
\shortauthors{Kim \& Woo}

\begin{document}
\title{A Modified 3D Biconical Outflow Model: Spatial Constraints on AGN-driven Outflows}

\correspondingauthor{Jong-Hak Woo}
\email{woo@astro.snu.ac.kr}

\author[0000-0002-2156-4994, gname='Changseok', sname='Kim"]{Changseok Kim}\snu
\email[show]{kcs1996kcs@snu.ac.kr}
\author[0000-0002-8055-5465, gname='Jong-Hak', sname='Woo']{Jong-Hak Woo}\snu
\affil{SNU Astronomy Research Center, Seoul National University, 1 Gwanak-ro, Gwanak-gu, Seoul 08826, Republic of Korea}
\email{woo@astro.snu.ac.kr}

\begin{abstract}
We present a modified outflow model and its application to constrain ionized outflow properties of active galactic nuclei (AGNs).
By adding a rotating disk component to the biconical outflow model of \citet{Bae16}, we find that models with a rotating disk require faster launching velocities ($\lesssim$ 1500 \kms) than outflow-only models to be consistent with the observed gas kinematics of local type 2 AGNs.
We perform Monte Carlo simulations to reproduce the observed distribution of gas kinematics of a large sample ($\sim$ 39,000), 
constraining the launching velocity and opening angle.
While the launching velocity is moderate for the majority of the local AGNs, the notable cases of 2 - 5 \% show strong outflows with $V_{max} \sim 1000-1500$ \kms. 
By examining the seeing effect based on the mock integral field unit data, we find that the outflow sizes measured based on velocity widths tend to be overestimated when the angular size of the outflow is comparable to or smaller than the seeing. This result highlights the need for more careful treatments of the seeing effect in the outflow size measurement, yet it still supports the lack of global feedback by gas outflows for local AGNs.
\end{abstract}



\section{Introduction} \label{sec:intro}
Accreting supermassive black holes, or active galactic nuclei (AGNs), emit a vast amount of energy through various channels, i.e., radiation, gas outflows, and relativistic jets, which are thought to interact with the interstellar medium (ISM) and eventually affect star formation in host galaxies
\citep[e.g., ][]{Silk98, Fabian12}. Recent theoretical studies commonly implement AGN feedback to reproduce primary observables, i.e., galaxy luminosity function \citep[e.g., ][]{Hopkins08, Weinberger17, Wellons23}. Nonetheless, the overall role and detailed mechanism of AGN feedback remain unclear in observations \citep[see reviews by][]{Harrison17, Harrison24}.

AGN-driven gas outflows are widely considered as one of the primary candidates for a feedback channel. It is now well-known that these outflows are multiphase \citep[e.g., ][]{Fiore17, Fluetsch19, Riffel23}, ubiquitous \citep[e.g., ][]{Mullaney13, Woo16, Rakshit18}, and kpc-scales \citep[e.g., ][]{Harrison14, Kim23}. In particular, the broad wing component of the [\OIII] $\lambda5007$ emission line is frequently utilized to investigate kpc-scale warm ionized gas outflows \citep[][]{Boroson05, Greene05}. Large spectral surveys such as Sloan Digital Sky Survey (SDSS) enabled the scrutiny of integrated outflow properties within the central region for several tens of thousands of AGNs and their correlations with host galaxy properties \citep[e.g., ][]{Woo17, Woo20}. 

On the other hand, using spatially resolved spectroscopy, many other studies have focused on the detailed outflow structures for a limited number of targets ($\lesssim$ several tens). One of the primary findings in early studies is that the ionized outflows are well-described by a biconical model with two axisymmetric cones \citep{Crenshaw00a, Crenshaw00b, Veilleux01, Das05, Das06}. 
Subsequent studies took into account the dust extinction by the galactic plane \citep[e.g, ][]{Crenshaw10a, Crenshaw10b, Fischer10, Fischer11, Fischer13, Bae16} or a rotating disk \citep[e.g., ][]{Muller-Sanchez11} to reproduce the [\OIII] line profile and diverse narrow line region (NLR) kinematics. \citet{Marconcini23} recently developed a new modeling tool for analyzing 3D gas kinematics based on the 3D distribution of numerous gas clouds. 
By using a sample of nearby AGNs, these studies have provided various constraints on the geometry, extent, and energetics of outflows, which are crucial for quantifying outflow energetics and their impact on the host galaxy evolution.

Despite the importance of estimating the outflow parameters, it is challenging to expand these works to a large sample due to observational limitations. For instance, spatially resolved spectroscopic observations are still too expensive to be applied to a large sample. Furthermore, while integral field unit (IFU) surveys such as MaNGA or SAMI provide spatially resolved kinematics for $\lesssim$ thousands of AGNs \citep{Wylezalek20, Deconto-Machado22, Gatto24, Oh24}, they typically have a seeing-limited resolution (FWHM $\sim$ 1\arcsec - 2\arcsec), which is often insufficient to resolve outflow structures at moderate redshifts.
For example, a few kpc scales of outflows at z $\sim$ 0.1 (1 kpc $\sim$ 0.5\arcsec) are barely resolved even under the best seeing conditions on the ground. While space telescopes such as Hubble Space Telescope (HST) and James Webb Space Telescope (JWST) or adaptive optics (AO) can improve the spatial resolution down to a diffraction-limit (FWHM $\sim$ 0.1\arcsec), the detailed geometry and kinematics can only be resolved in rare cases such as nearby AGNs \citep[e.g., ][]{Fischer13, Polack24} or galactic scale outflows \citep[e.g., ][]{Vayner21a, Wylezalek22, Cresci23}. 

Nevertheless, detailed outflow models and simulated mock data can be efficiently utilized to overcome the limited spatial resolution and constrain the physical
parameters of the outflows, which are not directly measurable. 
\citet{Bae16} constructed a 3D biconical outflow model combined with a simple extinction by a thin dust plane and investigated the effect of each parameter on the gas kinematics. Their main findings are as follows: (1) velocity dispersion is higher for larger intrinsic outflow velocity or bicone inclination; (2) velocity offset with respect to systemic velocity increases with higher dust extinction; (3) a maximum velocity of 500 - 1000 \kms\ can explain most of the observed gas kinematics of local SDSS type 2 AGNs from \citet{Woo16}; and (4) higher intrinsic velocities tend to have wider bicone opening angles. However, this model did not account for rotating disk emission and seeing convolution, which are critical for reproducing 2D gas kinematics.

In this study, we present a 3D biconical outflow model modified from \citet{Bae16} along with applications of mock SDSS and IFU data. Comparing with observations, we provide spatial constraints on general properties of gas outflows and discuss the effect of seeing and the rotating disk component.
 The paper is structured as follows. In Section~\ref{sec:outflowmodel}, we summarize the original outflow model of \citet{Bae16} and describe three main modifications of this study. Section~\ref{sec:mockdata} describes procedures for generating mock SDSS and IFU data and their applications. In Section~\ref{sec:result}, we present the main results of the mock data analysis, including constraints on outflow properties. We discuss the implications of the main results and comparisons with other outflow models in Section~\ref{sec:discussion}. Lastly, we summarize the key conclusions in Section~\ref{sec:summary}. In this paper, we adopt a standard $\Lambda$CDM cosmology with H$_0$ = 67.8 km/s/Mpc, $\Omega_{\Lambda}$=0.692, and $\Omega_{m}$=0.308.

\section{Modified 3D Biconical Outflow Model} \label{sec:outflowmodel}
We summarize the original outflow model from \citet{Bae16} in Section~\ref{subsec:Bae16} and demonstrate three major modifications in Section~\ref{subsec:rotatinggas}, \ref{subsec:seeing}, and \ref{subsec:mockdata}, respectively.

\subsection{3D Biconical Outflow Model}\label{subsec:Bae16}
\begin{figure}[]
    \includegraphics[width=0.46\textwidth]{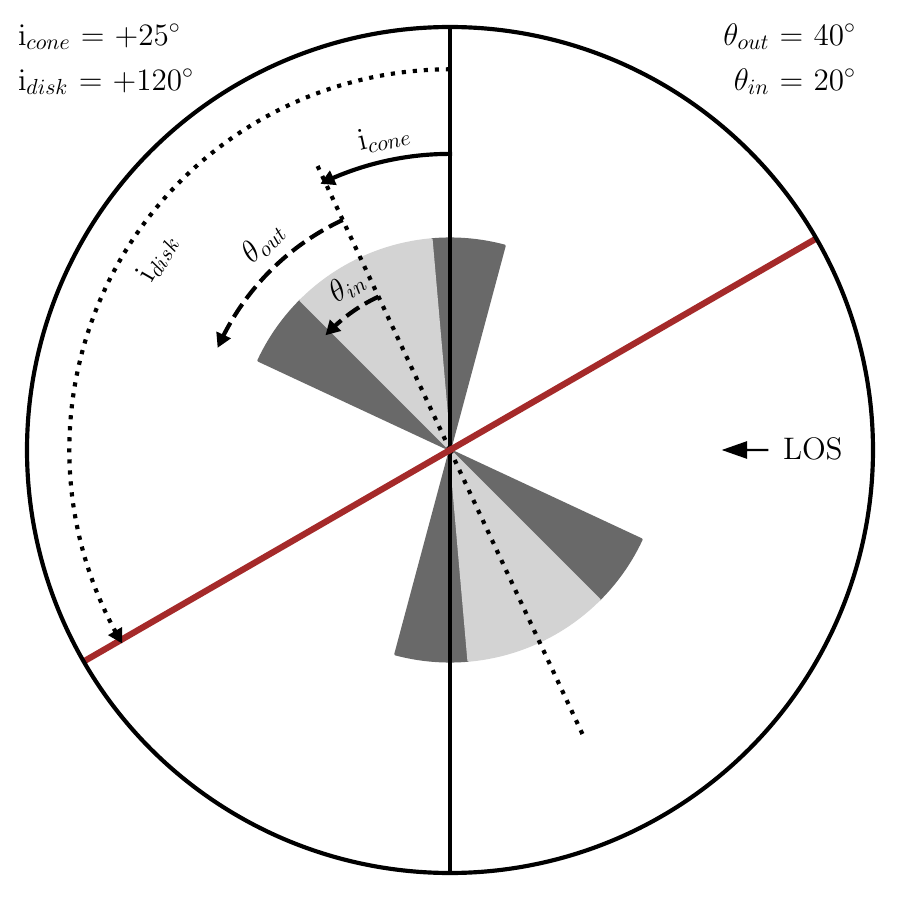}
    \caption{Example biconical outflow model from \citet{Bae16}. The dust plane is shown as a red line between the two cones. The dark gray region represents the volume occupied by the outflowing gas (i.e., between inner and outer opening angles), while the light gray region denotes the hollow interior. The LOS direction is indicated by the black arrow. The definitions and adopted values of the cone and disk inclination angles are also shown.}
    \label{model_plots}
\end{figure}

\citet{Bae16} presented a 3D outflow model adopted from \citet{Crenshaw10a, Crenshaw10b}, which consists of two axisymmetric outflow cones and a simple extinction factor by a thin galactic dust plane (see Figure~\ref{model_plots}). Once a set of 11 parameters is given, the model constructs 3D distributions of flux and velocity. By collapsing the 3D model along the line-of-sight (LOS), 2D distributions of flux, velocity, and velocity dispersion are generated. \citet{Bae16} also calculated flux-weighted velocity and velocity dispersion using the 2D maps for a direct comparison with the observed [\OIII] velocity and velocity dispersion (VVD) distribution of SDSS type 2 AGNs from \citet{Woo16}.


The input parameters define the bicone structure, 3D distributions of flux and kinematics, and the orientations of the bicone and the dust plane. 
For the bicone structure, three parameters determine the outflow size ($R_{out}$) and inner and outer half opening angles ($\theta_{in}$ and $\theta_{out}$). As illustrated by the dark gray region in Figure~\ref{model_plots}, we model gas outflows using a hollow bicone. 
For simplicity, we fix the difference between the outer and inner half opening angles to 20 $^\circ$ ($\theta_{out} = \theta_{in} + 20^\circ$).
While the opening angle differences estimated by \citet{Fischer13} are mostly in the 5$^\circ$--20$^\circ$ range, they adopted the minimum opening angle difference required to match the observations, meaning that the larger difference is also plausible.
Thus, we adopted 20$^\circ$ as a representative value from observations.
We also note that the inner opening angle has an opposite, but less significant effect on the integrated kinematics compared to the outer opening angle \citep{Bae16}.

The flux distribution is determined by $e$-folding radius ($\tau$), i.e., a radius where the flux decreases by a factor of $e$, and dust extinction factor ($A$). Since we only focus on relative flux distributions, we adopt an arbitrary central flux value ($f_0$). We assume that the intrinsic 3D flux distribution follows an exponential form based on the distance from the center ($R$). Thus, the intrinsic 3D flux distribution can be written as $f(R)=f_0e^{-R/\tau}$. Following \citet{Bae16}, we assume the $e$-folding radius is 20\% of the outflow size ($\tau=R_{out}/5$), which gives approximately two orders of magnitude fainter flux at the outflow edge than at the center. The dust extinction factor is multiplied when collapsing the 3D flux distributions to 2D flux maps. If a voxel is located behind the galactic dust plane along the LOS, its flux will be dimmed by the extinction factor in the 2D map. The opening angles and the relative inclination compared to the galactic plane decide which voxels are subject to dust extinction. 

The 3D velocity distributions are determined by two parameters, i.e., maximum velocity ($V_{max}$) and velocity profile. There are three options for the velocity profile as a function of the distance from the center ($R$): (1) a linear increase case ($v(R)=V_{max}(R/R_{out})$); (2) a linear decrease case ($v(R)=V_{max}(1-R/R_{out})$); and (3) a constant velocity ($v(R)=V_{max}$). In all three cases, the velocity is directed radially outward from the center. Following \citet{Bae16}, we adopt the linear decrease case for all models in this study. Note that the maximum velocity can be regarded as a launching velocity, as it represents the intrinsic outflow velocity at the center. This 3D velocity distribution is then collapsed to the 2D velocity and velocity dispersion maps in a flux-weighted manner as follows:

\begin{equation}
V(x,y) = \frac{\int v_p(x,y,z)f(x,y,z)\, dz}{\int f(x,y,z)\, dz},
\label{eq:vmap}
\end{equation}

\begin{equation}
\sigma^2(x,y) = \frac{\int v^2_p(x,y,z)f(x,y,z)\, dz}{\int f(x,y,z)\, dz} - V^2(x,y)
\label{eq:dmap}
\end{equation}
where $V(x,y)$ and $\sigma(x,y)$ are 2D distributions of velocity and velocity dispersion in the $(x,y)$ sky plane (see example 2D maps in the top row of Figure~\ref{2D_map_example}). $f(x,y,z)$ and $v_p(x,y,z)$ are 3D flux and projected velocity values at $(x,y,z)$ voxel. The LOS is defined as the z-axis in this coordinate system. 

Lastly, the orientations of the model are determined by position angles (PAs; PA$_{cone}$ and PA$_{disk}$) and inclinations ($i_{cone}$ and $i_{disk}$) of the bicone and the dust plane (or disk in Section~\ref{subsec:rotatinggas}). The orientations of the bicone and the dust plane are independent of each other, allowing them to rotate freely on the sky plane. However, for simplicity, we fix the PAs of the bicone and dust plane to 0$^\circ$ and 90$^\circ$, respectively, meaning that the bicone axis and the major axis of the plane extend along the north-south and east-west directions, respectively. 
Since our mock data analysis focuses on 3\arcsec-extracted kinematics and radial profiles of velocity widths, the effect of PAs is negligible.

Note that we do not consider various relative PA values between the disk and bicone for simplicity. We expect that the relative PA has no significant effect on mock SDSS data since we integrate all pixels within the 3\arcsec-aperture. In addition, velocity widths in mock IFU analysis are dominated by the bicone component in cases of strong outflows, thereby reducing the impact of the rotational orientation of the disk on the kinematics.
The effect of relative PAs on mock IFU data will be investigated in future work.

Using this biconical model, \citet{Bae16} tested the impact of each parameter on the velocity offset and dispersion, and the line profile of the [\OIII] emission line. Nevertheless, the model lacked several components crucial for comparison with observational data. For instance, it did not consider rotating gas in a galactic disk, which is thought to mainly contribute to the narrow/core component of the typical [\OIII] line profile. Furthermore, to accurately compare the spatial properties of the simulated outflows with observations, the seeing convolution should be taken into account. Therefore, we add a rotating gas component (Section~\ref{subsec:rotatinggas}) and the seeing effect (Section~\ref{subsec:seeing}) to the \citet{Bae16} model. In addition, to improve mock data compatibility, we switch from a coordinate system normalized by the outflow size to a physical scale in arcsec units based on the redshift and outflow size parameters (Section~\ref{subsec:mockdata}).

\subsection{Rotating Component}\label{subsec:rotatinggas}
While outflows are the dominant nongravitational component in AGN kinematics, the gas motions in the NLR are more complex due to additional dynamical effects. In particular,  gas motion due to the gravitational potential of the host galaxy is represented by a core component of narrow emission lines, which show similar line width and offset compared to stellar absorption lines \citep[$\sigma_{[\mathrm{OIII}], narrow}\sim \sigma_*$][]{Nelson96, Greene05}. If there is no nongravitational effect, gas kinematics are governed by a host galaxy potential, i.e., a rotating disk \citep[e.g., ][]{Luo19}.  
Given that a majority of AGNs in the VVD diagram (see Figure~\ref{VVD_model_grid}(a)) show relatively small velocity offset and dispersion, the disk component should be included to reproduce the observed distribution of the gas kinematics properly. Even for AGNs with strong outflows, 
a rotational pattern is often observed in IFU data \citep{Kang18, Kim23}. Thus, a disk component is required to reproduce the characteristics in the spatially resolved kinematics.

To generate disk models, we use the \texttt{GalPak$^\mathrm{3D}$} package \citep[Galaxy Parameters and Kinematics, ][]{Bouche15}, which was originally developed for the morphokinematics of 3D observational data. Since our goal is to investigate the effect of a rotating disk on the observed gas kinematics, we employ only the basic models generated by the package without adjusting details. Using the default models of \texttt{GalPak$^\mathrm{3D}$}, we assume radial and perpendicular flux profiles as exponential ($n=1$) and Gaussian, respectively, and set the rotation curve to a hyperbolic tan (tanh) profile.


We regulate two flux parameters ($f_{disk}/f_{cone}$ and $r_{1/2}$) and three kinematic parameters ($V_{max, rot}$, $r_t$, and $\sigma_0$) of the disk model. The inclination and PA of the disk are assumed to be the same as those of the dust plane in the outflow model. For the flux, the total disk flux within the SDSS fiber size (3\arcsec) is first determined by the disk-to-cone flux ratio parameter ($f_{disk}/f_{cone}$) and the integrated flux of the outflow. We then determine the disk flux distribution by assuming an exponential disk with a given half-light radius parameter ($r_{1/2}$).

For the gas kinematics, the maximum rotational velocity ($V_{max, rot}$) and the turnover radius ($r_t$) determine the velocity profile with a tanh function as follows:
\begin{equation}
v(r)=V_{max, rot}\;\mathrm{tanh}(r/r_t), 
\label{eq:disk_vel_profile}
\end{equation}
where $r$ is the radius on the galactic plane. At $r>r_t$, the rotational velocity begins to follow a flat radial profile. In addition, we adjust the intrinsic velocity dispersion ($\sigma_0$), which corresponds to the dynamical hotness of the disk. 

Once the five disk parameters and the disk orientation (PA$_{disk}$ and $i_{disk}$) are given, \texttt{GalPak$^\mathrm{3D}$} computes flux, velocity, and velocity dispersion maps. We then sum the 2D flux maps of the outflow and disk models. For velocity and velocity dispersion maps, we combine them by flux-weighted means. In Figure~\ref{2D_map_example}, the second row shows mock 2D maps of the disk model for an example with $i_{disk}\sim120^\circ$ at $z=0.05$. We assume a stellar mass of $\sim10^{10.5}\;\rm{M}_{\odot}$, which corresponds to the following parameters: $ V_{max,rot}=198$ \kms, $r_{1/2}=5.2$ kpc, $\sigma_0\sim$ 117 \kms, $r_t\sim$ 2.5 kpc. The third row of Figure~\ref{2D_map_example} shows the combined maps of the outflow and disk models with $f_{disk}/f_{cone}=1$. These combined maps demonstrate that the biconical structure is properly embedded in the background disk model.

\subsection{Seeing Effect}\label{subsec:seeing}

The beam smearing effect by atmospheric seeing has been pointed out as a significant obstacle in spatially resolved studies of AGNs \citep[e.g., ][]{Husemann16, Villar-Martin16, Singha22}. Given that the seeing effect is more pronounced at higher redshifts, we need to include it in our model to account for outflows in a majority of AGN host galaxies. 

\begin{figure*}[ht]
    \centering
    \includegraphics[width=0.90\textwidth]{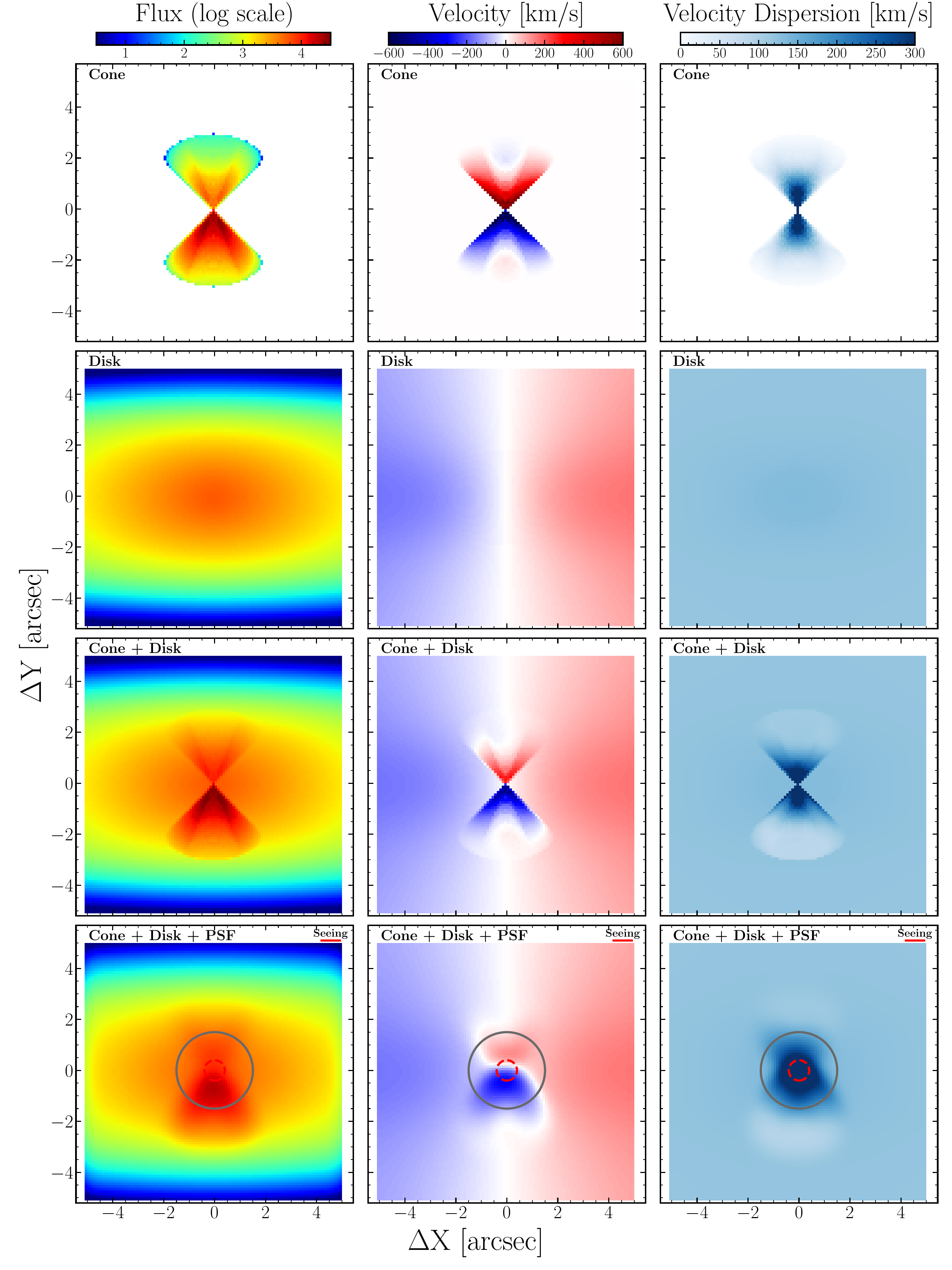}
   \caption{Examples of 2D maps produced by the modified model. From left to right, each column shows the flux, velocity, and velocity dispersion maps. 
   The first row presents the biconical outflow model of \citet{Bae16}.
   The second row shows the rotating disk model generated with \texttt{GalPak$^\mathrm{3D}$}. 
   The third row displays the combined bicone + disk model, and the last row presents the combined model after applying the seeing convolution. 
   Red dashed circles at the center, and red bars in the upper-right corner indicate the FWHM of the seeing. 
   Gray circles denote the 3\arcsec\ aperture corresponding to the SDSS fiber size.}
    \label{2D_map_example}
\end{figure*}

To reproduce more realistic 2D maps, we convolve the combined 2D maps with a Gaussian kernel as follows:
\begin{equation}
G(x,y)=\frac{1}{2\pi\sigma^2_g}\mathrm{exp}(-\frac{x^2+y^2}{2\sigma^2_g}) 
\label{eq:Gkernel}
\end{equation}
where $\sigma_g$ is a seeing size parameter ($\sigma_g=\mathrm{FWHM}/(2\sqrt{2\mathrm{ln}2})$). In this study, we adopt two cases of the seeing size, 0.8\arcsec$\;$ and 1.5\arcsec. The former is a typical seeing size of high resolution optical IFU data, while the latter is a median seeing size in SDSS. The convolved 2D maps are calculated as follows:

\begin{equation}
F_{c}(x,y) = F(x,y)*G(x,y),
\label{eq:fmap_conv}
\end{equation}

\begin{equation}
V_{c}(x,y) = \frac{(V(x,y)\times F(x,y))*G(x,y)}{F_{c}(x,y)},
\label{eq:vmap_conv}
\end{equation}

\begin{equation}
    \begin{split}
    &\sigma_{c}^2(x,y) \\ 
     = & \frac{((V^2(x,y)+\sigma^2(x,y))\times F(x,y))*G(x,y)}{F_{c}(x,y)} \\
      & -V^2_{c}(x,y) 
    \end{split}
\label{eq:dmap_conv}
\end{equation}
where $F_c(x,y), V_c(x,y)$, and $\sigma_c(x,y)$ are convolved flux, velocity, and velocity dispersion, respectively. In the bottom row of Figure~\ref{2D_map_example}, we show an example of convolved 2D maps. The red horizontal bar on the upper right represents the applied seeing size in this example (FWHM $=0.8$\arcsec). The seeing convolution blurs the central bicone structure, so the final images are more like a circular shape in all 2D maps. We will discuss the seeing effect in Section~\ref{subsec:outflowsizes} and \ref{subsec:seeingeff}, focusing on implications for the outflow size measurements in IFU data.

\subsection{Mock Data Compatibility}\label{subsec:mockdata}
 
We update our model to calculate 2D maps in an (x, y) coordinate system defined in arcsec units based on the input redshift and physical size of the outflow, while the \citet{Bae16} model adopted relative scales normalized by the outflow size.
This approach offers several advantages: (1) the outflow model uses the same pixel units as the disk model, making it convenient to combine the two models; (2) mock data can be directly compared with real observational data. Accordingly, we specify the mock data by pixel size and field-of-view (FoV) parameters, and all calculations are performed using this pixel scale from given redshift and physical length parameters ($R_{out}$, $r_{1/2}$, and $r_t$).

\section{Mock Data Analysis} \label{sec:mockdata}

In this Section, we describe our main analysis procedures based on mock data generated from the modified outflow model. First, we generate mock SDSS data extracted from the central 3\arcsec-aperture, and compare them with the observed [\OIII] kinematics of a large sample of SDSS type 2 AGNs in Section~\ref{subsec:mockSDSS}. In Section~\ref{subsec:mockIFU}, we produce mock IFU data to analyze outflow sizes and 2D distributions of flux, velocity, and velocity dispersion. 

\subsection{Mock SDSS Data}\label{subsec:mockSDSS}

\citet{Bae16} integrated the [\OIII] velocity offset and dispersion within the outflow radius and compared them with the observed VVD distribution of $\sim$39,000 SDSS type 2 AGNs from \citet{Woo16}. Here, we use our modified model to reexamine the comparison with the observed VVD distribution. 

To simulate SDSS spectra, we calculate the integrated velocity offset and dispersion ($v_{int}$, $\sigma_{int}$) within the central 3\arcsec-aperture shown as gray circles in Figure~\ref{2D_map_example}. 
In this example ($z=0.05$ and $R_{out}$=3 kpc), the projected size of the bicone exceeds the SDSS fiber region, so its outer part is not included in the mock SDSS data. While this rarely occurs for models with low redshifts and large outflow sizes, it provides a representative example of how modifications affect the mock data.

As we integrate the kinematics within an aperture twice as large as the FWHM of the seeing (1.5\arcsec), the seeing convolution does not significantly impact the VVD distributions. On the other hand, the rotating gas component influences the integrated velocity offset and dispersion since it has a velocity offset near zero and a narrow width with comparable flux to the outflow component. Thus, we mainly investigate the effect of the rotating gas component on the VVD distribution.

\subsubsection{Model Grids for the Observed VVD}\label{subsubsec:obsVVD}
To investigate the effect of outflow properties on the observed kinematics, we overlay various model grids on the observed VVD diagram. We mainly investigate five parameters: the dust extinction factor ($A$), opening angles ($\theta_{out}$ and $\theta_{in}$), bicone inclination ($i_{cone}$), launching velocity ($V_{max}$), and disk-to-cone flux ratio ($f_{disk}/f_{cone}$). We fix the disk inclination as 60$^\circ$, the outflow size as 1 kpc, and the stellar mass as $10^{10.5}\;\mathrm{M}_{\odot}$. 

We first compare the models with and without the rotating disk component. Then we adjust the bicone inclination while keeping the opening angle fixed, and vice versa. We change the bicone inclination from -40$^\circ$ to 0$^\circ$ and the outer half opening angle from 30$^\circ$ to 60$^\circ$ with a 10$^\circ$ step. For each model grid, we also adjust the dust extinction rate from 0 to 100 \% (i.e., $A$ = 1 to 0) and launching velocities by 500, 1000, and 2000 \kms. 

\subsubsection{Mock VVD from Monte Carlo Simulation}\label{subsubsec:MCsimul}

We perform Monte Carlo (MC) simulations and generate mock VVD distributions by assuming a probability distribution function (PDF) of each parameter. The primary purpose of these MC simulations is to identify the best PDFs of the launching velocity and opening angle parameters that reproduce the observed VVD distribution, and ultimately, to constrain these outflow properties for the type 2 AGN sample.

\begin{figure*}[ht]
    \includegraphics[width=0.24\textwidth]{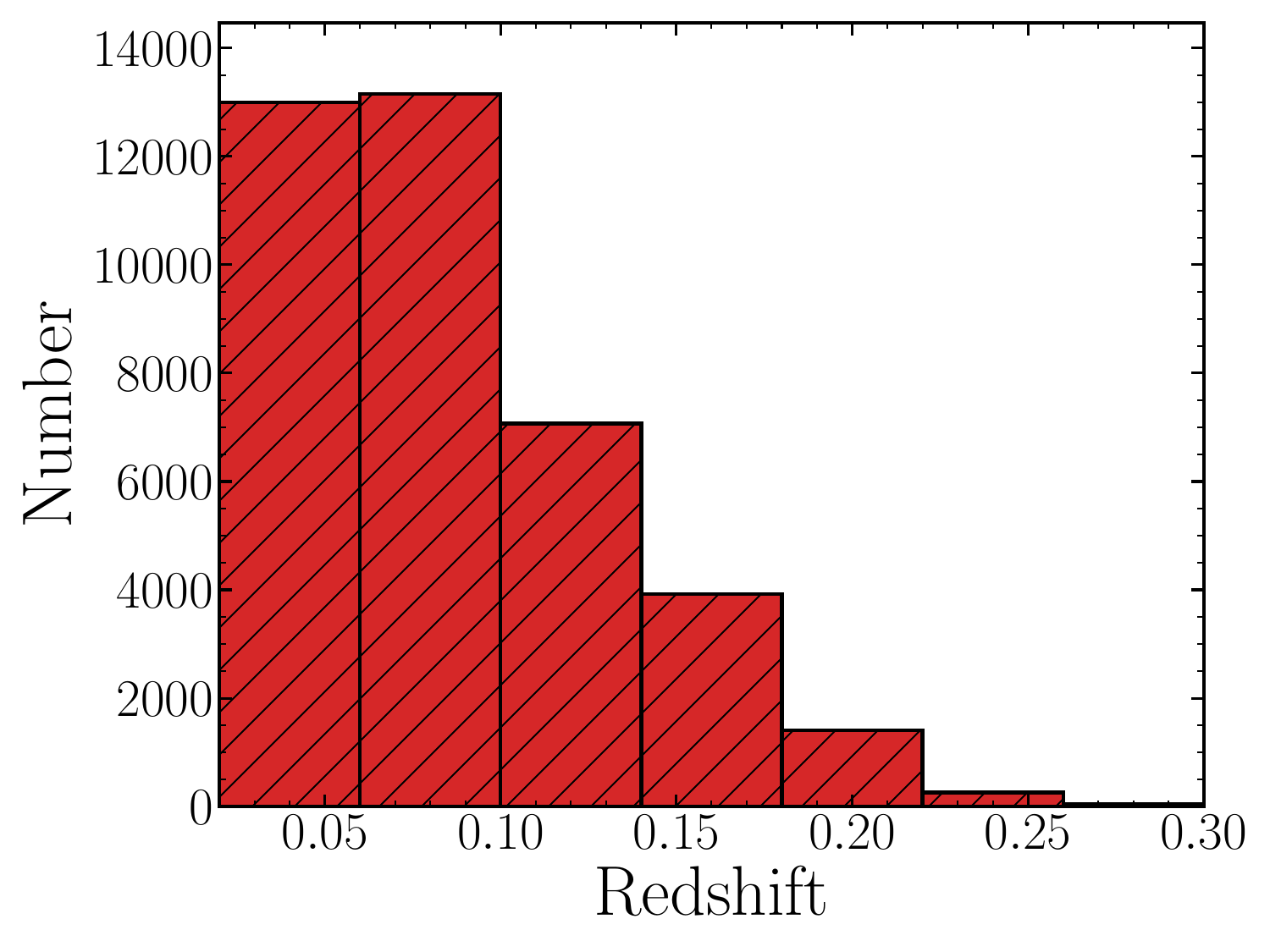}
    \includegraphics[width=0.24\textwidth]{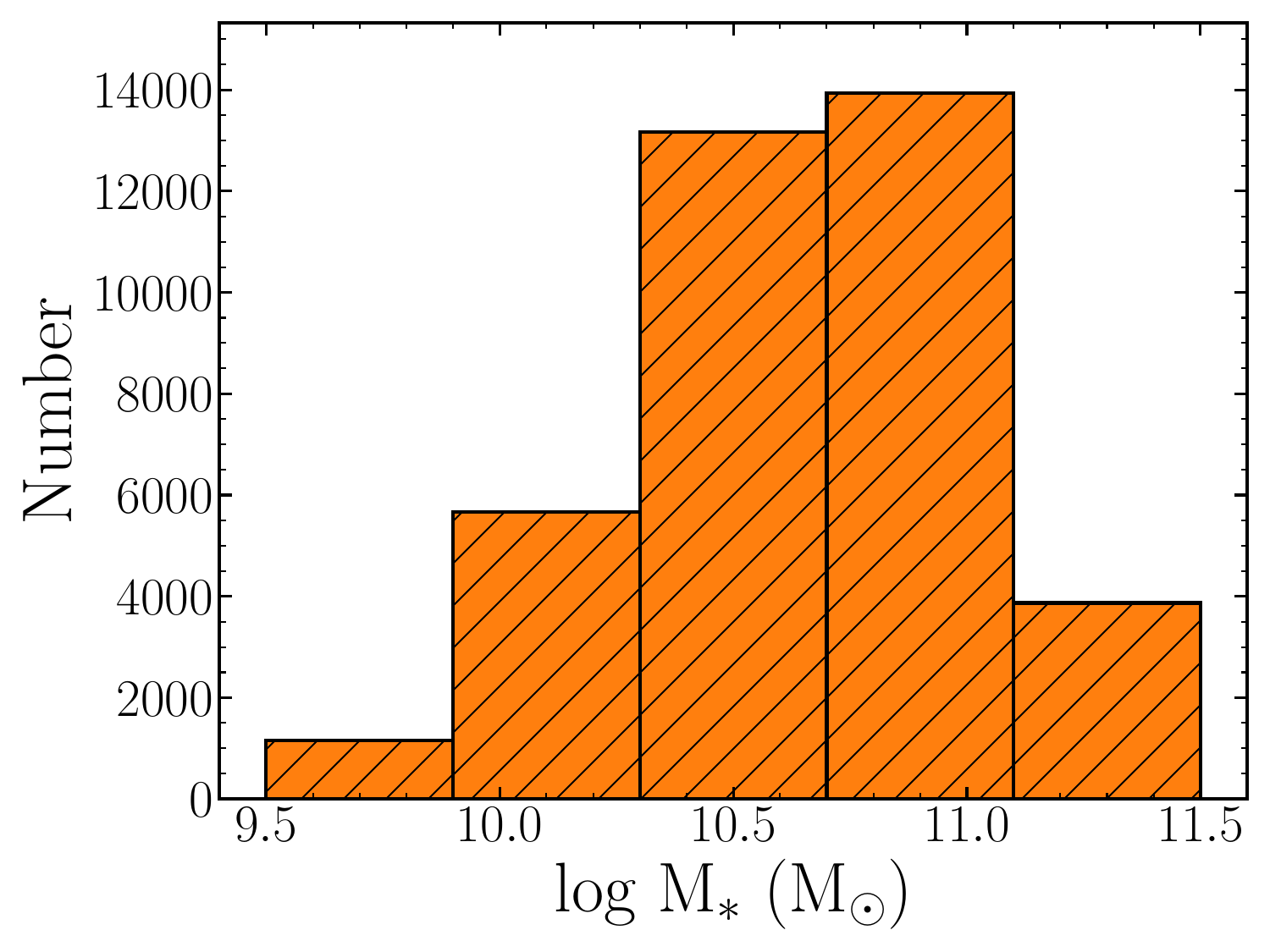}
    \includegraphics[width=0.24\textwidth]{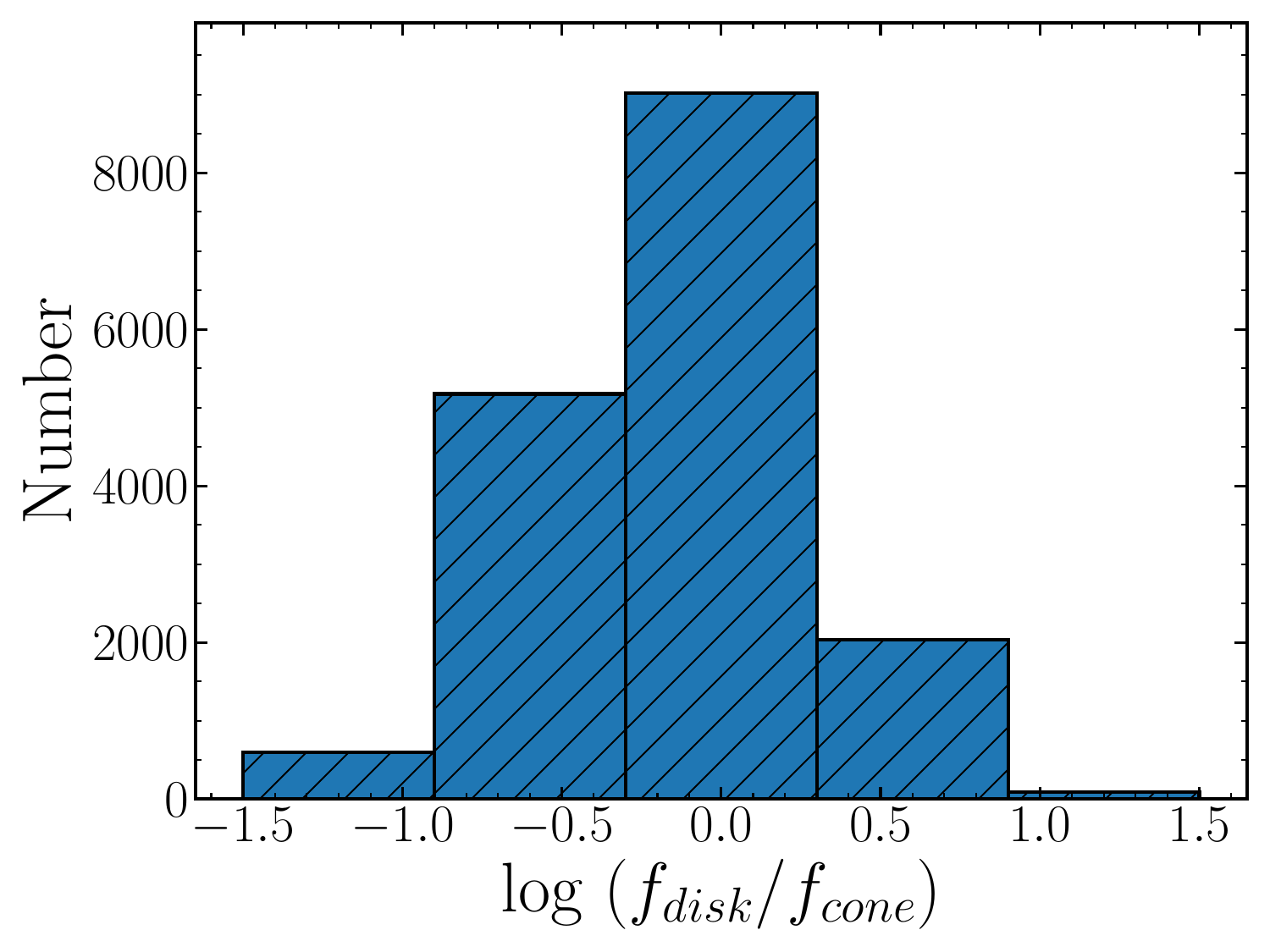}
    \includegraphics[width=0.24\textwidth]{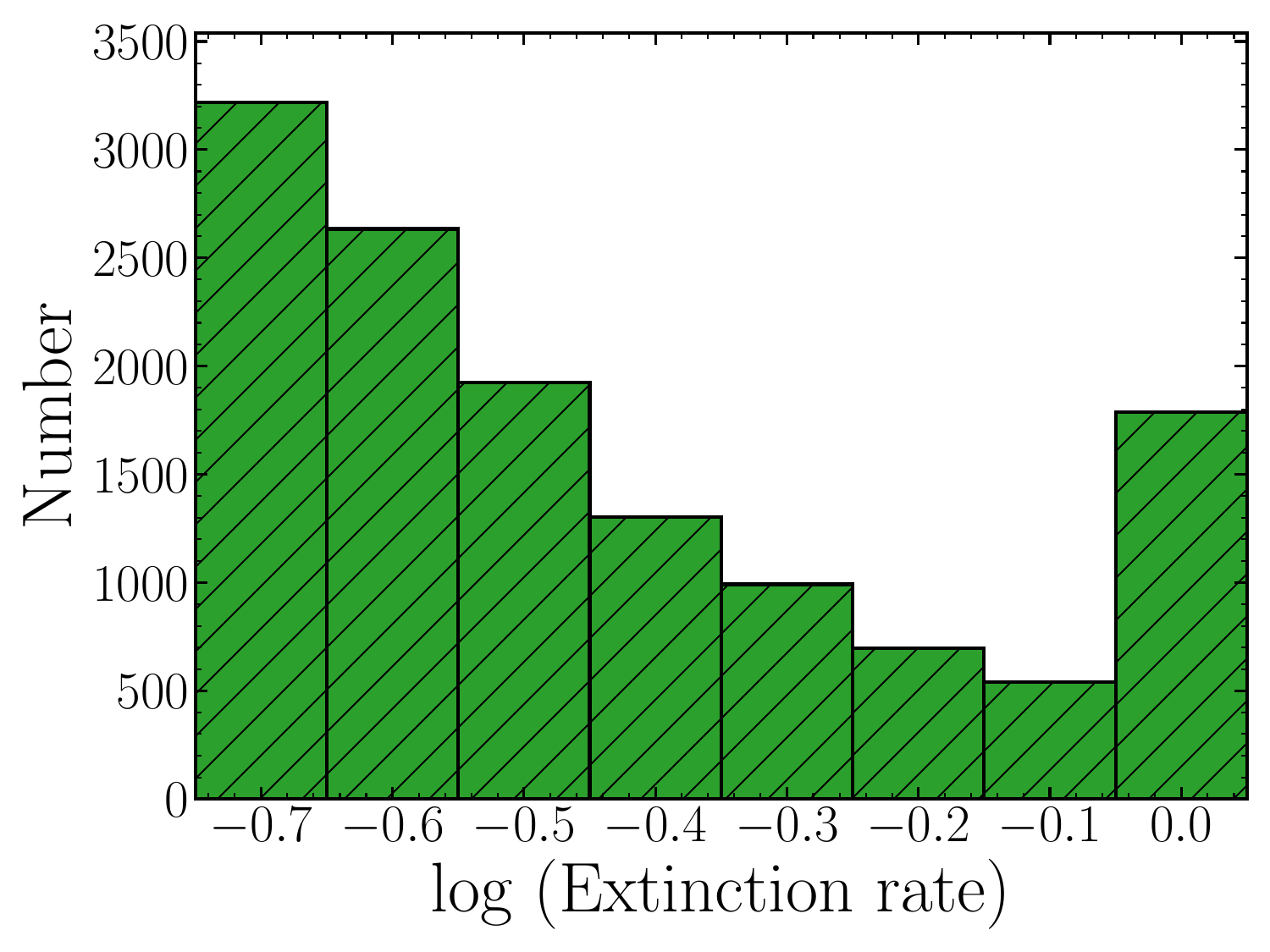}
    \caption{Histograms of physical properties of the type 2 AGN sample from \citet{Woo16}. From left to right, the panels show the distributions of redshift, stellar mass, flux ratio between disk and outflow components, and extinction factor derived from the Balmer decrement.}
    \label{Woo16_properties_dist}
\end{figure*}

We first take PDFs of various parameters, such as redshift, stellar mass, disk-to-cone flux ratio, and extinction, from the observed SDSS type 2 AGNs (see Figure~\ref{Woo16_properties_dist}). The PDF of redshift (0.04 to 0.28 with 0.04 steps) is directly applied to the MC simulations. The observed disk-to-cone flux ratio is determined by the flux ratio between broad/wing and narrow/core components when the [\OIII] line is fitted with a double Gaussian. If the [\OIII] line is fitted with a single Gaussian, we classify the emission as either disk-only or outflow-only based on the difference between stellar and [\OIII] velocity dispersion. Specifically, when $|\sigma_\mathrm{[OIII]}-\sigma_*|$ is less than 30 \kms, we classify this single Gaussian as disk-only emission, i.e., $f_{disk}/f_{cone}=\infty$, otherwise as outflow-only emission, i.e., $f_{disk}/f_{cone}=0$. We also estimate the PDF of the dust extinction factor by using observed H$\alpha$/H$\beta$ with $R^A_V=4.5$ \citep[][]{Vogt13}. Note that we restrict the dust extinction rate to smaller than 80\% (or $A>0.2$) since a higher extinction rate results in too broad mock VVD distributions along the V$_\mathrm{[OIII]}$ axis, which are inconsistent with the observed distribution. Furthermore, an extinction rate higher than 80\% is an extreme case, corresponding to $A_{\lambda 5007} \gtrsim $ 1.75 mag.

The stellar mass is taken from the MPA-JHU catalog \footnote{\url{https://wwwmpa.mpa-garching.mpg.de/SDSS/}} and divided into five bins ($\rm{log}\;M_*=$9.7 to 11.3 with 0.4 step).
We calculate the four remaining disk parameters, i.e., $r_{1/2}$ \citep{Dutton11}, $V_{max, rot}$ \citep{Torres-Flores11}, $\sigma_0$ \citep{Oh22}, and $r_t$ \citep{Amorisco10}, using empirical relations with stellar mass. Thus, once the stellar mass is selected among the five values, these four disk parameters are determined accordingly. Note that we restrict our sample to type 2 AGNs in the MC simulations due to the large uncertainties in stellar mass for type 1 AGNs.

For the outflow model, we first adopt random inclinations and fixed PAs (PA$_{cone}$=0$^\circ$ and PA$_{disk}$=90$^\circ$) for both the bicone and dust plane. The inclination of the dust plane is given from 7.5$^\circ$ to 172.5$^\circ$ with a 15$^\circ$ step, and its PDF is proportional to the solid angle of the normal vector integrated from 0$^\circ$ to 360$^\circ$ of PA. 
Since we focus on type 2 AGNs, we restrict the bicone inclination to less than 40$^\circ$ (7.5$^\circ$, 22.5$^\circ$, and 37.5$^\circ$), and its PDF is also proportional to the solid angle of the bicone axis. 
Models with higher bicone inclinations, i.e., more aligned bicone axis with the LOS, are likely to be observed as type 1 AGNs, as their accretion disk and broad line region (BLR) are readily exposed to the observer \citep[e.g., ][]{Marin14}.
Then we assume that the PDF for five outflow size values (0.3, 1, 3, 6, 9 kpc) follows a Gaussian distribution with a mean of zero and a standard deviation of 2 kpc. Note that the outflow size has little effect on the kinematics in mock SDSS data.

Our main goal is to constrain two parameters, i.e., the launching velocity and opening angle of the outflow model. For these two parameters, we assume that their PDFs follow a truncated Gaussian function. We first set the range of the launching velocity to $[100,\;\infty)$ \kms\ and restrict the outer opening angle values from 30$^\circ$ to 70$^\circ$ with a 10$^\circ$ step based on \citet{Fischer13}. Note that larger outer opening angles are excluded due to their physically extreme geometry, while smaller opening angles are excluded since they do not improve the main results. We then run multiple MC simulations by changing the mean and standard deviation of the Gaussian to find the best $V_{max}$ and $\theta_{out}$ distributions that generate the best-matching mock VVD with the observed VVD. Each MC simulation generates 100,000 pairs of velocity offset and dispersion.

\subsubsection{Kullback-Leibler Divergence and $D_{max}$}\label{subsubsec:KLdiv_Dmax}

To quantitatively compare the mock VVDs with the observed VVD, we apply two statistical quantities that assess proximity between two 2D distributions. We first calculate Kullback-Leibler (KL) divergence \citep[$D_{KL}(P||Q)$, ][]{KLDIV} as follows:
\begin{equation}
D_{KL}(P||Q) = \sum P(v_{obs}, \sigma_{obs})\;\mathrm{log}\frac{P(v_{obs}, \sigma_{obs})}{Q(v_{int}, \sigma_{int})},
\label{eq:KLDiv}
\end{equation}
where $P(v_{obs}, \sigma_{obs})$ and $Q(v_{int}, \sigma_{int})$ are probability distributions of the observed and mock VVD, respectively. In brief, the KL divergence is a measure of a statistical distance between two PDFs derived from information theory, i.e., the smaller the KL divergence is, the more similar the two PDFs are.

The second quantity is the $D_{max}$ from the 2D Kolmogorov-Smirnov test, calculated from the public code \textsc{ndtest}\footnote{Written by Zhaozhou Li, \url{https://github.com/syrte/ndtest}}. Like the 1D Kolmogorov-Smirnov test, it calculates the maximum absolute difference between two cumulative distribution functions ($D_{max}$) and performs a p-value test of whether two PDFs originate from the same distribution. Although all p-values are too small to conclude that the observed and mock VVDs are identical distributions, we utilize the $D_{max}$ value to compare the relative superiority between the mock VVDs.

Note that we add random error smoothing to all mock VVDs based on typical SDSS uncertainties of velocity offset and dispersion to account for the observational uncertainties. To remove variations induced by this error smoothing, we repeat this process 100 times in each grid of $V_{max}$ and $\theta_{out}$ PDFs and take the mean and standard deviation of KL divergence and $D_{max}$ values. If the mean KL divergence (or $D_{max}$) for given parameter distributions is notably smaller than for the others, by more than its standard deviation, then that mock VVD distribution is considered a better reproduction of the observed data.

\subsection{Mock IFU Data}\label{subsec:mockIFU}
We simulate mock 2D maps using various redshifts and outflow sizes to test the validity of previous methods for outflow size measurements. In particular, we focus on the dependence of the measured sizes on the seeing size normalized by the angular size of the outflow.

In our previous IFU studies, we developed a kinematic approach to measure the outflow size and applied it to AGNs at $z<0.3$ \citep{Karouzos16a, Kang18, Bae17, Luo21, Kim23}. The outflow region is defined as where the [\OIII] velocity dispersion is more enhanced than stellar velocity dispersion, which traces the gravitational potential of the host galaxies. Specifically, we first plot a radial profile of ratios between the [\OIII] and stellar velocity dispersion ($\sigma_\mathrm{[OIII]}/\sigma_*$) and define the outflow size as a radius where the ratio becomes unity. Then we quadratically subtract the seeing size ($\sigma_g$) from the outflow size to correct the seeing effect.

We apply the same procedure to mock IFU data and compare the measured outflow size with the input outflow size. Since we do not have stellar velocity dispersion in our model, we alternatively adopt the [\OIII] velocity dispersion extracted within a 3\arcsec-aperture of the convolved disk-only maps. This velocity dispersion of the disk model mimics the [\OIII] narrow/core component, which is an optimal proxy of stellar velocity dispersion \citep[e.g.,][]{Greene05}.

Another widely used method for outflow size measurements is W$_{80}$-based method \citep[e.g.,][]{Harrison14, Wylezalek20, Ruschel-Dutra21}, where W$_{80}$ is defined as a [\OIII] line width including 80\% of the total line flux. Similar to the $\sigma_\mathrm{[OIII]}$-based method, the W$_{80}$-based method defines the outflow region where the W$_{80}$ is broader than a certain criterion, frequently set to 600 \kms. Thus, the outflow size is determined as the radius of the furthest pixel from the center with W$_{80}$ larger than 600 \kms.

To perform the W$_{80}$-based analysis to mock IFU data, we create mock 3D cube data by adding a spectral axis. We assume that the bicone and disk components are observed as individual Gaussians that have the same flux, velocity, and velocity dispersion at each position in the convolved 2D maps. As a result, we obtain the double Gaussian line profile of each pixel, consisting of a narrow/core component from the convolved disk map and a broad/wing component from the convolved bicone map. Then we calculate W$_{80}$ values of all pixels and determine the outflow size as the radius of the furthest pixel from the center with W$_{80} >$ 600 \kms.

\section{Result}\label{sec:result}

\subsection{Location of Model Grids in VVD Diagram}\label{subsec:VVD}

\begin{figure*}[]
    \includegraphics[width=0.98\textwidth]{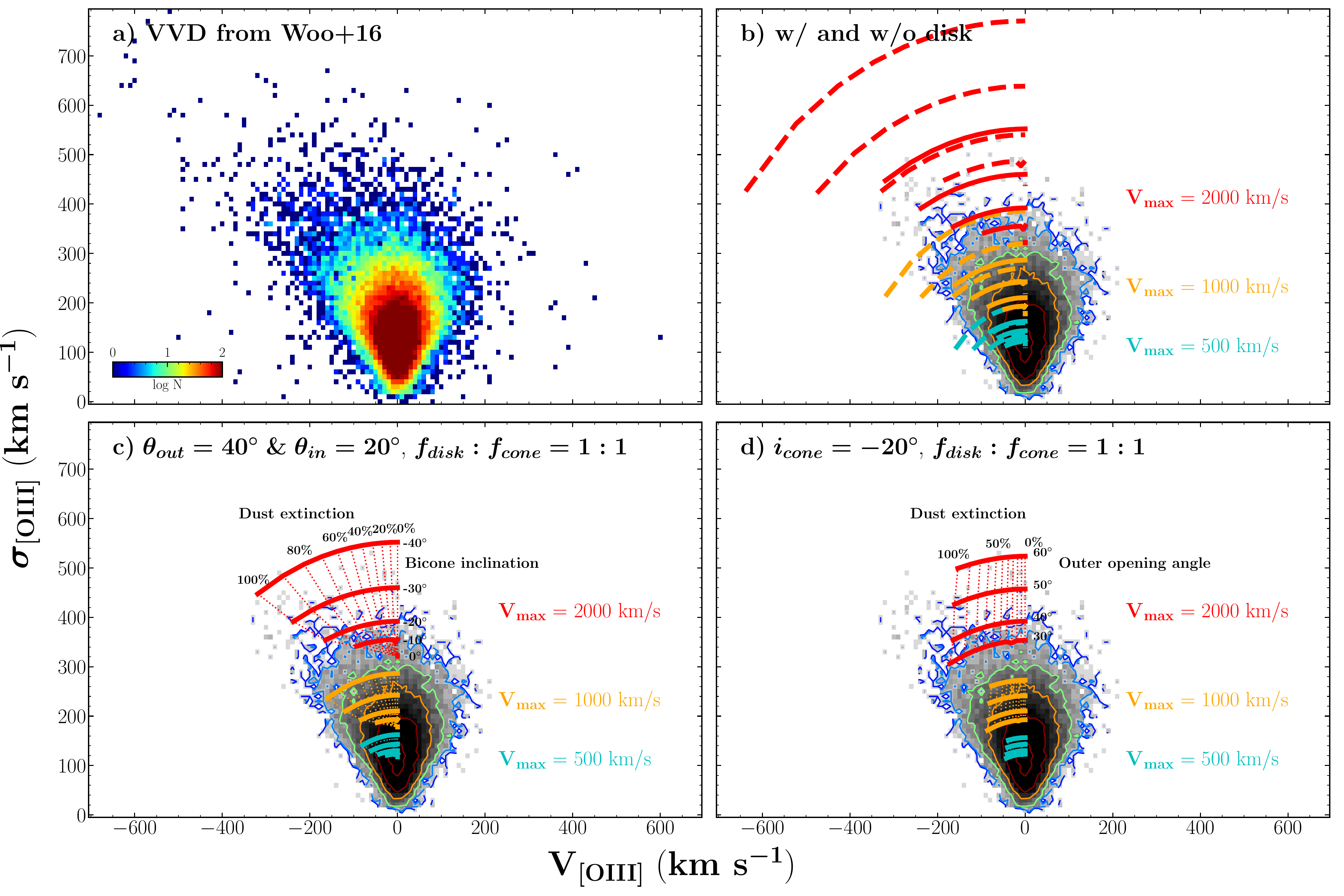}
    \caption{
Model grids with a fixed launching velocity of 500 km s$^{-1}$ (cyan), 1000 km s$^{-1}$ (yellow), and 2000 km s$^{-1}$ (red).
(a) Observed [\OIII] VVD distribution of 39,000 type 2 AGNs from \citet{Woo16}; the color scale indicates the number of AGNs in each bin. 
This distribution is also shown as contours and gray heat maps in panels (b)–(d).  
(b) Model grids with (solid lines) and without (dashed lines) the disk component. 
Each line traces the change in kinematics as a function of the extinction at a fixed inclination.  
(c) Model grids showing the effects of bicone inclination and extinction when the disk component is included. 
Solid lines indicate variations with extinction at a fixed inclination, while dotted lines indicate variations with inclination at a fixed extinction.  
(d) Model grids with different opening angles and extinction values. 
For clarity, solid and dotted lines denote models with fixed opening angles and fixed extinction values, respectively.
}
    \label{VVD_model_grid}
\end{figure*}

We investigate the dependency of outflow parameters on the location in the VVD diagram by comparing model grids with observations. 
The observed VVD distribution of local type 2 AGNs (Figure~\ref{VVD_model_grid}(a)) is characterized by a V-shaped envelope and a larger number of blueshifted than redshifted [\OIII] sources \citep{Woo16}.

First, we examine the influence of the rotating disk component by comparing model grids with and without disk emission (Figure~\ref{VVD_model_grid}(b)). 
Including the disk component reduces both the integrated velocity offset and dispersion ($v_{\rm int}$ and $\sigma_{\rm int}$), because the disk typically exhibits smaller velocity offsets and dispersions than the outflow. 
As a result, the combined model requires a higher launching velocity up to 1500 \kms\ to reproduce the observed VVD distribution, whereas the original outflow-only model \citep{Bae16} could explain most AGNs with $V_{\rm max}$ up to 1000 \kms\ (orange dashed lines).

Second, we revisit the effects of bicone inclination and dust extinction (Figure~\ref{VVD_model_grid}(c)). 
Both $v_{\rm int}$ and $\sigma_{\rm int}$ increase with larger inclination angles, consistent with \citet{Bae16}. 
Increasing extinction decreases $\sigma_{\rm int}$ but increases $v_{\rm int}$, owing to the reduced contribution from the receding cone. 
Notably, the $|v_{\rm int}|/\sigma_{\rm int}$ ratio is generally smaller than in the outflow-only model because the disk component has zero intrinsic velocity offset.

Lastly, we investigate the effect of the opening angle (Figure~\ref{VVD_model_grid}(d)). 
A wider opening angle produces a larger $\sigma_{\rm int}$ due to the broader distribution of LOS velocities, while $v_{\rm int}$ remains nearly constant. 
Overall, the dependence on opening angle is consistent with the trends reported by \citet{Bae16}.

\subsection{Constraints on $V_{max}$ and $\theta_{out}$ Distributions}\label{subsec:VVDMC}

We constrain the distributions of $V_{\rm max}$ and $\theta_{\rm out}$ by comparing the simulated VVD distributions with the observed VVD distribution. 
We perform multiple MC realizations, adopting Gaussian PDFs in which the mean and standard deviation of $V_{\rm max}$ range from 150 to 450 km s$^{-1}$ and from 250 to 450 km s$^{-1}$, respectively, in steps of 50 km s$^{-1}$. 
For $\theta_{\rm out}$, we use mean values of $30^\circ$, $40^\circ$, and $50^\circ$, and standard deviations of $5^\circ$, $10^\circ$, and $20^\circ$. 
We note that the adopted means and standard deviations do not correspond exactly to those of the resulting PDFs because the distributions are truncated Gaussians.
We then calculate 4D arrays containing 315 values ($7 \times 5 \times 3 \times 3$) for the KL divergence and $D_{max}$, and identify the best PDFs where these values are minimized.
As a result, KL divergence is minimized when the launching velocity follows ($\mu_V, \sigma_V$)=(250 \kms, 400 \kms) and the opening angle follows ($\mu_\theta, \sigma_\theta$)=(30$^\circ$, 10$^\circ$). $D_{max}$ exhibits a minimum value when the launching velocity follows ($\mu_V , \sigma_V$)=(350 \kms, 300 \kms) and the opening angle follows ($\mu_\theta, \sigma_\theta$)=(30$^\circ$, 5$^\circ$).

\begin{figure*}[t]
    \centering
    \includegraphics[width=0.49\textwidth]{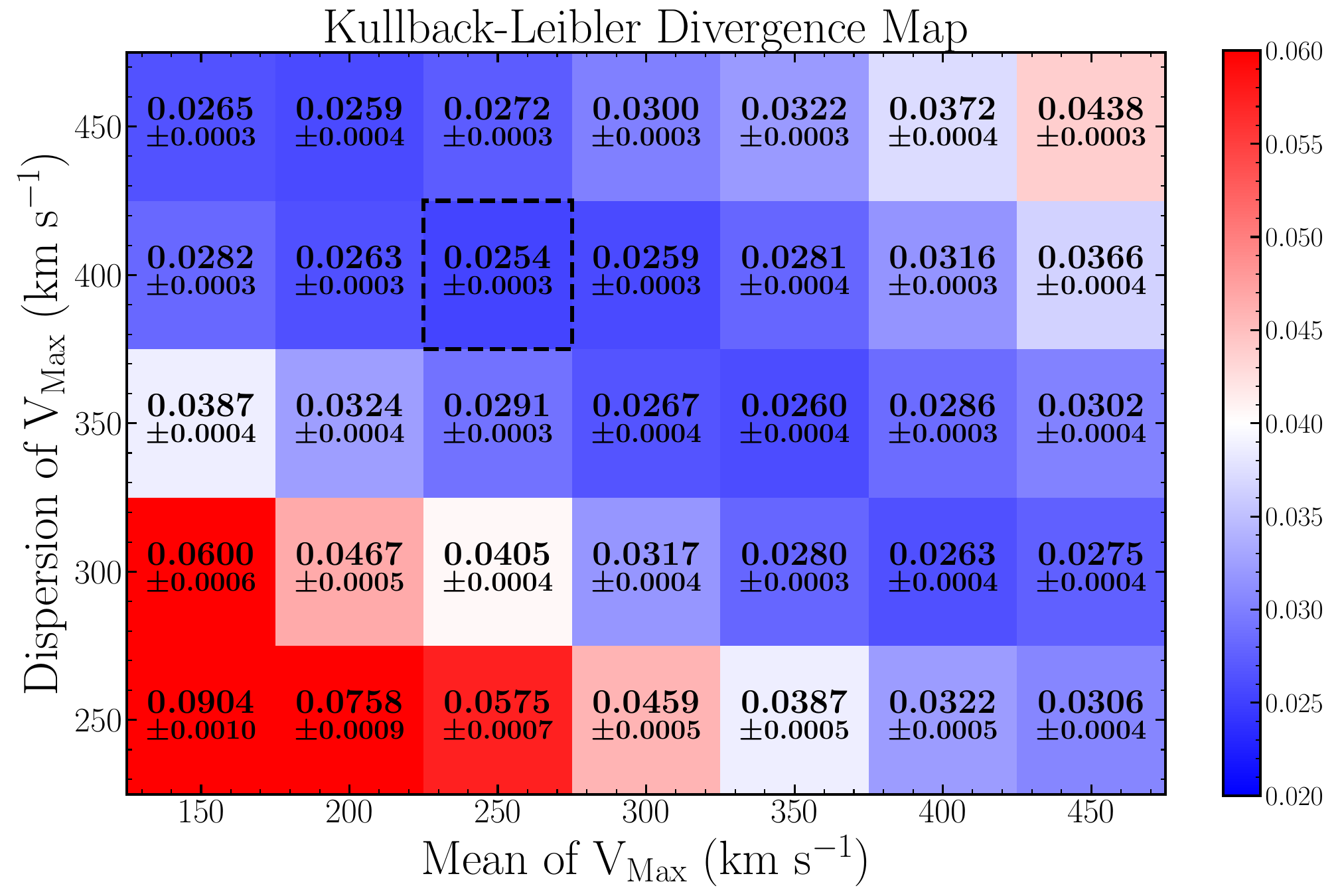}
    \includegraphics[width=0.49\textwidth]{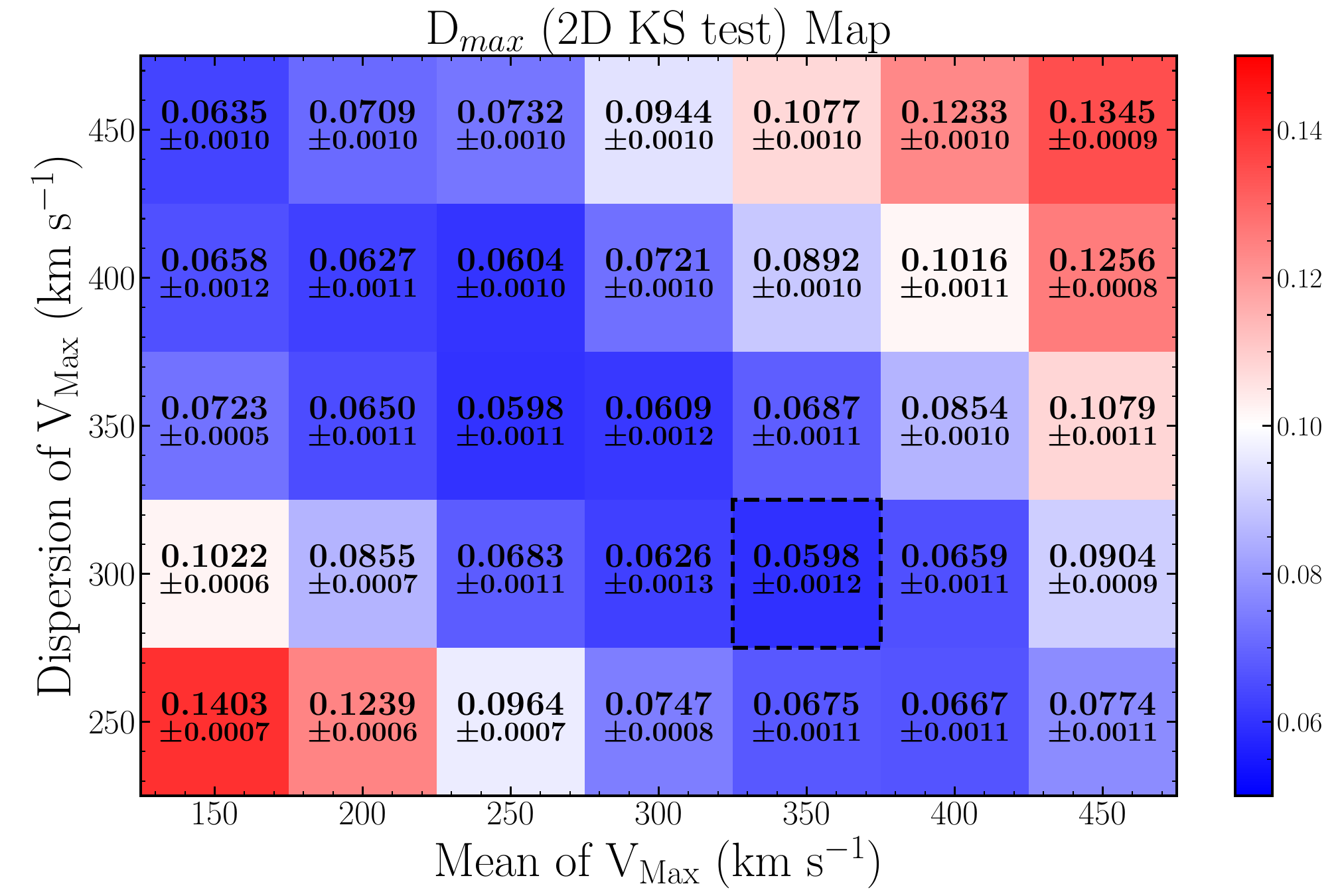}
    \caption{
Left: KL divergence values for differently assumed PDFs of the launching velocity $V_{\rm max}$, while keeping the opening angle distribution fixed to its best-fit value.  
The Gaussian parameters (mean and standard deviation) for $V_{\rm max}$ are varied and truncated below 100 km s$^{-1}$.  
The minimum KL divergence is found at a mean of 250 km s$^{-1}$ and a standard deviation of 400 km s$^{-1}$.  
Right: same plot but showing $D_{\rm max}$ values instead of KL divergence.  
In this case, the minimum occurs at a mean of 350 km s$^{-1}$ and a standard deviation of 300 km s$^{-1}$.  
In each panel, the location of the lowest value is highlighted by a dashed-line box.
}
    \label{Vmax_KLDiv_D_max_maps}
\end{figure*}

\begin{figure*}[]
    \centering
    \includegraphics[width=0.49\textwidth]{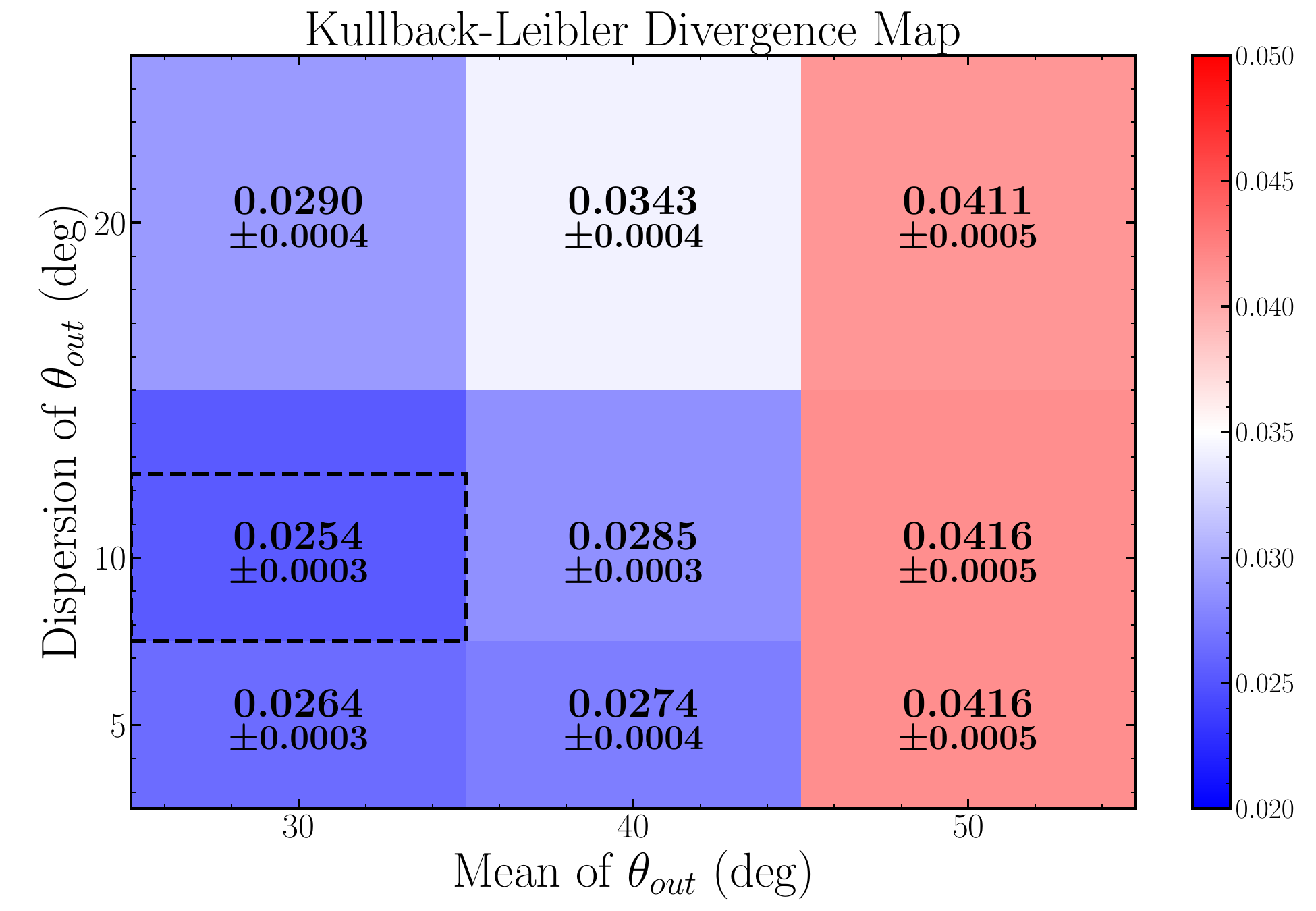}
    \includegraphics[width=0.49\textwidth]{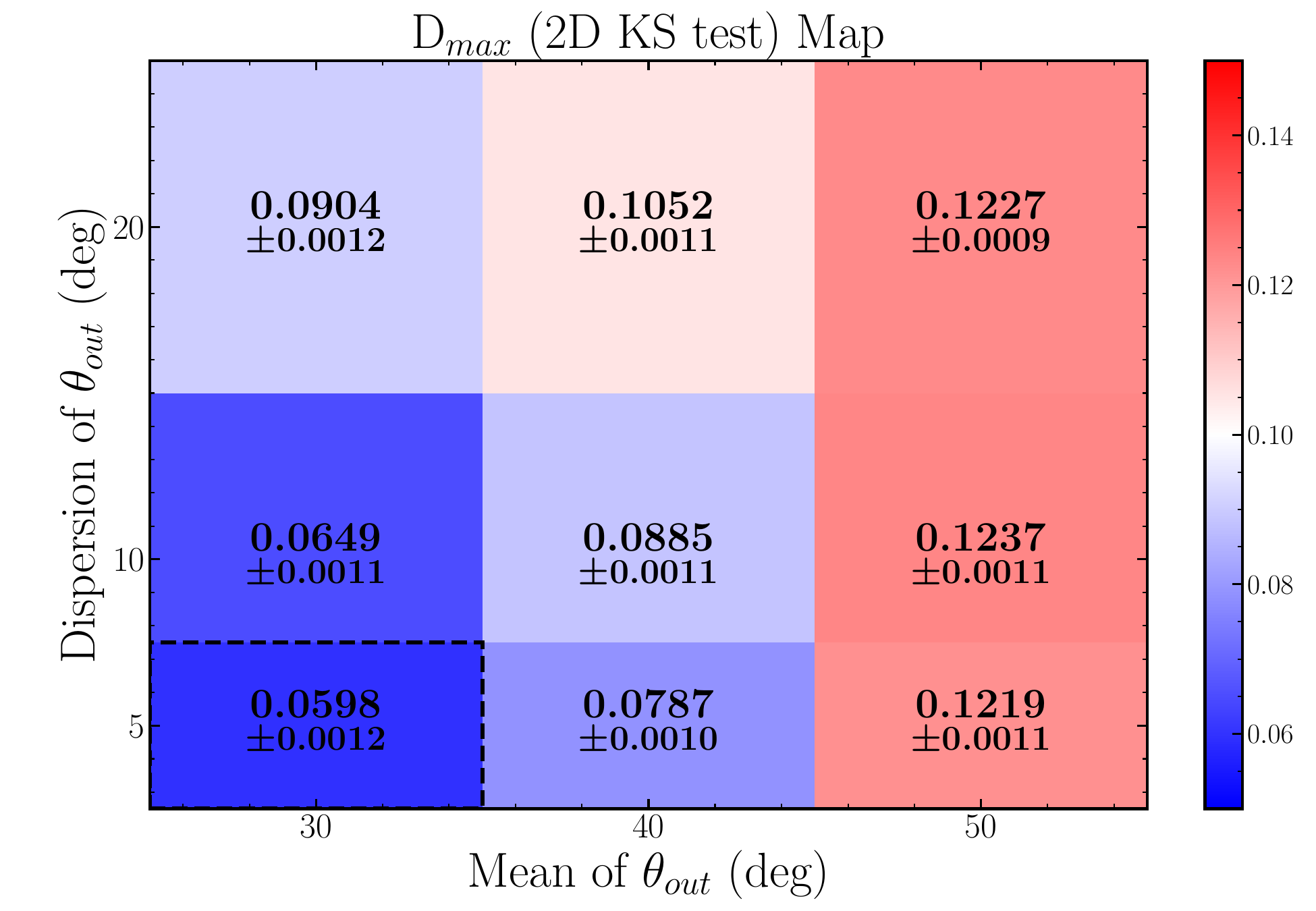}
    \caption{
Left: KL divergence values computed for differently assumed PDFs of the opening angle $\theta_{\rm out}$.  The minimum KL divergence is found when the mean and standard deviation are $30^\circ$ and $10^\circ$, respectively.  
Right: Same as left, but showing the $D_{\rm max}$ values from the 2D KS test.  The minimum occurs at a mean of $30^\circ$ and a standard deviation of $5^\circ$. 
In each panel, the location of the lowest value is highlighted by a dashed-line box.
}
    \label{Oangle1_KLDiv_D_max_maps}
\end{figure*}

We first investigate variations in the KL divergence and $D_{max}$ values depending on the input launching velocity and outer opening angle PDFs in Figures~\ref{Vmax_KLDiv_D_max_maps} and \ref{Oangle1_KLDiv_D_max_maps}.
For visualization, we collapse the 4D arrays into 2D maps by fixing the other parameter to its best PDF.
The KL divergence and $D_{max}$ maps of the launching velocity show local minima near the center, ruling out the lower left and upper right corners. This indicates that the observed VVD distribution generally prefers a moderate range of launching velocity. On the other hand, the opening angle PDF favors small values (i.e., lower left corner regions in Figure~\ref{Oangle1_KLDiv_D_max_maps}). 

\begin{figure*}[]
    \centering
    \includegraphics[width=0.49\textwidth]{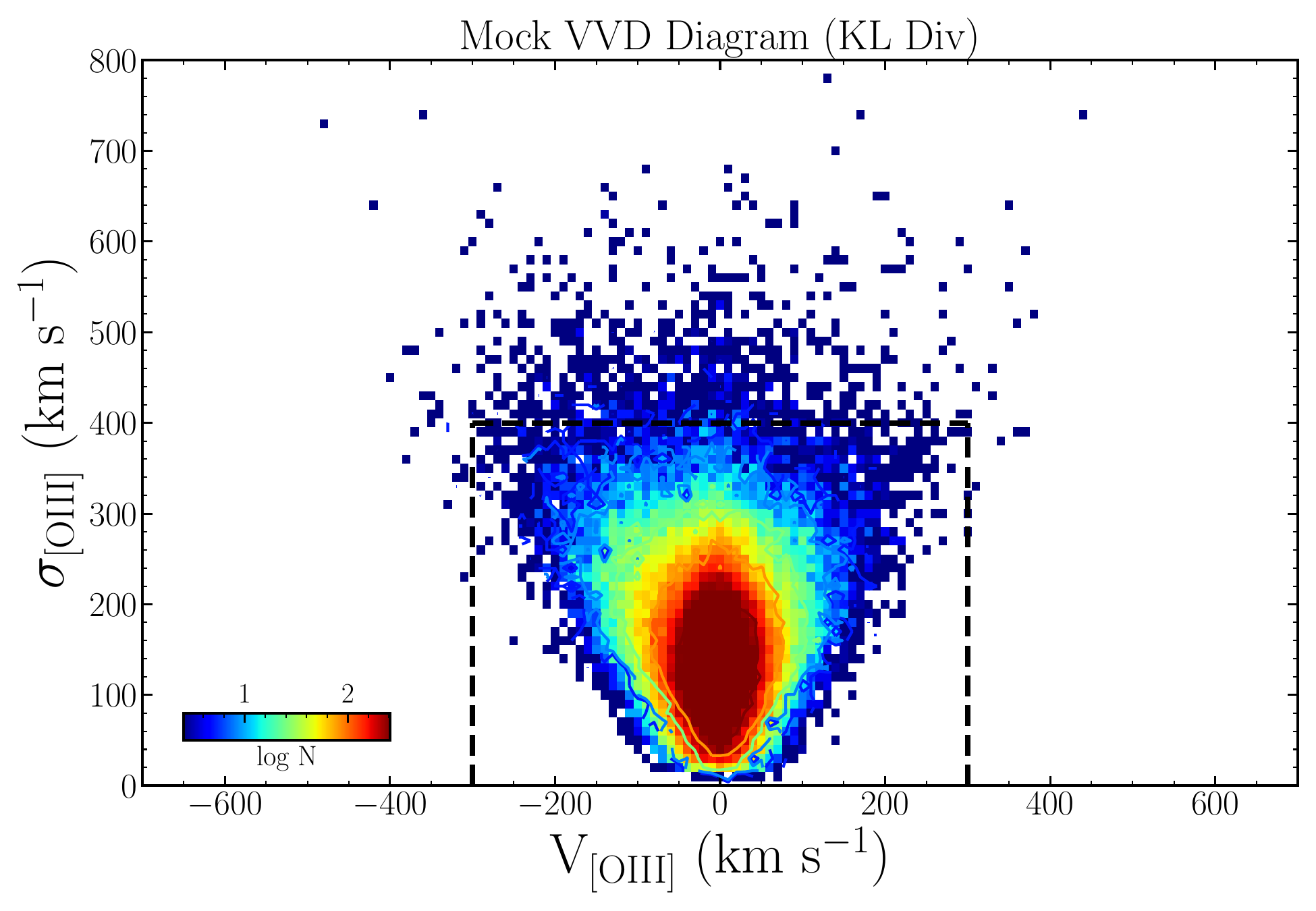}
    \includegraphics[width=0.49\textwidth]{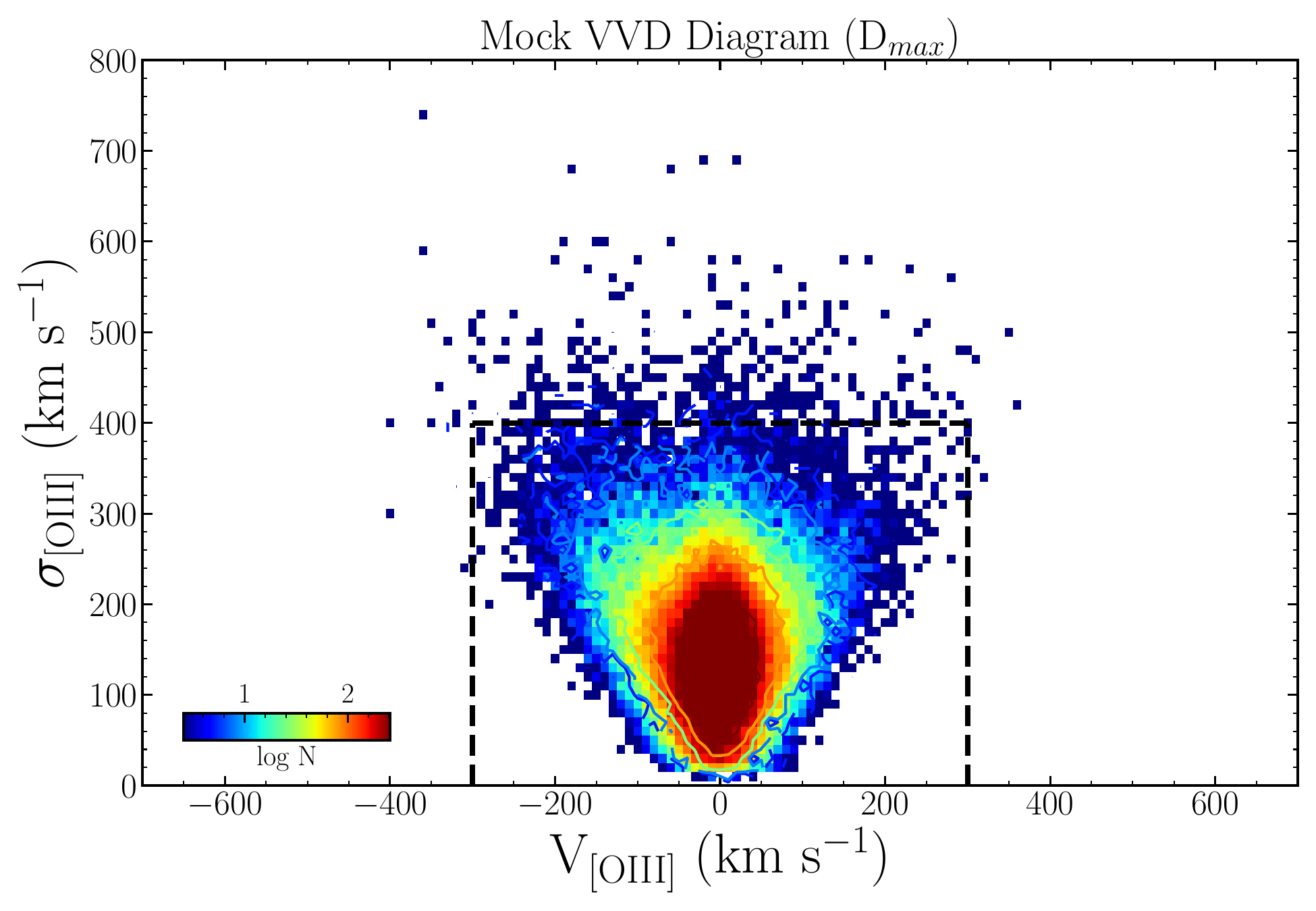}
\caption{
Best-match mock VVD distributions selected based on KL divergence (left) and $D_{\rm max}$ (right).  
The contours from the observed VVD distribution are overlaid for comparison.  
The dashed-box outlines the selection criterion for “strong outflow” AGNs, defined as $|\mathrm{V}_{\rm [OIII]}| > 300\ \mathrm{km\ s^{-1}}$ or $\sigma_{\rm [OIII]} > 400\ \mathrm{km\ s^{-1}}$.
}
    \label{Best_mock_VVDs}
\end{figure*}

We present the best mock VVD distributions based on KL divergence (left) and $D_{max}$ (right) in Figure~\ref{Best_mock_VVDs}.
The two best mock VVD distributions are generally consistent with the observed one, reproducing both the V-shape feature and the overall range of the V$_\mathrm{[OIII]}$ and $\sigma_\mathrm{[OIII]}$. The best mock VVD distribution based on KL divergence exhibits a slightly larger $\sigma_\mathrm{[OIII]}$ range than $D_{max}$. Despite the extinction limit ($A\geq0.2$), the V-shape envelope of the mock VVD distribution is slightly broader than the observed one, especially in the low $\sigma_\mathrm{[OIII]}$ regime. This may result from several factors, such as large observational uncertainties for weak outflow AGNs.

\begin{figure*}[t]
    \centering
    \includegraphics[width=0.49\textwidth]{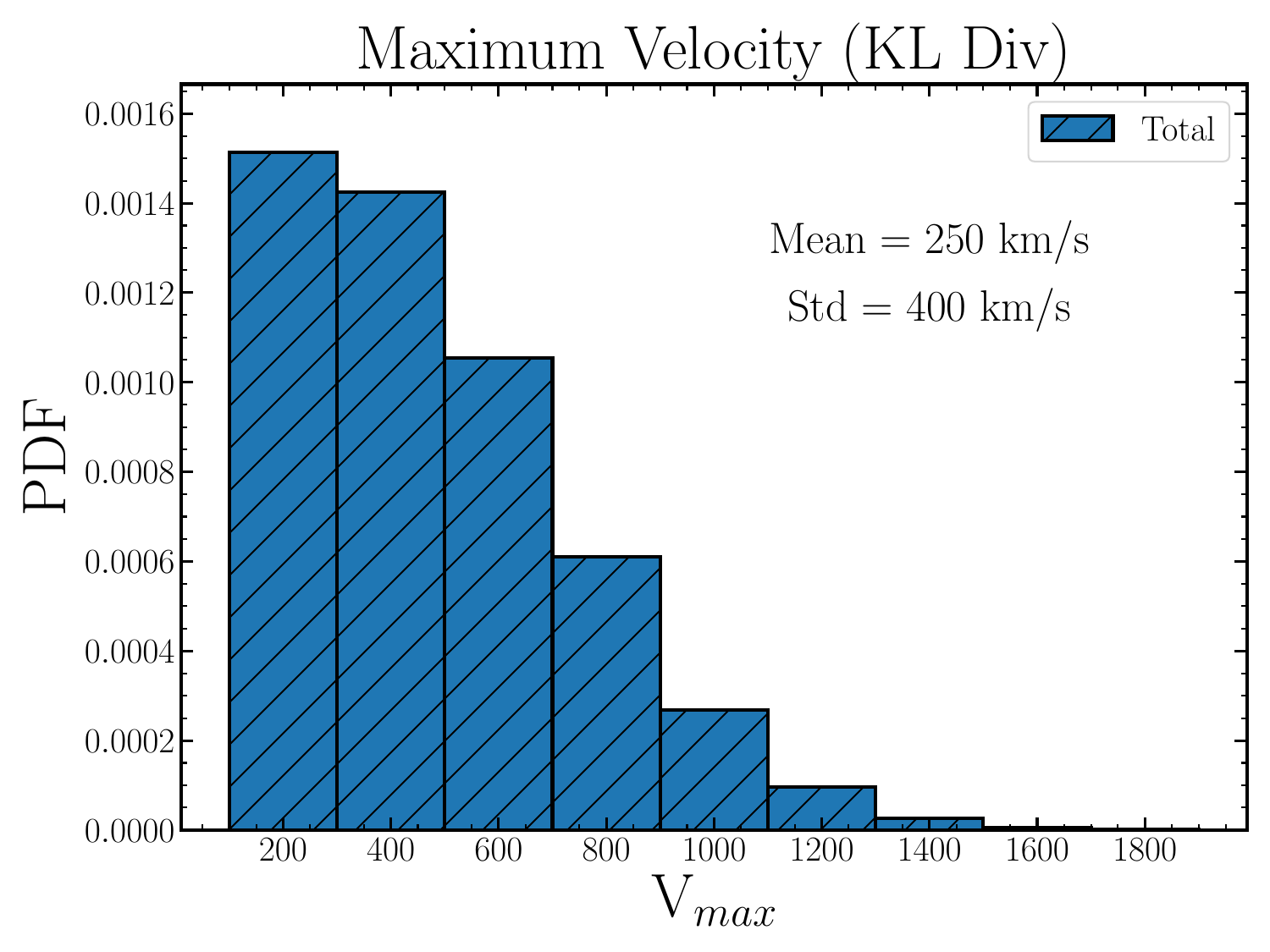}
    \includegraphics[width=0.49\textwidth]{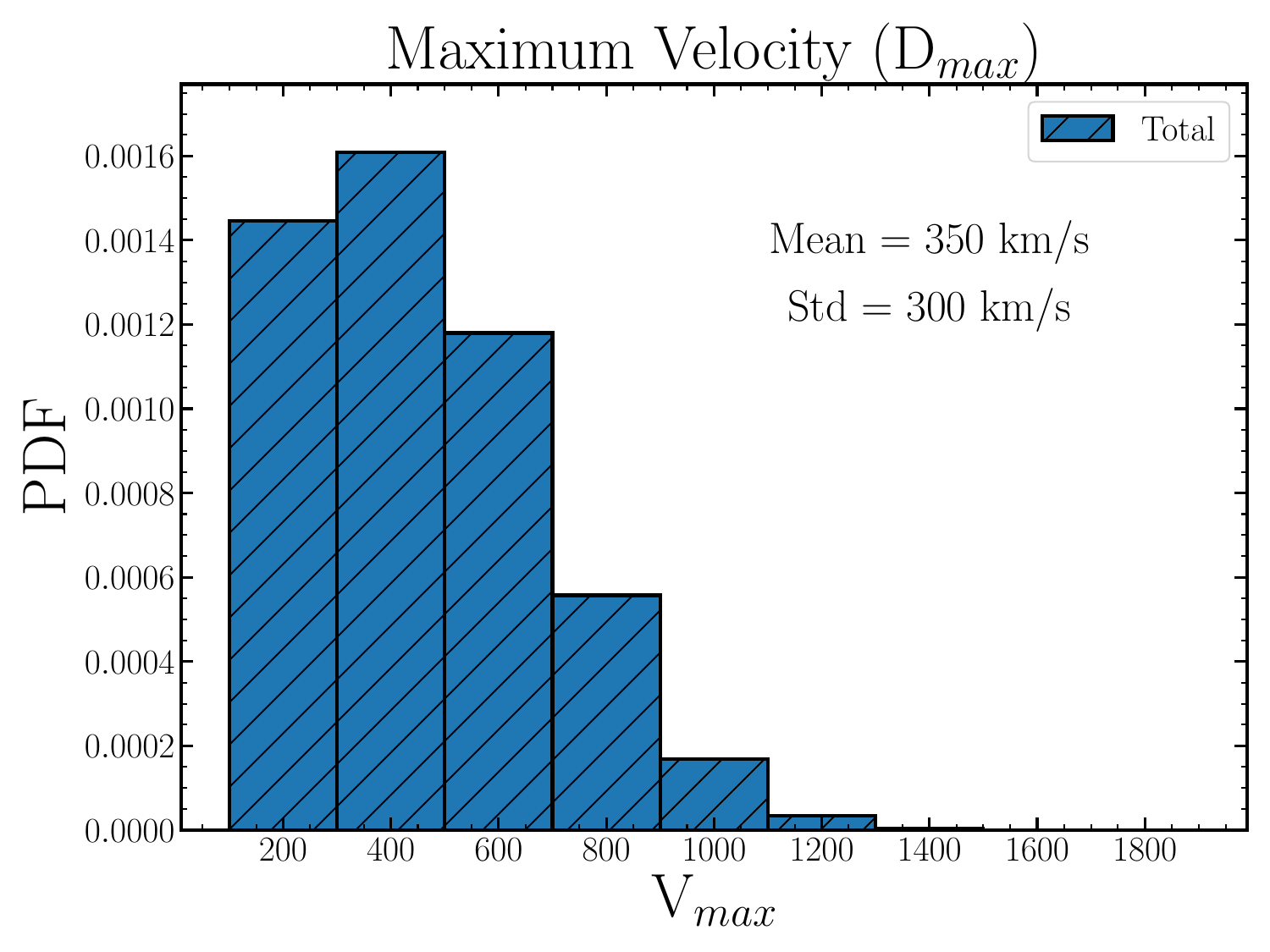}
    \caption{Best-fitting PDFs of the launching velocity based on the KL divergence (left) and $D_{max}$ (right). Histograms show the adopted Gaussian PDFs for the full mock VVD distributions with truncation at 100 km s$^{-1}$. The mean and standard deviation of each Gaussian are indicated in their respective panels.
}
    \label{Vmax_dist_best}
\end{figure*}

\begin{figure*}[]
    \centering
    \includegraphics[width=0.49\textwidth]{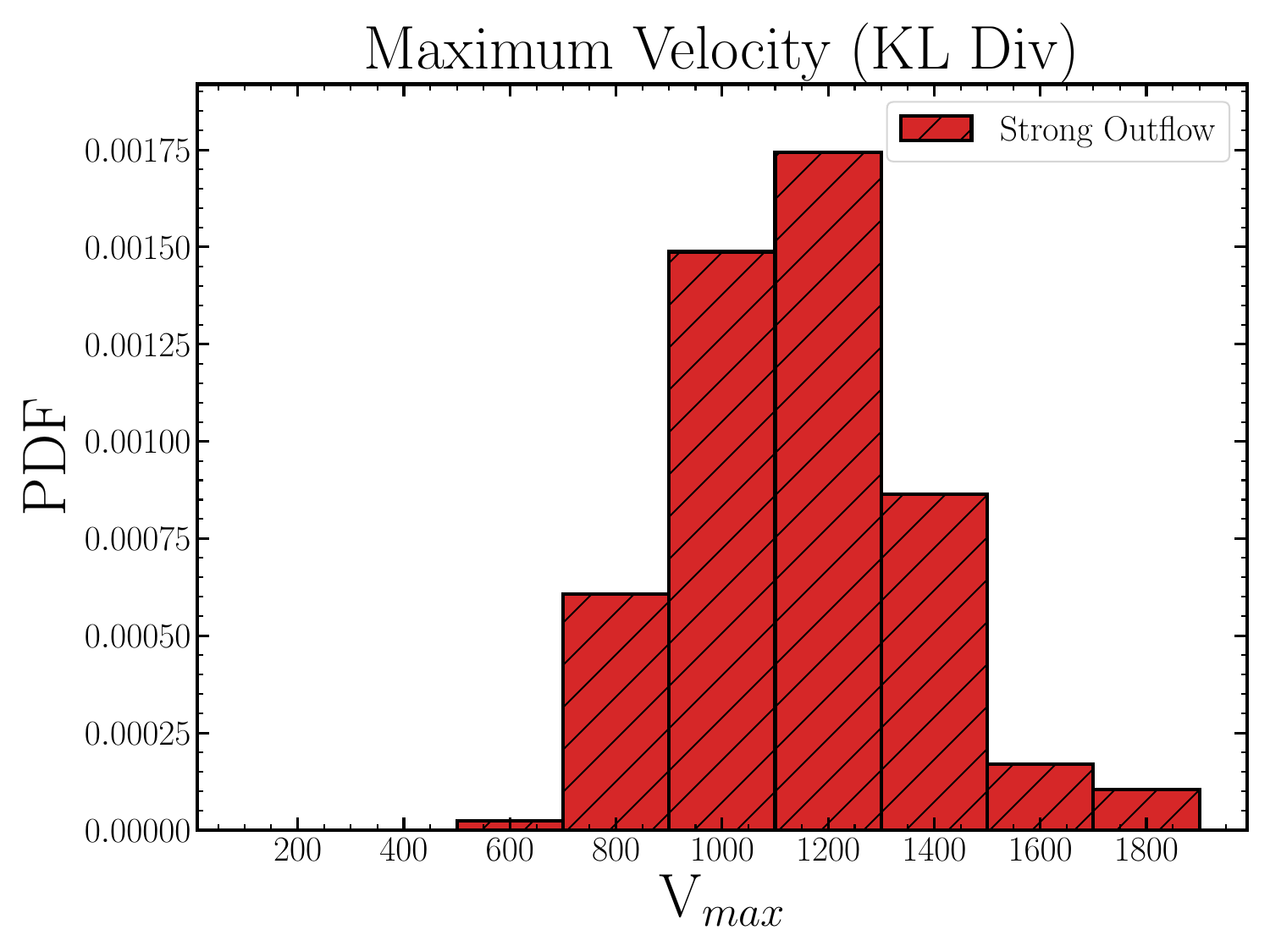}
    \includegraphics[width=0.49\textwidth]{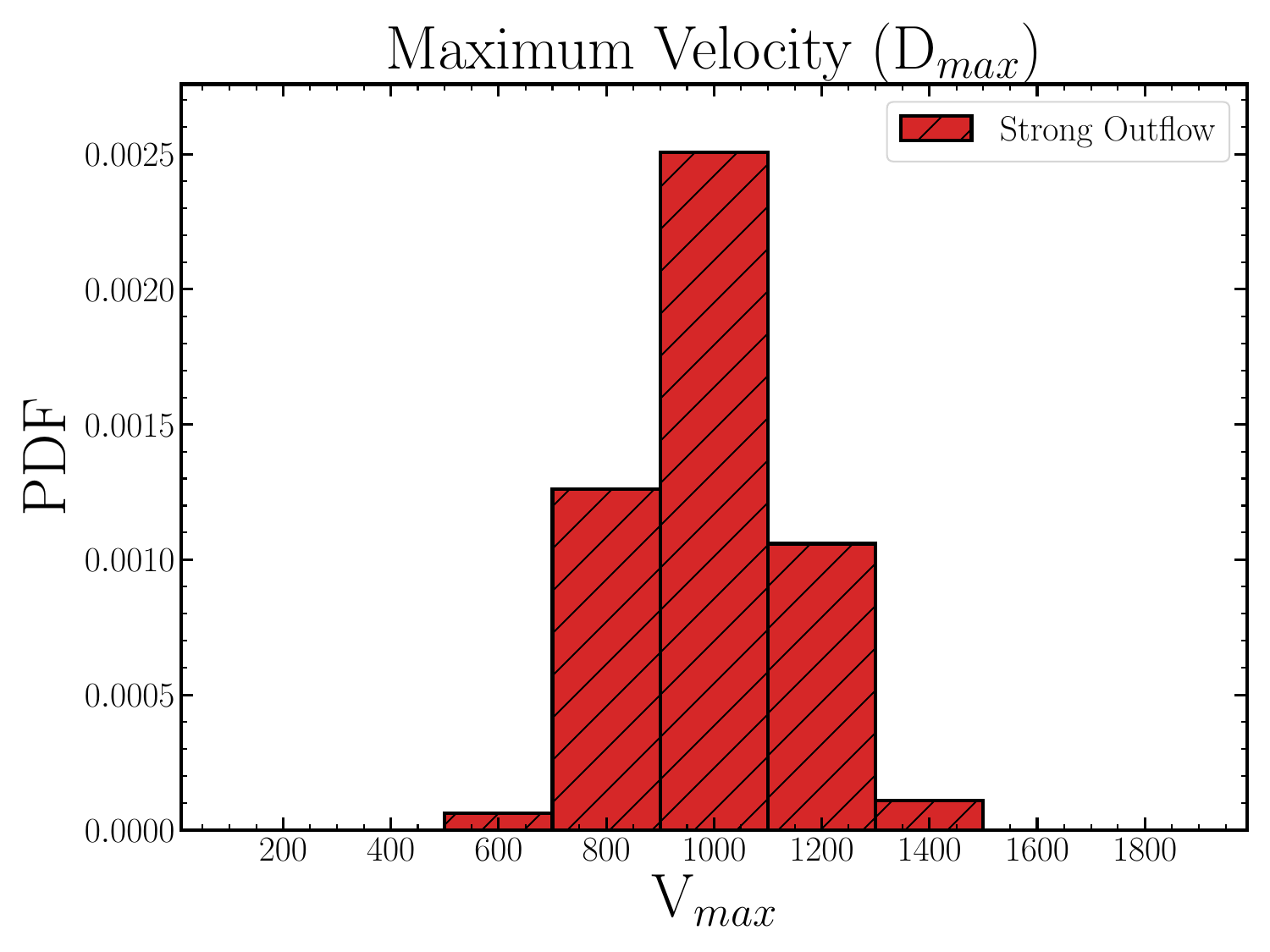}
    \caption{
Distributions of the launching velocity for a subsample of strong-outflow AGNs (those lying outside the dashed boxes in Figure~\ref{Best_mock_VVDs}).  
Histograms are derived from the best-matching mock VVD distributions based on  KL divergence (left) and $D_{\rm max}$ (right).  
}
    \label{Vmax_dist_strong}
\end{figure*}

\begin{figure*}[]
    \centering
    \includegraphics[width=0.49\textwidth]{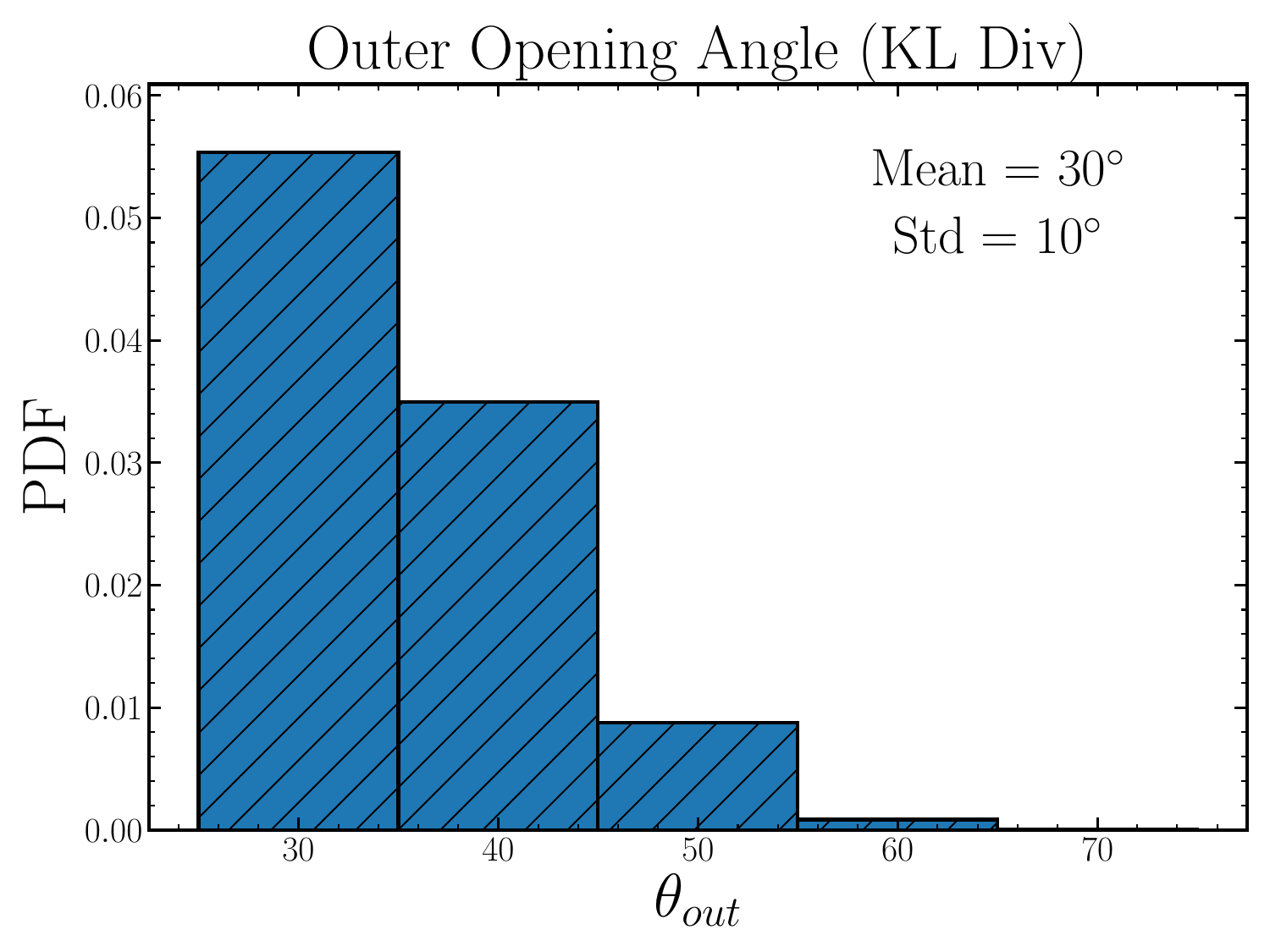}
    \includegraphics[width=0.49\textwidth]{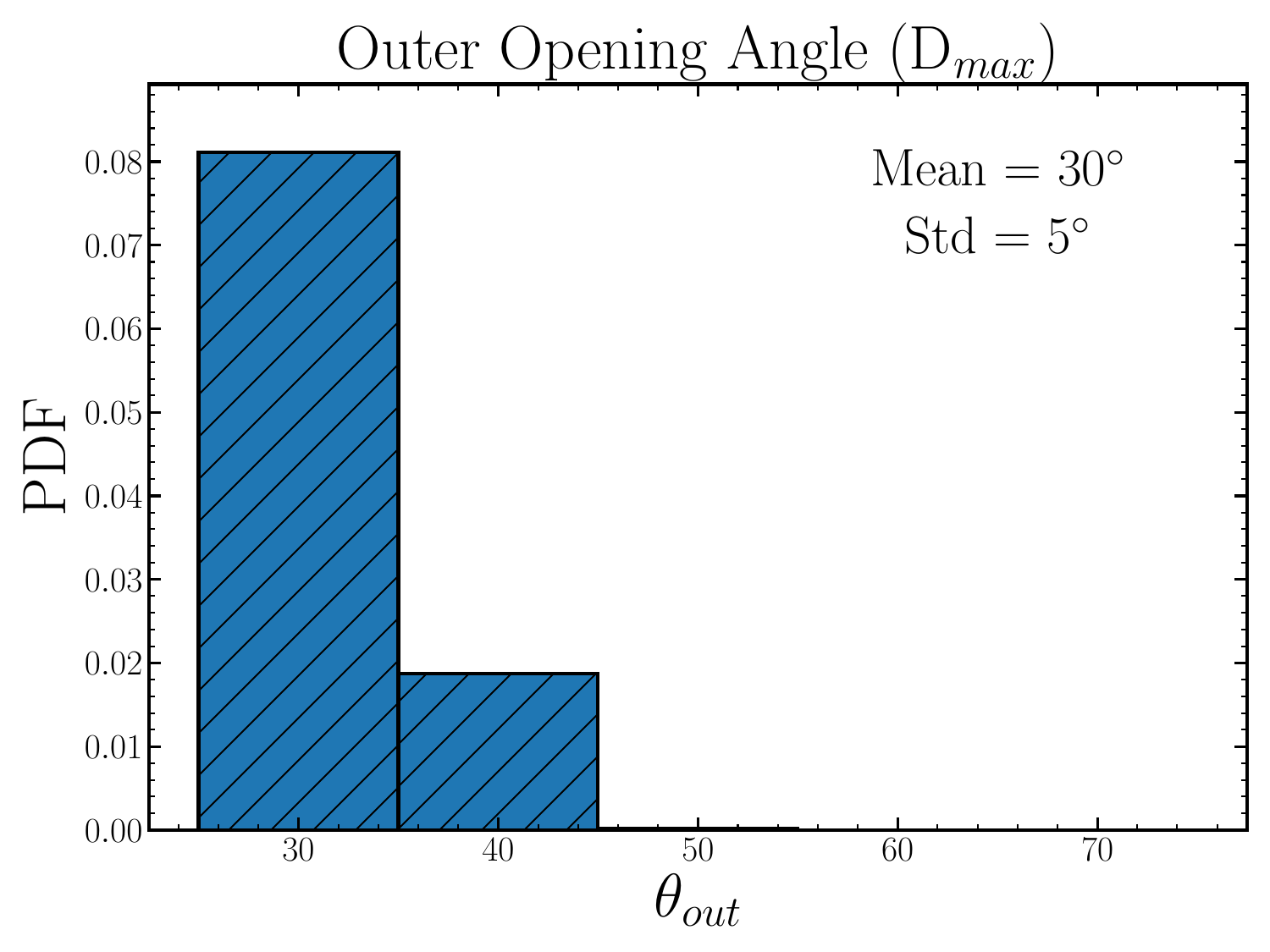}
    \caption{
Best-fit PDFs for the outflow opening angle ($\theta_{\rm out}$) based on the KL divergence (left) and $D_{max}$ (right).
 All distributions are truncated Gaussians defined over the range $30^\circ$–$70^\circ$. }
    \label{Onalge1_dist_best}
\end{figure*}


The histograms in Figure~\ref{Vmax_dist_best} show the best PDFs of the launching velocity based on the KL divergence (left) and $D_{max}$ (right). The two quantities adopt similar best PDFs for the launching velocity. As predicted in Section~\ref{subsec:VVD}, the launching velocities are mostly smaller than 1000 \kms, while 2 - 5\% exhibit higher launching velocities (1000 - 1500 \kms).

In contrast, strong outflow AGNs exhibit significantly different $V_{max}$ distribution from the total sample. The histograms in Figure~\ref{Vmax_dist_strong} represent $V_{max}$ distributions of strong outflow AGNs, which are located outside the dashed boxes in Figure~\ref{Best_mock_VVDs} ($|\mathrm{V}_\mathrm{[OIII]}| > 300$ \kms\ or $\sigma_\mathrm{[OIII]}>400$ \kms). Compared to the total distributions, strong outflow cases have fast launching velocities, mostly 800 - 1500 \kms. Since these AGNs are primary targets in spatially resolved studies, this result constrains a typical range of their launching velocities.

Lastly, the best PDFs of the opening angle (Figure~\ref{Onalge1_dist_best}) are slightly different depending on the statistical quantities. Specifically, $\theta_{out}\geq50^\circ$ is only exhibited in the best PDF based on the KL divergence, while the best PDF based on $D_{max}$ is concentrated on $\theta_{out}=30^\circ-40^\circ$. Nevertheless, the two best PDFs consistently favor relatively narrow outer opening angles, indicating that the opening angle of the local type 2 AGN 
is likely to be narrow, mostly 30$^\circ$ - 40$^\circ$.


\subsection{Testing Outflow Sizes}\label{subsec:outflowsizes}
\begin{deluxetable*}{cccccccccccc}
    \tablecolumns{12}
    \tablecaption{Model Parameters for Mock IFU Analysis}\label{T1_mock_IFU_parameters}
    \tablehead{
       z & $R_{out}$ & $V_{max}$ & $\theta_{out}$ & $i_{cone}$ & $i_{disk}$ & $A$ & $f_{disk}/f_{cone}$ & $r_t$ & $\sigma_0$ & $V_{max, rot}$ & $r_{1/2}$ \\
       {[}1{]} & {[}2{]} & {[}3{]} & {[}4{]} & {[}5{]} & {[}6{]} & {[}7{]} & {[}8{]} & {[}9{]} & {[}10{]} & {[}11{]} & {[}12{]}
       }
        \startdata
        0.05 vs 0.3  & 1 kpc & 1250 \kms &  40$^\circ$ & 25$^\circ$ & 120$^\circ$ & 0.2 & 1 & 2.5 kpc & 117 \kms & 198 \kms &  5.2 kpc\\
        \enddata
    \tablecomments{(1) Redshift; (2) outflow size; (3) maximum (launching) velocity; (4) outer opening angle; (5) bicone inclination; (6) disk (and dust plane) inclination; (7) extinction factor (smaller value means more extinction); (8) flux ratio between disk and bicone components; (9) velocity turnover radius (disk); (10) intrinsic velocity dispersion (disk); (11) maximum rotational velocity (disk); (12) effective radius.}
\end{deluxetable*}

\begin{figure*}[t]
    \includegraphics[width=0.45\textwidth]{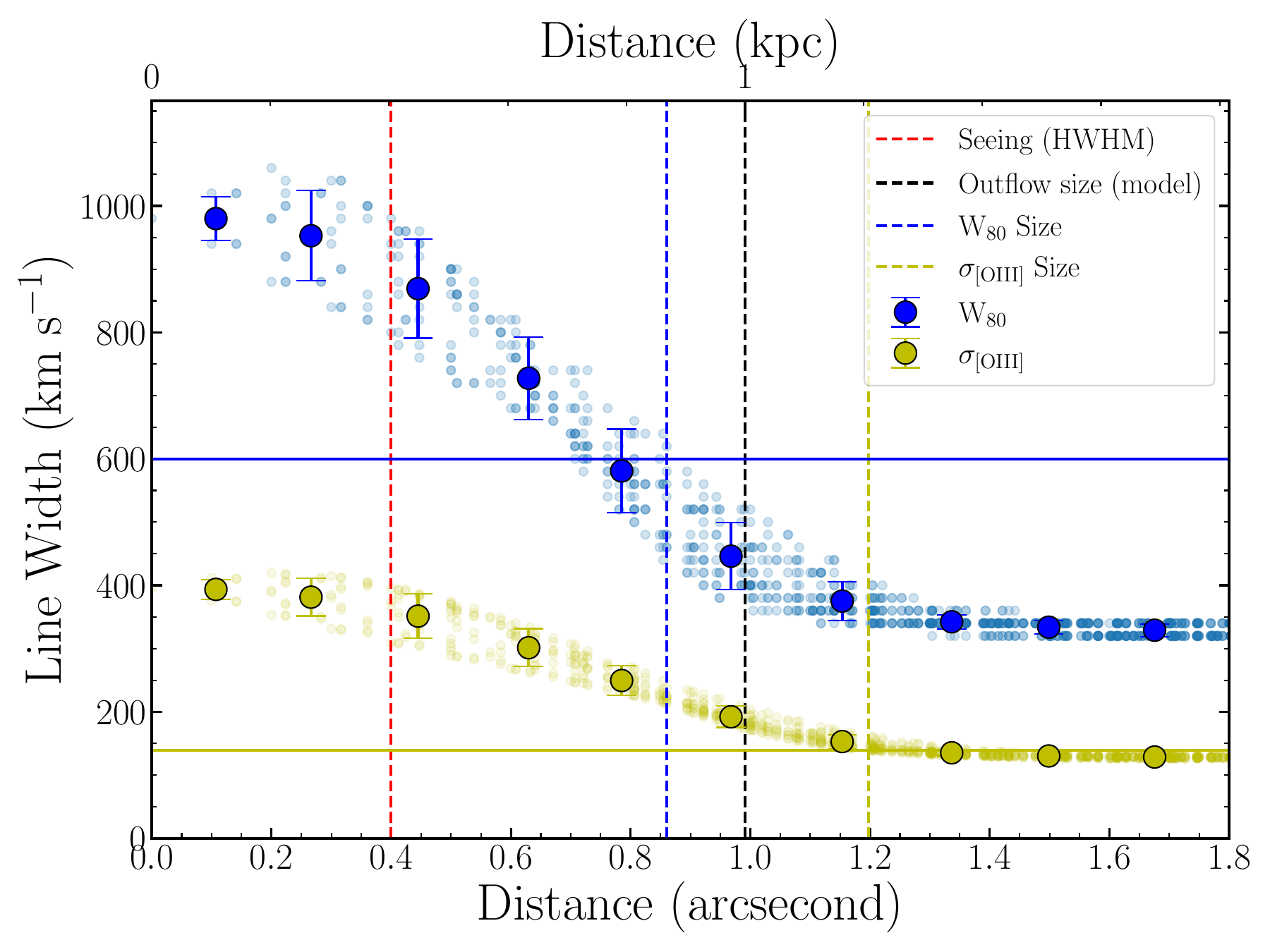}
    \includegraphics[width=0.53\textwidth]{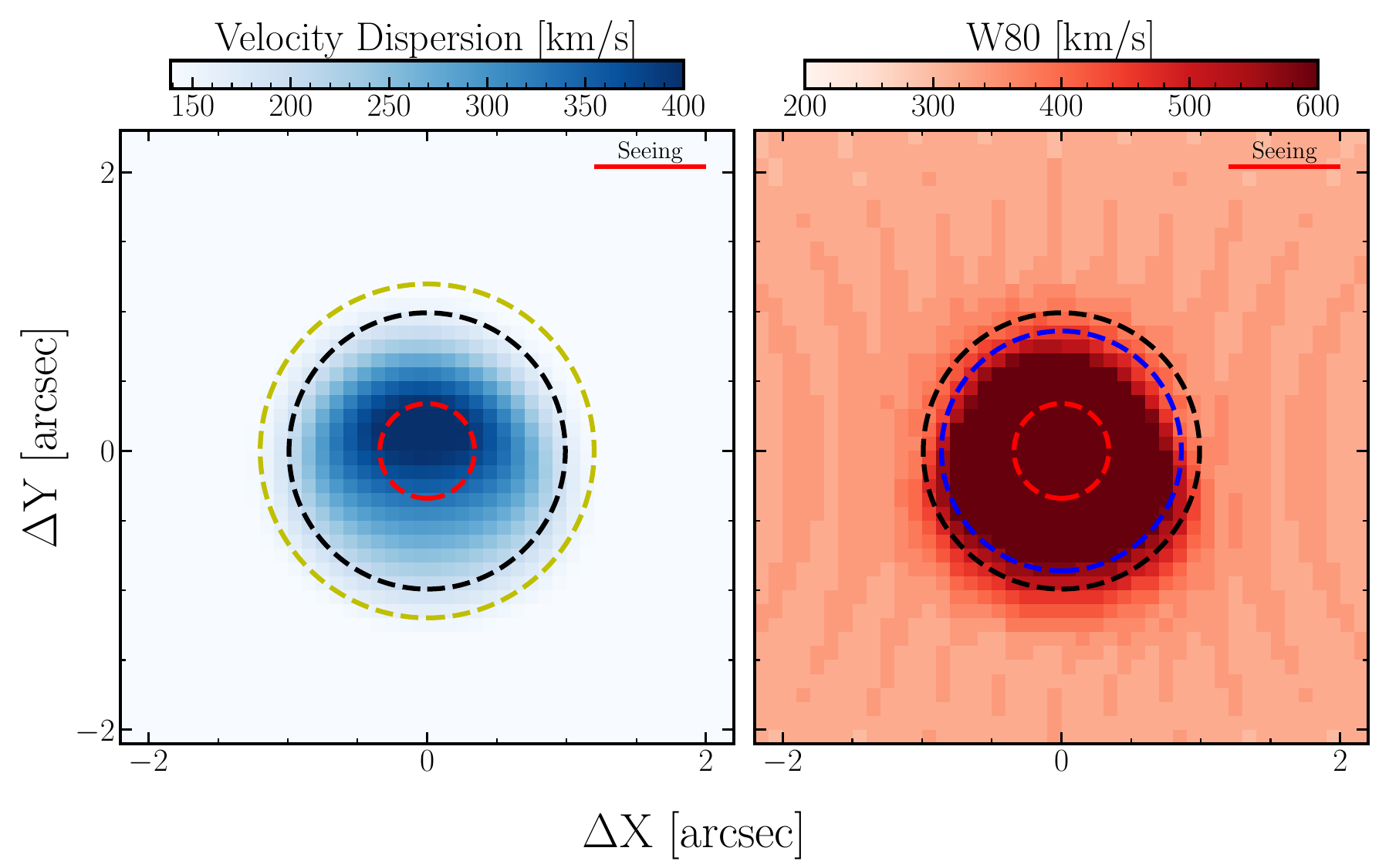}
    \caption{Example of mock IFU-analysis result for an AGN at $z = 0.05$.  
Left: Radial profiles of $\sigma_\mathrm{[OIII]}$ (yellow markers) and W$_{80}$ (blue markers). Large colored symbols indicate the mean velocity widths within individual radial bins.  
Horizontal lines mark the outflow-region criteria: W$_{80} = 600\ \mathrm{km\ s^{-1}}$ (blue) and $\sigma_{\rm [OIII]} = \sigma_*$ (yellow).  
Vertical dashed lines denote, from left to right, the half-width at half-maximum (HWHM) seeing size (red), the input outflow radius (black), the W$_{80}$-based outflow size (blue), and the $\sigma_{\rm [OIII]}$-based size (yellow).  
Right: 2D maps of velocity dispersion and W$_{80}$, with dashed circles following the same color scheme as the vertical lines in the left panel.
}
    \label{Vdisp_W80_radial_profile_maps_resolved}
\end{figure*}

\begin{figure*}[]
    \includegraphics[width=0.45\textwidth]{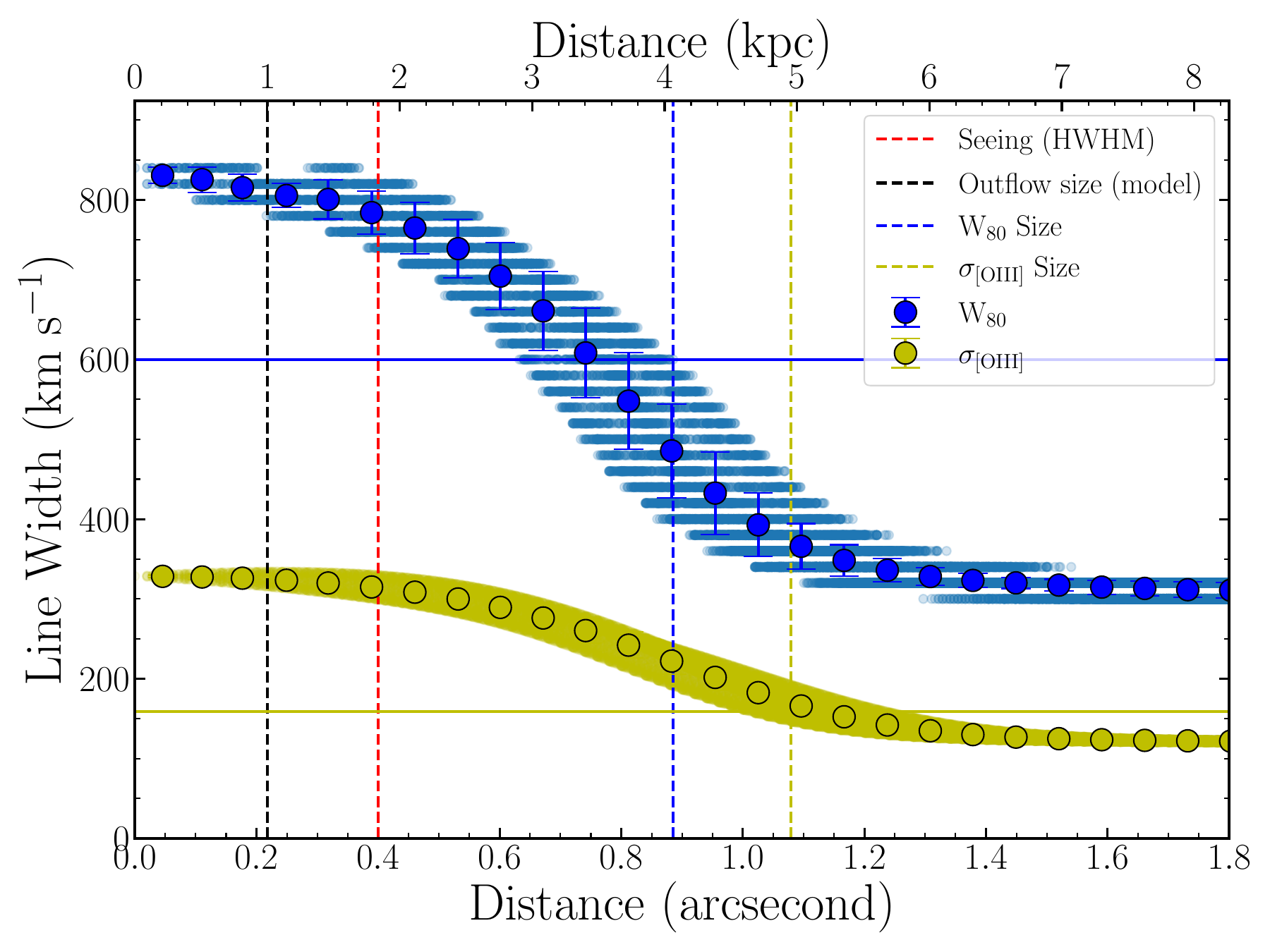}
    \includegraphics[width=0.53\textwidth]{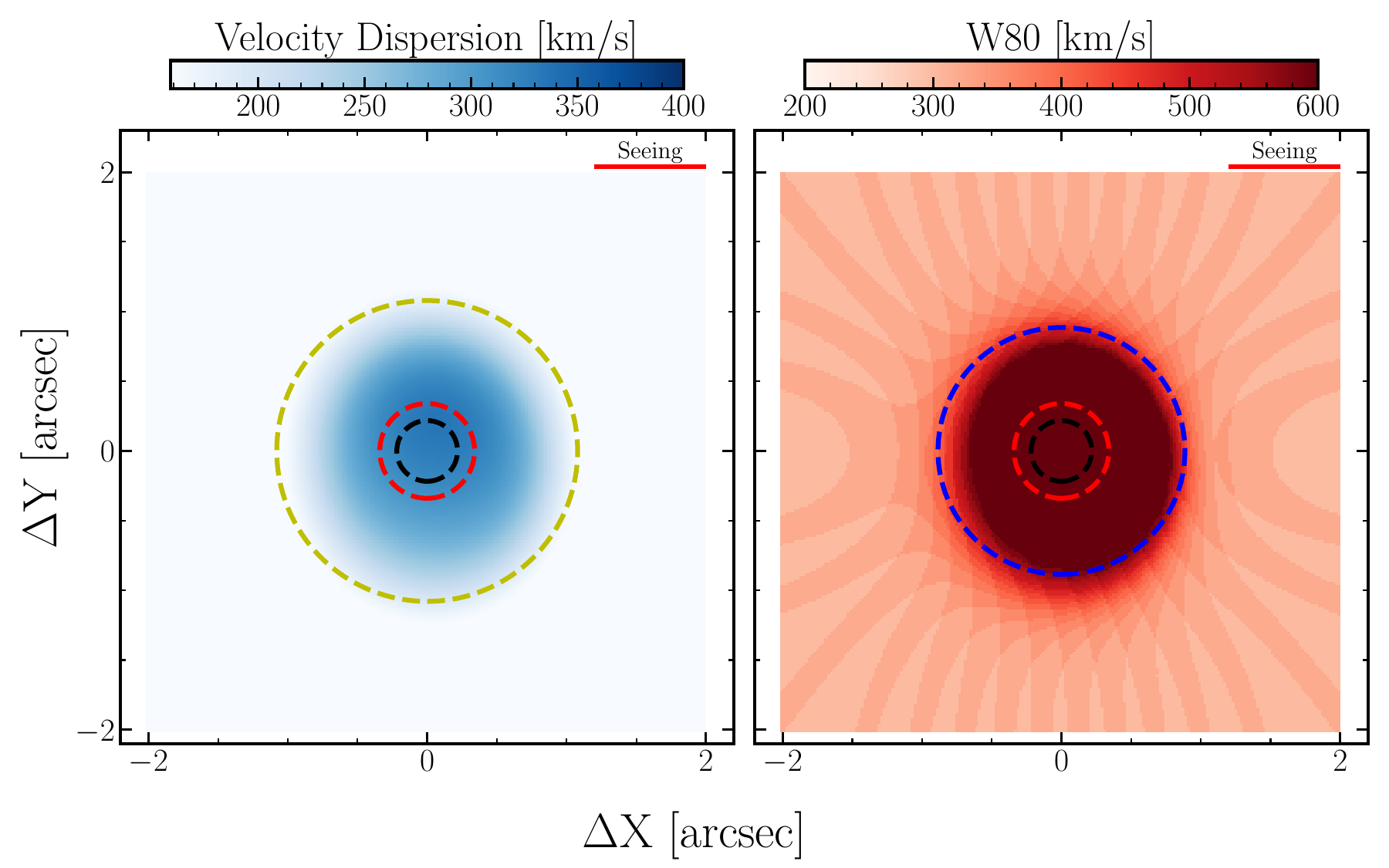}
    \caption{Same as Figure~\ref{Vdisp_W80_radial_profile_maps_resolved} but for a higher redshift of $z= 0.3$.}
    \label{Vdisp_W80_radial_profile_maps_unresolved}
\end{figure*}

We investigate the validity of the outflow size measurements based on mock IFU data by testing the effect of seeing. 
We set two examples of the same physical properties except for their redshifts (0.05 vs 0.3, see Table~\ref{T1_mock_IFU_parameters}) to contrast the angular sizes of the outflows by a factor of $\sim$ five for a given seeing (FWHM = 0.8\arcsec). 

We present the spatial distributions of velocity widths of the two examples in Figure~\ref{Vdisp_W80_radial_profile_maps_resolved} and \ref{Vdisp_W80_radial_profile_maps_unresolved}. 
In our previous IFU studies, radial profiles of velocity dispersion often exhibit constant values in the central region (i.e., initial plateau) before a smooth decrease \citep[e.g., ][]{Karouzos16a, Kang18}. 
This initial plateau is well reproduced in our mock IFU data, for both $\sigma_\mathrm{[OIII]}$ (yellow) and W$_{80}$ (blue).
Moreover, the radius of the initial plateau is comparable to the HWHM of the seeing (red dashed line), indicating that it mainly results from the seeing effect. 

In the low redshift case (Figure~\ref{Vdisp_W80_radial_profile_maps_resolved}), since the angular outflow size is larger than the seeing by more than a factor of two, the measured outflow size is consistent with the input value, indicating that the outflow size is well recovered by applying the velocity-width-based methods. 
In contrast, in the high redshift case (Figure~\ref{Vdisp_W80_radial_profile_maps_unresolved}), the angular outflow size becomes smaller than the seeing, leading to a significant overestimation of the outflow size even after quadratically subtracting the seeing size. Note that the measured size is also several times larger than the HWHM of the seeing size, as the seeing profile does not converge to zero even outside the FWHM. For example, if an outflow is unresolved, its seeing-convolved flux will follow the seeing profile. Thus, the outflow flux broadens the line profile at a radius further than the HWHM of the seeing, although its emission is dimmed by more than half compared to the center.


These results indicate that the seeing effect has to be carefully considered for interpreting IFU observations, i.e., initial plateau and outflow size. While the outflow size measured based on the emission line width is reliable when the seeing size is relatively small, the outflow size can be overestimated when the seeing size is similar to or larger than the angular size of the outflow. 

\section{Discussion}\label{sec:discussion}

\subsection{Comparison with Other Outflow Models}\label{subsec:modelcomparison}

In this section, we discuss the characteristics of our model in comparison with the previous models developed over the last two decades.
First, early outflow models did not include a rotating disk. For example, \citet{Crenshaw00a, Crenshaw00b} constructed a hollow bicone model to reproduce position-velocity (PV) diagrams of nearby AGNs \citep[e.g., ][]{Das05, Das06}. Subsequently, \citet{Crenshaw10a, Crenshaw10b} added a dust extinction component to the model to apply for dozens of nearby AGNs \citep[e.g., ][]{Fischer10, Fischer11, Fischer13, Meena21, Falcone24}. Since these models were developed for the HST observations of nearby AGNs, it is not essential to include the effect of the point spread function (PSF). 

On the other hand, \citet{Carniani15} developed a model with a single approaching cone for testing their spectroastrometric analysis of powerful outflows in quasars at cosmic noon. Since the seeing effect is substantial for their high-z AGNs, they convolved each velocity channel map with the PSF, reproducing consistent [\OIII] flux, velocity, and velocity dispersion maps with the observations. Although their model lacks a disk component, the contribution of the disk component is presumably weak in these extremely luminous AGNs.

Second, other models were also reported for accommodating various kinematical considerations. For example, \citet{Muller-Sanchez11} constructed a model to analyze the coronal line regions (CLRs) using AO-assisted near-infrared IFU data. By combining a bicone model with a rotating disk, they successfully fitted CLR kinematics at the central several hundred pc, while no seeing effect was included.

Recently, \citet{Marconcini23} developed a new model (\texttt{MOKA$^\mathrm{3D}$}) based on \citet{Venturi17} by constructing a 3D distribution of point-like clouds to reproduce the observed IFU data. The model also convolves the 3D data and assigns a proper weight to each cloud.
Since outflows often exhibit clumpy regions within their bicone geometry \citep[e.g., ][]{Kakkad23}, their model has a major advantage in reproducing various substructures, showing excellent agreements with the observations of both nearby and high-z AGNs \citep{Cresci23, Marconcini23, Marconcini25a, Marconcini25b}.

Compared to the previous models, which are mainly focused on fitting 3D data cubes of individual AGNs, our model is better suited for investigating the distributions of outflow parameters for a large sample, including the effect of seeing size and the disk component. Note that our model can also be utilized for reproducing the observation of individual AGNs \citep[e.g., ][]{Shin19}. Thus, our analysis with the modified model is complementary to the previous outflow models.

\subsection{Constraints on the AGN-driven outflows}\label{subsec:constraints}

The launching velocity is the key parameter for quantifying outflow energetics (see Section~\ref{subsec:energetics}), while the opening angle is related to how strongly gas outflows interact with the ISM. Thus, it is essential to constrain the launching velocity and opening angle of outflows for a large sample to investigate AGN feedback. 


We report launching velocities of $\lesssim$ 1500 \kms\ and typical opening angles of 30$^\circ$-40$^\circ$ (Section~\ref{subsec:VVDMC}). 
For a consistency check, we discuss the previous estimations of these two parameters, mostly from nearby AGNs.
For example, \citet{Crenshaw00a, Crenshaw00b} reported the inner/outer opening angles as 26$^\circ$/40$^\circ$ and the maximum outflow velocity as 1300 \kms\ for NGC 1068, and inner/outer opening angles as 20$^\circ$/36$^\circ$ and the maximum outflow velocity as 750 \kms\ for NGC 4151 \citep[see also][]{Das05, Das06}. After including dust extinction, the outer opening angles are slightly changed to 40$^\circ$ for NGC 1068 and 33$^\circ$ for NGC 4151 \citep{Crenshaw10b}. 

Their following studies also reported outflow parameters of other nearby AGNs \citep[][]{Crenshaw10a, Fischer10, Fischer11, Meena21, Falcone24}. In particular, among the cases of 17 Seyfert galaxies investigated by \citet{Fischer13}, the maximum velocity agrees with our range of $V_{max}\leq1500$\kms, except for NGC 1068. On the other hand, the outer opening angle distribution is slightly different. While a majority of their sample shows consistent opening angles ($<$50$^\circ$) with our results, 7 AGNs have wider opening angles. 
The difference can be caused by a couple of effects. First, AGNs with higher luminosity and stronger outflows are likely to have wider opening angles \citep{Bae16}. For example, only 13\% of our type 2 AGN sample have bolometric luminosity higher than $10^{44.5}$ erg/s, while more than half of their sample (9/17) are more luminous than $10^{44.5}$ erg/s. Second, their sample was selected based on clear signs of extended outflows, whereas we do not apply any criterion on the outflow extent. The inclusion of the disk component in our model can also induce differences. 

\citet{Shin19} also simulated 2D flux, velocity, and velocity dispersion maps of NGC 5728 by adding seeing convolution to the \citet{Bae16} model. The outer half opening angle was estimated as 28$^\circ$, which is the most prevalent value in our best PDFs. In contrast, \citet{Durre19} constrained the projected full opening angle to 71$^\circ$ based on [\FeII] 1.644 $\mu$m line. As discussed in \citet{Shin19}, this discrepancy may originate from different analytic models between the two studies.

The maximum velocity and the outer opening angle estimated by \citet{Muller-Sanchez11, Muller-Sanchez16} are also consistent with our constraints. On the other hand, for the best-fit parameters from \citet{Nevin18}, while the maximum velocities ($V_{max}=230-720$ \kms) also agree with our study, most of the opening angle values are much higher than the typical range of our constraint. This discrepancy also appears to be due to different biconical model and/or the characteristics of their sample (i.e., AGNs with double-peaked narrow emission lines). Other studies that constrained the opening angles of individual galaxies \citep[e.g.,][]{Baron18, Riffel21, Costa-Souza24} are in agreement with our constraint.


\subsection{True Outflow Size}\label{subsec:seeingeff}

\begin{figure}[]
    \includegraphics[width=0.47\textwidth]{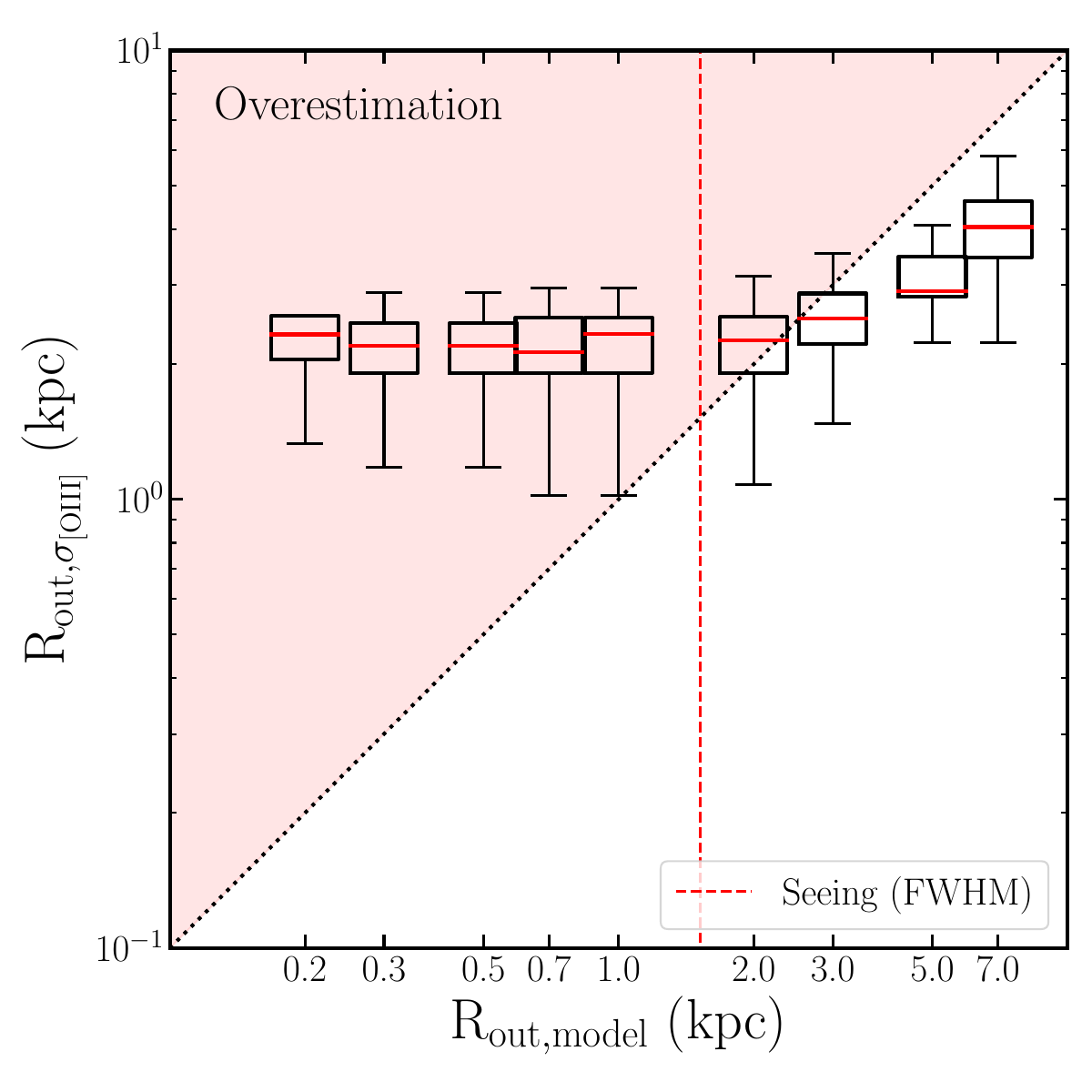}
    \caption{Comparison between the measured outflow sizes based on $\sigma_\mathrm{[OIII]}$ and the input outflow sizes. The red vertical line indicates the seeing (FWHM=0.8\arcsec $\sim$ 1.52 kpc at z=0.1) while the black dotted line shows a one-to-one relation.}
    \label{Rout_obs_model_comparison}
\end{figure}

In Section~\ref{subsec:outflowsizes}, we verified that the outflow size measured based on velocity widths can be significantly overestimated by the seeing, especially when a true outflow size is comparable to or smaller than the seeing size. To properly constrain the spatial scale of outflows, it is essential to investigate the seeing-limited cases in detail.

Previous studies reported notable spatial features in luminous type 2 AGNs, i.e., circularly extended NLR morphologies or enhanced W$_{80}$ values with flat radial profiles \citep[e.g.,][]{Liu13a, Liu13b, Harrison14}. Interestingly, our model reproduces a similar flat W$_{80}$ profile when the outflow is compact and the disk emission is much weaker than the outflow. Seeing convolution of a compact outflow generates a broad wing component with an almost constant velocity width across the entire FoV, while the disk emission is too faint to affect the velocity width, even at pixels far from the center. Thus, given that their samples consist of luminous type 2 quasars, implying relatively weak disk emission, the flat W$_{80}$ profiles may indicate smaller outflow sizes. 

Furthermore, \citet{Villar-Martin16} found that only 2 out of 15 AGNs in their sample show clearly resolved outflows by comparing the spatial profiles of seeing with the [\OIII] wing emission. Based on spectroastrometry and seeing sizes, they constrained lower and upper limits of the outflow sizes as several hundred pc and a few kpc, respectively. These results are in agreement with our constraints and represent observational evidence of the seeing effect in local AGNs.

For type 1 AGNs, the BLR emission (mostly H$\beta$) can serve as a reference PSF for deblending methods \citep{Husemann13, Husemann14, Husemann16, Ibarra-Medel25}. 
For example, \citet{Husemann13} reported that a significant fraction of their type 1 AGN sample shows no/weak extended emission after a PSF subtraction. 
They also found that the resolved [\OIII] component often exhibits a significantly narrower velocity width, indicating no extended outflows. 
Despite the effectiveness of their method, it cannot be applied to type 2 AGNs since it requires a reference PSF from BLR emission.
In addition, a compact but extended outflow can remain even after a PSF subtraction, leading to an overestimation of the size. 

\citet{Singha22} tested whether the [\OIII] wing component is spatially resolved for type 1 AGNs at $z<0.06$ by comparing the 2D flux map with a PSF from the broad H$\beta$, finding that 23 out of 36 AGNs show unresolved [\OIII] wing components.
They also reported a median spatial offset of $\sim$ 27 pc between the center of AGN and flux centroid of [\OIII] wing component for the unresolved cases, which can be considered as lower limits of the outflow size.  
In contrast, based on the $\sigma_\mathrm{[OIII]}$-based method of \citet{Karouzos16a}, they obtained an order of magnitude larger sizes than their lower limits, suggesting that the seeing effect can significantly overestimate the outflow sizes.  

In our previous IFU studies \citep{Karouzos16a, Bae17, Kang18, Luo21, Kim23}, the outflow sizes were determined based on the $\sigma_\mathrm{[OIII]}$-based method. Our main conclusion was that the outflow sizes are mostly less than several kpc scales, indicating a lack of galactic scale feedback. We also reported a kinematic outflow size--luminosity relation between the outflow size and the [\OIII] luminosity with a slope of $\sim$0.3. 
For this work, we subtracted the HWHM of the seeing size from the outflow size in quadrature to correct for the seeing effect. 
However, this correction may not be sufficient as shown in Section~\ref{subsec:outflowsizes}. Thus, our measurements can be conservatively considered as upper limits, particularly for higher redshift targets among our sample ($z>0.1$). 

To test the reliability of the size measurements, we apply the $\sigma_\mathrm{[OIII]}$-based method to various mock data.  
For this test, we set the redshift z=0.1 as a representative case and restrict $V_{max}$ from 750 to 2000 \kms\ to focus on strong outflows (see Section~\ref{subsec:VVDMC}). 
In Figure~\ref{Rout_obs_model_comparison}, we present the measured outflow size as a function of the input (true) size.
As expected, the outflow size is overestimated by more than a factor of two when the true outflow size is smaller than the FWHM of the seeing ($\sim$1.52 kpc at z = 0.1).
On the other hand, when the seeing is smaller than the true size, the measured size is generally consistent with the input size with slight underestimations ($<$ 0.3 dex). 
Therefore, the $\sigma_\mathrm{[OIII]}$-based method provides a good size estimation within 0.3 dex, or at least can be used as a reliable upper limit, depending on the relative seeing size.

We emphasize that although the $\sigma_\mathrm{[OIII]}$-based method overestimates the sizes of compact outflows, the main conclusion of \citet{Kim23}, i.e., lack of global feedback, remains valid. Since the true size can be more compact than the measured one, the upper limit of the size still provides strong constraints on the spatial scales of AGN feedback.

It is also noteworthy that the measured sizes are broadly consistent within $\sim$ 0.3 dex at a given input size, regardless of other input parameters. 
There are some outliers with small measured sizes, which mostly have slow launching velocities ($<$ 1000 \kms). 
Thus, if the launching velocity is fast enough, the $\sigma_\mathrm{[OIII]}$-based size has no significant dependence on other parameters.

%

\subsection{Outflow Energetics}\label{subsec:energetics}

\begin{figure}[]
    \includegraphics[width=0.47\textwidth]{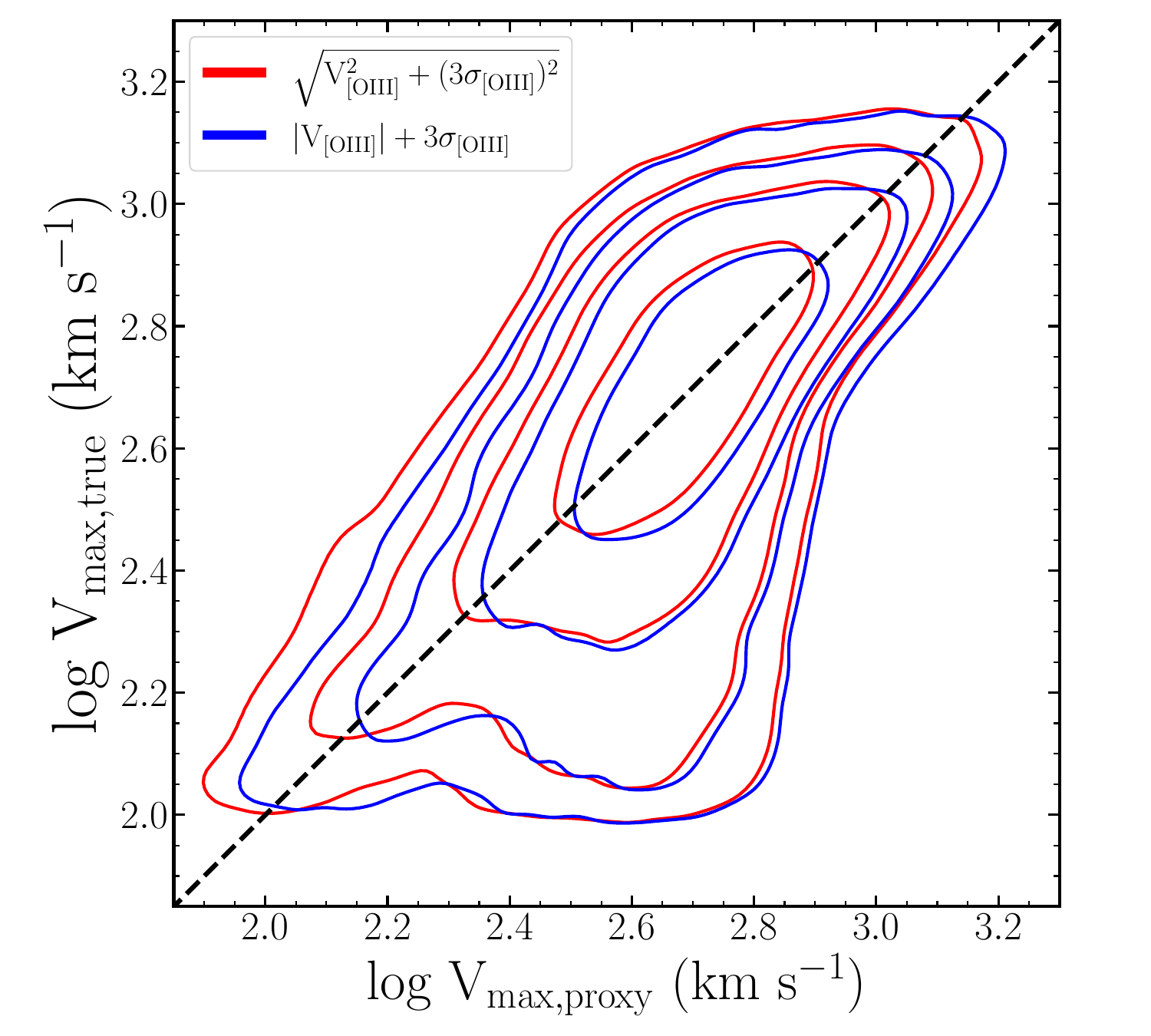}
    \caption{Comparison of the launching velocity with its proxies derived from mock kinematics 
    in the MC simulation. The red and blue contours show 0.5 - 2$\sigma$ levels when we use a quadratic sum ($\sqrt{\mathrm{V}^2_\mathrm{[OIII]} + (3\sigma_\mathrm{[OIII]})^2}$) \citep[][]{Bae16} and a direct summation ($|\mathrm{V}_\mathrm{[OIII]}| + 3\sigma_\mathrm{[OIII]}$) \citep{Rupke13} as a proxy, respectively. The dashed line indicates a one-to-one relation.}
    \label{Vmax_estimation}
\end{figure}

In this section, we present estimates of the mass outflow rate ($\dot{\mathrm{M}}_\mathrm{out}$), momentum rate ($\dot{\mathrm{P}}_\mathrm{out}$), and kinetic power ($\dot{\mathrm{E}}_\mathrm{out}$) for local type 2 AGNs to assess the efficiency of AGN feedback. 
For conical outflows, these quantities are computed as \citep[e.g.,][]{Maiolino12, Fiore17}
\begin{gather}
\dot{\mathrm{M}}_\mathrm{out} = 3\,\mathrm{M}_\mathrm{out}\frac{v_\mathrm{out}}{\mathrm{R}_\mathrm{out}}, \\
\dot{\mathrm{P}}_\mathrm{out} = \dot{\mathrm{M}}_\mathrm{out}\,v_\mathrm{out}, \\
\dot{\mathrm{E}}_\mathrm{out} = \frac{1}{2}\,\dot{\mathrm{M}}_\mathrm{out}\,v_\mathrm{out}^2,
\label{eq:energetics}
\end{gather}
where $\mathrm{M}_\mathrm{out}$, $v_\mathrm{out}$, and $\mathrm{R}_\mathrm{out}$ denote the ionized gas mass, outflow velocity, and outflow size, respectively.  
We estimate the ionized gas mass from the extinction-corrected [\OIII] luminosity using Eq.~29 of \citet{Veilleux20}, adopting a fixed electron density of $1000\ \mathrm{cm^{-3}}$.

A variety of velocity definitions have been used in previous studies to approximate the characteristic outflow velocity, including W$_{80}$ \citep[e.g.,][]{Wylezalek20, Ruschel-Dutra21}, $\sqrt{v_{\rm [OIII]}^2 + \sigma_{\rm [OIII]}^2}$ \citep[e.g.,][]{Rakshit18, Kim23}, and $v_{\rm broad} + 2\sigma_{\rm broad}$ \citep[e.g.,][]{Rupke13, Fiore17}.  
In our analysis, we derive a mass-weighted mean outflow velocity based on the radial profiles of the intrinsic velocity ($v(R)=V_{\rm max}(1-R/R_{\rm out})$) and flux ($f(R)=f_0 e^{-5R/R_{\rm out}}$) in our model, assuming an electron density profile $n_e(R) \propto R^{-2}$.  
Since the mass is proportional to L$_\mathrm{[OIII]}/n_e$, the mass-weighted mean velocity becomes
\begin{equation} \label{mean_vout}
\begin{split}
v_\mathrm{out} 
 &= \frac{\int_0^{R_{\rm out}} v(R)\,\mathrm{M}_\mathrm{out}(R)\,R^2\,dR}
         {\int_0^{R_{\rm out}} \mathrm{M}_\mathrm{out}(R)\,R^2\,dR} \\
 &= \frac{\int_0^{R_{\rm out}} V_{\rm max}\!\left(1-\frac{R}{R_{\rm out}}\right)
         f_0 e^{-5R/R_{\rm out}} n_e^{-1}(R)\,R^2\,dR}
         {\int_0^{R_{\rm out}} f_0 e^{-5R/R_{\rm out}} n_e^{-1}(R)\,R^2\,dR} \\
 &= V_{\rm max}\,
    \frac{\int_0^1 t^4 (1-t) e^{-5t}\,dt}
         {\int_0^1 t^4 e^{-5t}\,dt}
    \simeq 0.31\,V_{\rm max}.
\end{split}
\end{equation}
Thus, the mass-weighted mean outflow velocity is approximately $v_\mathrm{out} = 0.31\,V_{\rm max}$.  
Since the momentum and kinetic power scale with $v_\mathrm{out}^2$ and $v_\mathrm{out}^3$, respectively, 
their mass-weighted mean values correspond to $\sim 0.14\,V_{\rm max}^2$ and $\sim 0.07\,V_{\rm max}^3$.
These relations are used to calculate $\dot{\mathrm{M}}_\mathrm{out}$, $\dot{\mathrm{P}}_\mathrm{out}$, and $\dot{\mathrm{E}}_\mathrm{out}$.

Note that the launching velocity cannot be directly measured from observational data. 
Instead, we test several potential proxies for $V_{\rm max}$ that can be computed from the observed V$_{\rm [OIII]}$ and $\sigma_{\rm [OIII]}$, using the mock SDSS sample (Section~\ref{subsec:VVDMC}). 
We find that two commonly used prescriptions show good agreement with the input $V_{\rm max}$. 
The first is the quadratic sum of $|\mathrm{V}_{\rm [OIII]}|$ and $3\,\sigma_{\rm [OIII]}$, similar to that adopted by \citet{Bae16} and \citet{RodriguezZaurin13}. 
The second is the direct sum $|\mathrm{V}_{\rm [OIII]}| + 3\,\sigma_{\rm [OIII]}$, following \citet{Rupke13} and \citet{Fiore17}. 
Figure~\ref{Vmax_estimation} demonstrates that both proxies recover the launching velocity with a scatter of $\sim$0.2 dex. 
In the following analysis, we adopt the first proxy (i.e., the quadratic sum).

Lastly, we use two different outflow size estimates to set upper and lower limits on the energetics. 
First, we adopt the $\sigma_{\rm [OIII]}$-based sizes inferred from the kinematic outflow size--luminosity relation as upper limits (see Section~\ref{subsec:seeingeff}), and compute the corresponding lower limits on the outflow rates. 
Second, we employ the NLR sizes from the NLR size--luminosity relation derived for nearby AGNs \citep{Polack24}. 
We convert these NLR sizes to outflow sizes by dividing by three, following the typical ratio of $\mathrm{R_{out}}/\mathrm{R_{[OIII]}}$ reported in \citet{Fischer18} and \citet{Polack24}, and then calculate the corresponding upper limits on the outflow rates. 
These NLR-based sizes are typically 0.4--0.6 dex smaller than the $\sigma_{\rm [OIII]}$-based sizes in the dynamic range of the [\OIII] luminosity.

\begin{figure}[]
    \includegraphics[width=0.47\textwidth]{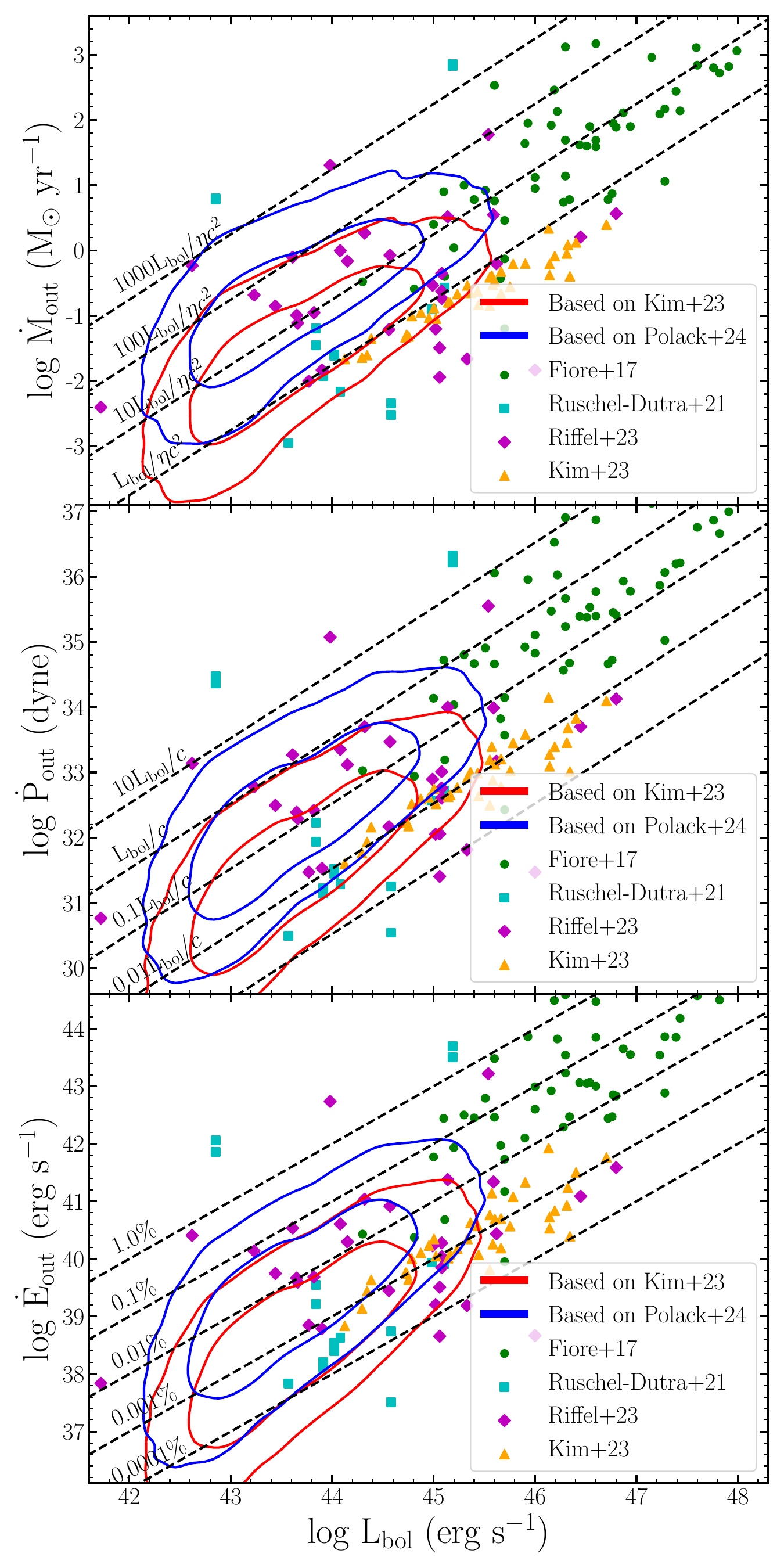}
    \caption{Mass outflow rate (log $\dot{\mathrm{M}}_\mathrm{out}$, top), momentum rate  (log $\dot{\mathrm{P}}_\mathrm{out}$, middle), and kinetic power (log $\dot{\mathrm{E}}_\mathrm{out}$, bottom) as a function of the bolometric luminosity of the SDSS type 2 AGNs. The red and blue contours show the energetics based on the upper limits (red contours) and lower limits (blue contours) of the outflow sizes. The energetics of individual AGNs from the literature are denoted with various symbols.}
    \label{Outflow Energetics}
\end{figure}

In Figure~\ref{Outflow Energetics}, we compare the upper and lower limits of the outflow rates as a function of the bolometric luminosity to evaluate whether the outflows are sufficiently energetic to drive efficient feedback. 
Despite the large scatter, the estimated outflow rates (shown as contours) generally fall within the ranges 
$\dot{\mathrm{M}}_\mathrm{out} \sim 1$--$100\,\mathrm{L}_\mathrm{bol}/\eta c^2$ $(\eta = 0.1)$, 
$\dot{\mathrm{P}}_\mathrm{out} \sim 0.01$--$1\,\mathrm{L}_\mathrm{bol}/c$, 
and 
$\dot{\mathrm{E}}_\mathrm{out} \sim 10^{-6}$--$10^{-3}\,\mathrm{L}_\mathrm{bol}$. 
These ranges are broadly consistent with previous studies, though some systematic differences arise from the various definitions adopted for outflow parameters. 
For instance, the \citet{Fiore17} sample reports kinetic powers typically in the range of $0.01$--$1\%$ of $L_\mathrm{bol}$, largely because their characteristic outflow velocity is defined using the broad [\OIII] component ($|v_{\rm broad}| + 2\sigma_{\rm broad}$), which yields substantially larger $v_\mathrm{out}$ values. 
By contrast, our analysis computes $v_\mathrm{out}$ from the first and second moments of the total [\OIII] line profile, resulting in systematically smaller velocities at fixed luminosity. 
This discrepancy becomes even more pronounced in $\dot{\mathrm{E}}_\mathrm{out}$, as the kinetic power scales with $v_\mathrm{out}^3$.

Theoretical studies generally suggest that the kinetic power of AGN-driven outflows is required to be $0.5$--$5\%$ of the bolometric luminosity for efficient AGN feedback \citep[e.g.,][]{DiMatteo05, Hopkins08, Zubovas12}. 
Our estimated kinetic powers are mostly much lower than this threshold, indicating that the feedback may not be efficient for the bulk of the sample.
It is important to note, however, that our estimates are based solely on the ionized gas component, which represents only a small fraction of the total outflow mass \citep[e.g.,][]{Fluetsch19}. 
Nevertheless, a non-negligible subset of AGNs---$0.85\%$ or $0.14\%$ depending on whether the outflow sizes are estimated using the \citet{Polack24} or \citet{Kim23} relations, respectively---exhibits $\dot{\mathrm{E}}_\mathrm{out}/\mathrm{L}_\mathrm{bol} > 0.5\%$. These AGNs also show high outflow velocities, making them excellent targets for spatially resolved outflow studies in detail.


\subsection{Limitations of the current model}\label{subsec:limitation}

Despite these improvements, our model and mock data analysis still have several limitations. 
First, because we adopt analytic functions for the flux and velocity distributions, the model cannot capture the detailed substructures of real outflows. 
High-resolution studies of nearby AGNs have revealed clumpy and irregular features within the bicone \citep[e.g.,][]{Venturi18, Mingozzi19, Kakkad23}, which are not represented in our simplified prescription.

Second, our mock data do not include flux uncertainties. 
For instance, the $\sigma_{\rm [OIII]}$-based method defines the outflow size as the radius of the outermost [\OIII]-detected pixel when the $\sigma_\mathrm{[OIII]}/\sigma_*$ does not converge to unity; in such cases, the S/N becomes critical in determining the size. 
We also did not test spectroastrometry-based techniques \citep[e.g.,][]{Carniani15, Singha22}, which estimate the outflow size from the spatial offset between the center and the flux centroid of the [\OIII] emission. 
These methods require realistic flux errors to properly fit the 2D flux distribution and determine the centroid.

Third, we assume that all parameter PDFs in the MC simulations are independent, owing to the lack of well-established relations among them. 
In reality, several outflow parameters are expected to be correlated, implying that multivariate PDFs may be more appropriate. 
Previous studies have reported correlations between outflow properties and bolometric luminosity (or Eddington ratio).
The opening angle and outflow velocity are also suggested to have positive \citep{Bae16}, negative \citep{Muller-Sanchez11}, or no correlations \citep{Fischer13}.
Future MC simulations incorporating such parameter correlations may better reproduce the observed gas kinematics.

\section{Summary}\label{sec:summary}
In this study, we present the modified outflow model and its applications to local type 2 AGNs. Here we summarize the main results of this study.

\begin{enumerate}
\item We find that including a rotating disk component in the model reduces the gas velocity and velocity dispersion within the SDSS fiber size. Thus, the outflow launching velocity needs to be faster to compensate for the disk contribution. The observed VVD distribution of the local type 2 AGNs can be reproduced by $V_{max}\lesssim1500$ \kms.

\item Based on the MC simulations to reproduce the observed VVD distribution, we find that the launching velocity is $\lesssim$1000 \kms\ for a majority of the type 2 AGNs except for several percent of strong outflow cases ($>$ 1000 - 1500 \kms), and that the outer half opening angle is less than 50$^\circ$.

\item By generating mock IFU data, we find similar features reported in previous IFU studies, e.g., circular morphology in the flux map and the initial plateau of the radial profiles of velocity widths. These results suggest that the biconical model with a seeing convolution can successfully reproduce observational data. 


\item By investigating the seeing effect on the outflow size measurement, we confirmed that the outflow size measured based on velocity widths can be overestimated, particularly when the angular size of the outflow is comparable to or smaller than the seeing size. Nevertheless, this still supports the lack of global feedback since the previously reported sizes can be considered as upper limits.
\end{enumerate}


\begin{acknowledgments}
This work has been supported by the Basic Science Research Program through the National Research Foundation of the Korean Government (2021R1A2C3008486). 
\end{acknowledgments}




\software{astropy \citep{Astropy22}, matplotlib \citep{matplotlib}, numpy \citep{numpy}, scipy \citep{scipy}}




\bibliography{ref}{}

@ARTICLE{Kim23,
       author = {{Kim}, Changseok and {Woo}, Jong-Hak and {Luo}, Rongxin and {Chung}, Aeree and {Baek}, Junhyun and {Le}, Huynh Anh N. and {Son}, Donghoon},
        title = "{Unraveling the Complex Structure of AGN-driven Outflows. VI. Strong Ionized Outflows in Type 1 AGNs and the Outflow Size-Luminosity Relation}",
      journal = {\apj},
         year = 2023,
        month = dec,
       volume = {958},
       number = {2},
          eid = {145},
        pages = {145},
          doi = {10.3847/1538-4357/acf92b},
archivePrefix = {arXiv},
       eprint = {2310.06928},
 primaryClass = {astro-ph.GA},
       adsurl = {https://ui.adsabs.harvard.edu/abs/2023ApJ...958..145K}
}

@ARTICLE{Luo21,
       author = {{Luo}, Rongxin and {Woo}, Jong-Hak and {Karouzos}, Marios and {Bae}, Hyun-Jin and {Shin}, Jaejin and {McConnell}, Nicholas and {Shih}, Hsin-Yi and {Kim}, Yoo Jung and {Park}, Songyoun},
        title = "{Unraveling the Complex Structure of AGN-driven Outflows. V. Integral-field Spectroscopy of 40 Moderate-luminosity Type-2 AGNs}",
      journal = {\apj},
         year = 2021,
        month = feb,
       volume = {908},
       number = {2},
          eid = {221},
        pages = {221},
          doi = {10.3847/1538-4357/abd5ac},
archivePrefix = {arXiv},
       eprint = {2012.10065},
 primaryClass = {astro-ph.GA},
       adsurl = {https://ui.adsabs.harvard.edu/abs/2021ApJ...908..221L}
}

@ARTICLE{Woo20,
       author = {{Woo}, Jong-Hak and {Son}, Donghoon and {Rakshit}, Suvendu},
        title = "{The Correlation of Outflow Kinematics with Star Formation Rate. VI. Gas Outflows in AGNs}",
      journal = {\apj},
         year = 2020,
        month = sep,
       volume = {901},
       number = {1},
          eid = {66},
        pages = {66},
          doi = {10.3847/1538-4357/abad97},
archivePrefix = {arXiv},
       eprint = {2008.04919},
 primaryClass = {astro-ph.GA},
       adsurl = {https://ui.adsabs.harvard.edu/abs/2020ApJ...901...66W}
}

@ARTICLE{Shin19,
       author = {{Shin}, Jaejin and {Woo}, Jong-Hak and {Chung}, Aeree and {Baek}, Junhyun and {Cho}, Kyuhyoun and {Kang}, Daeun and {Bae}, Hyun-Jin},
        title = "{Positive and Negative Feedback of AGN Outflows in NGC 5728}",
      journal = {\apj},
         year = 2019,
        month = aug,
       volume = {881},
       number = {2},
          eid = {147},
        pages = {147},
          doi = {10.3847/1538-4357/ab2e72},
archivePrefix = {arXiv},
       eprint = {1907.00982},
 primaryClass = {astro-ph.GA},
       adsurl = {https://ui.adsabs.harvard.edu/abs/2019ApJ...881..147S}
}

@ARTICLE{Luo19,
       author = {{Luo}, Rongxin and {Woo}, Jong-Hak and {Shin}, Jaejin and {Kang}, Daeun and {Bae}, Hyun-Jin and {Karouzos}, Marios},
        title = "{Unraveling the Complex Structure of AGN-driven Outflows. IV. Comparing AGNs with and without Strong Outflows}",
      journal = {\apj},
         year = 2019,
        month = mar,
       volume = {874},
       number = {1},
          eid = {99},
        pages = {99},
          doi = {10.3847/1538-4357/ab08e6},
archivePrefix = {arXiv},
       eprint = {1902.07560},
 primaryClass = {astro-ph.GA},
       adsurl = {https://ui.adsabs.harvard.edu/abs/2019ApJ...874...99L}
}

@ARTICLE{Rakshit18,
       author = {{Rakshit}, Suvendu and {Woo}, Jong-Hak},
        title = "{A Census of Ionized Gas Outflows in Type 1 AGNs: Gas Outflows in AGNs. V.}",
      journal = {\apj},
         year = 2018,
        month = sep,
       volume = {865},
       number = {1},
          eid = {5},
        pages = {5},
          doi = {10.3847/1538-4357/aad9f8},
archivePrefix = {arXiv},
       eprint = {1808.03415},
 primaryClass = {astro-ph.GA},
       adsurl = {https://ui.adsabs.harvard.edu/abs/2018ApJ...865....5R}
}

@ARTICLE{Kang18,
       author = {{Kang}, Daeun and {Woo}, Jong-Hak},
        title = "{Unraveling the Complex Structure of AGN-driven Outflows. III. The Outflow Size-Luminosity Relation}",
      journal = {\apj},
         year = 2018,
        month = sep,
       volume = {864},
       number = {2},
          eid = {124},
        pages = {124},
          doi = {10.3847/1538-4357/aad561},
archivePrefix = {arXiv},
       eprint = {1807.08356},
 primaryClass = {astro-ph.GA},
       adsurl = {https://ui.adsabs.harvard.edu/abs/2018ApJ...864..124K}
}

@ARTICLE{Woo17,
       author = {{Woo}, Jong-Hak and {Son}, Donghoon and {Bae}, Hyun-Jin},
        title = "{Delayed or No Feedback? Gas Outflows in Type 2 AGNs. III.}",
      journal = {\apj},
         year = 2017,
        month = apr,
       volume = {839},
       number = {2},
          eid = {120},
        pages = {120},
          doi = {10.3847/1538-4357/aa6894},
archivePrefix = {arXiv},
       eprint = {1702.06681},
 primaryClass = {astro-ph.GA},
       adsurl = {https://ui.adsabs.harvard.edu/abs/2017ApJ...839..120W}
}

@ARTICLE{Bae17,
       author = {{Bae}, Hyun-Jin and {Woo}, Jong-Hak and {Karouzos}, Marios and {Gallo}, Elena and {Flohic}, Helene and {Shen}, Yue and {Yoon}, Suk-Jin},
        title = "{The Limited Impact of Outflows: Integral-field Spectroscopy of 20 Local AGNs}",
      journal = {\apj},
         year = 2017,
        month = mar,
       volume = {837},
       number = {1},
          eid = {91},
        pages = {91},
          doi = {10.3847/1538-4357/aa5f5c},
archivePrefix = {arXiv},
       eprint = {1702.01900},
 primaryClass = {astro-ph.GA},
       adsurl = {https://ui.adsabs.harvard.edu/abs/2017ApJ...837...91B}
}

@ARTICLE{Bae16,
       author = {{Bae}, Hyun-Jin and {Woo}, Jong-Hak},
        title = "{The Prevalence of Gas Outflows in Type 2 AGNs. II. 3D Biconical Outflow Models}",
      journal = {\apj},
         year = 2016,
        month = sep,
       volume = {828},
       number = {2},
          eid = {97},
        pages = {97},
          doi = {10.3847/0004-637X/828/2/97},
archivePrefix = {arXiv},
       eprint = {1606.05348},
 primaryClass = {astro-ph.GA},
       adsurl = {https://ui.adsabs.harvard.edu/abs/2016ApJ...828...97B}
}

@ARTICLE{Woo16,
       author = {{Woo}, Jong-Hak and {Bae}, Hyun-Jin and {Son}, Donghoon and {Karouzos}, Marios},
        title = "{The Prevalence of Gas Outflows in Type 2 AGNs}",
      journal = {\apj},
         year = 2016,
        month = feb,
       volume = {817},
       number = {2},
          eid = {108},
        pages = {108},
          doi = {10.3847/0004-637X/817/2/108},
archivePrefix = {arXiv},
       eprint = {1511.05142},
 primaryClass = {astro-ph.GA},
       adsurl = {https://ui.adsabs.harvard.edu/abs/2016ApJ...817..108W}
}

@ARTICLE{Karouzos16a,
       author = {{Karouzos}, Marios and {Woo}, Jong-Hak and {Bae}, Hyun-Jin},
        title = "{Unraveling the Complex Structure of AGN-driven Outflows. I. Kinematics and Sizes}",
      journal = {\apj},
         year = 2016,
        month = mar,
       volume = {819},
       number = {2},
          eid = {148},
        pages = {148},
          doi = {10.3847/0004-637X/819/2/148},
archivePrefix = {arXiv},
       eprint = {1601.02621},
 primaryClass = {astro-ph.GA},
       adsurl = {https://ui.adsabs.harvard.edu/abs/2016ApJ...819..148K}
}

@ARTICLE{Marconcini25a,
       author = {{Marconcini}, C. and {Feltre}, A. and {Lamperti}, I. and {Ceci}, M. and {Marconi}, A. and {Ulivi}, L. and {Mannucci}, F. and {Cresci}, G. and {Belfiore}, F. and {Bertola}, E. and {Carniani}, S. and {D'Amato}, Q. and {Fernandez-Ontiveros}, J.~A. and {Fritz}, J. and {Ginolfi}, M. and {Hatziminaoglou}, E. and {Hern{\'a}n-Caballero}, A. and {Hirschmann}, M. and {Mingozzi}, M. and {Rojas}, A.~F. and {Sabatini}, G. and {Salvestrini}, F. and {Scialpi}, M. and {Tozzi}, G. and {Venturi}, G. and {Vidal-Garc{\'\i}a}, A. and {Vignali}, C. and {Zanchettin}, M.~V. and {Amiri}, A.},
        title = "{MIRACLE: I. Unveiling the multi-phase, multi-scale physical properties of the active galaxy NGC 424 with MIRI, MUSE, and ALMA}",
      journal = {\aap},
         year = 2025,
        month = sep,
       volume = {701},
          eid = {A113},
        pages = {A113},
          doi = {10.1051/0004-6361/202554797},
       adsurl = {https://ui.adsabs.harvard.edu/abs/2025A&A...701A.113M}
}

@ARTICLE{Marconcini25b,
       author = {{Marconcini}, Cosimo and {Marconi}, Alessandro and {Cresci}, Giovanni and {Mannucci}, Filippo and {Ulivi}, Lorenzo and {Venturi}, Giacomo and {Scialpi}, Martina and {Tozzi}, Giulia and {Belfiore}, Francesco and {Bertola}, Elena and {Carniani}, Stefano and {Cataldi}, Elisa and {Chakraborty}, Avinanda and {D'Amato}, Quirino and {Di Teodoro}, Enrico and {Feltre}, Anna and {Ginolfi}, Michele and {Moreschini}, Bianca and {Orientale}, Nicole and {Trefoloni}, Bartolomeo and {King}, Andrew},
        title = "{Evidence of the fast acceleration of AGN-driven winds at kiloparsec scales}",
      journal = {Nature Astronomy},
         year = 2025,
        month = jun,
       volume = {9},
        pages = {907-915},
          doi = {10.1038/s41550-025-02518-6},
archivePrefix = {arXiv},
       eprint = {2503.24359},
 primaryClass = {astro-ph.GA},
       adsurl = {https://ui.adsabs.harvard.edu/abs/2025NatAs...9..907M}
}

@ARTICLE{Ibarra-Medel25,
       author = {{Ibarra-Medel}, H. and {Negrete}, C.~A. and {Lacerna}, I. and {Hern{\'a}ndez-Toledo}, H.~M. and {Cortes-Su{\'a}rez}, E. and {S{\'a}nchez}, S.~F.},
        title = "{An iterative method to deblend AGN-Host contributions for Integral Field spectroscopic observations}",
      journal = {\mnras},
         year = 2025,
        month = jan,
       volume = {536},
       number = {1},
        pages = {752-776},
          doi = {10.1093/mnras/stae2623},
archivePrefix = {arXiv},
       eprint = {2411.13270},
 primaryClass = {astro-ph.GA},
       adsurl = {https://ui.adsabs.harvard.edu/abs/2025MNRAS.536..752I}
}

@ARTICLE{Costa-Souza24,
       author = {{Costa-Souza}, Jos{\'e} Henrique and {Riffel}, Rogemar A. and {Dors}, Oli L. and {Riffel}, Rog{\'e}rio and {da Rocha-Poppe}, Paulo C.},
        title = "{AGN feedback and star formation in the peculiar galaxy NGC 232: insights from VLT-MUSE observations}",
      journal = {\mnras},
         year = 2024,
        month = jan,
       volume = {527},
       number = {3},
        pages = {9192-9205},
          doi = {10.1093/mnras/stad3809},
archivePrefix = {arXiv},
       eprint = {2310.15842},
 primaryClass = {astro-ph.GA},
       adsurl = {https://ui.adsabs.harvard.edu/abs/2024MNRAS.527.9192C}
}

@ARTICLE{Gatto24,
       author = {{Gatto}, Lara and {Storchi-Bergmann}, T. and {Riffel}, Rogemar A. and {Riffel}, Rog{\'e}rio and {Rembold}, Sandro B. and {Schimoia}, Jaderson S. and {Mallmann}, Nicolas D. and {Ilha}, Gabriele S.},
        title = "{The extent and power of 'maintenance mode' feedback in MaNGA AGN}",
      journal = {\mnras},
         year = 2024,
        month = may,
       volume = {530},
       number = {3},
        pages = {3059-3074},
          doi = {10.1093/mnras/stae989},
archivePrefix = {arXiv},
       eprint = {2404.14502},
 primaryClass = {astro-ph.GA},
       adsurl = {https://ui.adsabs.harvard.edu/abs/2024MNRAS.530.3059G}
}

@ARTICLE{Oh24,
       author = {{Oh}, Sree and {Colless}, Matthew and {Barsanti}, Stefania and {Zovaro}, Henry R.~M. and {Croom}, Scott M. and {Yi}, Sukyoung K. and {Ristea}, Andrei and {van de Sande}, Jesse and {D'Eugenio}, Francesco and {Bland-Hawthorn}, Joss and {Bryant}, Julia J. and {Casura}, Sarah and {Jeong}, Hyunjin and {Sweet}, Sarah M. and {Zafar}, Tayyaba},
        title = "{The SAMI Galaxy Survey: impact of star formation and AGN feedback processes on the ionized gas velocity dispersion}",
      journal = {\mnras},
         year = 2024,
        month = jul,
       volume = {531},
       number = {4},
        pages = {4017-4032},
          doi = {10.1093/mnras/stae1382},
archivePrefix = {arXiv},
       eprint = {2405.20627},
 primaryClass = {astro-ph.GA},
       adsurl = {https://ui.adsabs.harvard.edu/abs/2024MNRAS.531.4017O}
}

@ARTICLE{Polack24,
       author = {{Polack}, Garrett E. and {Revalski}, Mitchell and {Crenshaw}, D. Michael and {Fischer}, Travis C. and {Schmitt}, Henrique R. and {Kraemer}, Steven B. and {Meena}, Beena and {Rafelski}, Marc},
        title = "{Determining the Extents, Geometries, and Kinematics of Narrow-line Region Outflows in Nearby Seyfert Galaxies}",
      journal = {\apj},
         year = 2024,
        month = nov,
       volume = {975},
       number = {1},
          eid = {129},
        pages = {129},
          doi = {10.3847/1538-4357/ad71c3},
       adsurl = {https://ui.adsabs.harvard.edu/abs/2024ApJ...975..129P}
}

@ARTICLE{Harrison24,
       author = {{Harrison}, Chris M. and {Ramos Almeida}, Cristina},
        title = "{Observational Tests of Active Galactic Nuclei Feedback: An Overview of Approaches and Interpretation}",
      journal = {Galaxies},
         year = 2024,
        month = apr,
       volume = {12},
       number = {2},
          eid = {17},
        pages = {17},
          doi = {10.3390/galaxies12020017},
archivePrefix = {arXiv},
       eprint = {2404.08050},
 primaryClass = {astro-ph.GA},
       adsurl = {https://ui.adsabs.harvard.edu/abs/2024Galax..12...17H}
}

@ARTICLE{Falcone24,
       author = {{Falcone}, Julia and {Crenshaw}, D. Michael and {Fischer}, Travis C. and {Meena}, Beena and {Revalski}, Mitchell and {Shea}, Maura Kathleen and {Riffel}, Rogemar A. and {Chapman}, Zo and {Ferree}, Nicolas and {Tutterow}, Jacob and {Davis}, Madeline},
        title = "{An Analysis of Active Galactic Nucleus{\textendash}driven Outflows in the Seyfert 1 Galaxy NGC 3227}",
      journal = {\apj},
         year = 2024,
        month = aug,
       volume = {971},
       number = {1},
          eid = {17},
        pages = {17},
          doi = {10.3847/1538-4357/ad5283},
archivePrefix = {arXiv},
       eprint = {2405.20162},
 primaryClass = {astro-ph.GA},
       adsurl = {https://ui.adsabs.harvard.edu/abs/2024ApJ...971...17F}
}

@ARTICLE{Cresci23,
       author = {{Cresci}, G. and {Tozzi}, G. and {Perna}, M. and {Brusa}, M. and {Marconcini}, C. and {Marconi}, A. and {Carniani}, S. and {Brienza}, M. and {Giroletti}, M. and {Belfiore}, F. and {Ginolfi}, M. and {Mannucci}, F. and {Ulivi}, L. and {Scholtz}, J. and {Venturi}, G. and {Arribas}, S. and {{\"U}bler}, H. and {D'Eugenio}, F. and {Mingozzi}, M. and {Balmaverde}, B. and {Capetti}, A. and {Parlanti}, E. and {Zana}, T.},
        title = "{Bubbles and outflows: The novel JWST/NIRSpec view of the z = 1.59 obscured quasar XID2028}",
      journal = {\aap},
         year = 2023,
        month = apr,
       volume = {672},
          eid = {A128},
        pages = {A128},
          doi = {10.1051/0004-6361/202346001},
archivePrefix = {arXiv},
       eprint = {2301.11060},
 primaryClass = {astro-ph.GA},
       adsurl = {https://ui.adsabs.harvard.edu/abs/2023A&A...672A.128C}
}

@ARTICLE{Marconcini23,
       author = {{Marconcini}, C. and {Marconi}, A. and {Cresci}, G. and {Venturi}, G. and {Ulivi}, L. and {Mannucci}, F. and {Belfiore}, F. and {Tozzi}, G. and {Ginolfi}, M. and {Marasco}, A. and {Carniani}, S. and {Amiri}, A. and {Di Teodoro}, E. and {Scialpi}, M. and {Tomicic}, N. and {Mingozzi}, M. and {Brazzini}, M. and {Moreschini}, B.},
        title = "{MOKA$^{3D}$: An innovative approach to 3D gas kinematic modelling. I. Application to AGN ionised outflows}",
      journal = {\aap},
         year = 2023,
        month = sep,
       volume = {677},
          eid = {A58},
        pages = {A58},
          doi = {10.1051/0004-6361/202346821},
archivePrefix = {arXiv},
       eprint = {2307.01854},
 primaryClass = {astro-ph.GA},
       adsurl = {https://ui.adsabs.harvard.edu/abs/2023A&A...677A..58M}
}

@ARTICLE{Riffel23,
       author = {{Riffel}, R.~A. and {Storchi-Bergmann}, T. and {Riffel}, R. and {Bianchin}, M. and {Zakamska}, N.~L. and {Ruschel-Dutra}, D. and {Bentz}, M.~C. and {Burtscher}, L. and {Crenshaw}, D.~M. and {Dahmer-Hahn}, L.~G. and {Dametto}, N.~Z. and {Davies}, R.~I. and {Diniz}, M.~R. and {Fischer}, T.~C. and {Harrison}, C.~M. and {Mainieri}, V. and {Revalski}, M. and {Rodriguez-Ardila}, A. and {Rosario}, D.~J. and {Sch{\"o}nell}, A.~J.},
        title = "{The AGNIFS survey: spatially resolved observations of hot molecular and ionized outflows in nearby active galaxies}",
      journal = {\mnras},
         year = 2023,
        month = may,
       volume = {521},
       number = {2},
        pages = {1832-1848},
          doi = {10.1093/mnras/stad599},
archivePrefix = {arXiv},
       eprint = {2302.11324},
 primaryClass = {astro-ph.GA},
       adsurl = {https://ui.adsabs.harvard.edu/abs/2023MNRAS.521.1832R}
}

@ARTICLE{Wellons23,
       author = {{Wellons}, Sarah and {Faucher-Gigu{\`e}re}, Claude-Andr{\'e} and {Hopkins}, Philip F. and {Quataert}, Eliot and {Angl{\'e}s-Alc{\'a}zar}, Daniel and {Feldmann}, Robert and {Hayward}, Christopher C. and {Kere{\v{s}}}, Du{\v{s}}an and {Su}, Kung-Yi and {Wetzel}, Andrew},
        title = "{Exploring supermassive black hole physics and galaxy quenching across halo mass in FIRE cosmological zoom simulations}",
      journal = {\mnras},
         year = 2023,
        month = apr,
       volume = {520},
       number = {4},
        pages = {5394-5412},
          doi = {10.1093/mnras/stad511},
archivePrefix = {arXiv},
       eprint = {2203.06201},
 primaryClass = {astro-ph.GA},
       adsurl = {https://ui.adsabs.harvard.edu/abs/2023MNRAS.520.5394W}
}

@ARTICLE{Kakkad23,
       author = {{Kakkad}, D. and {Mainieri}, V. and {Vietri}, G. and {Lamperti}, I. and {Carniani}, S. and {Cresci}, G. and {Harrison}, C. and {Marconi}, A. and {Bischetti}, M. and {Cicone}, C. and {Circosta}, C. and {Husemann}, B. and {Man}, A. and {Mannucci}, F. and {Netzer}, H. and {Padovani}, P. and {Perna}, M. and {Puglisi}, A. and {Scholtz}, J. and {Tozzi}, G. and {Vignali}, C. and {Zappacosta}, L.},
        title = "{SUPER VII. morphology and kinematics of H {\ensuremath{\alpha}} emission in AGN host galaxies at cosmic noon using SINFONI}",
      journal = {\mnras},
         year = 2023,
        month = apr,
       volume = {520},
       number = {4},
        pages = {5783-5802},
          doi = {10.1093/mnras/stad439},
archivePrefix = {arXiv},
       eprint = {2302.03039},
 primaryClass = {astro-ph.GA},
       adsurl = {https://ui.adsabs.harvard.edu/abs/2023MNRAS.520.5783K}
}

@ARTICLE{Wylezalek22,
       author = {{Wylezalek}, Dominika and {Vayner}, Andrey and {Rupke}, David S.~N. and {Zakamska}, Nadia L. and {Veilleux}, Sylvain and {Ishikawa}, Yuzo and {Bertemes}, Caroline and {Liu}, Weizhe and {Barrera-Ballesteros}, Jorge K. and {Chen}, Hsiao-Wen and {Goulding}, Andy D. and {Greene}, Jenny E. and {Hainline}, Kevin N. and {Hamann}, Fred and {Heckman}, Timothy and {Johnson}, Sean D. and {Lutz}, Dieter and {L{\"u}tzgendorf}, Nora and {Mainieri}, Vincenzo and {Maiolino}, Roberto and {Nesvadba}, Nicole P.~H. and {Ogle}, Patrick and {Sturm}, Eckhard},
        title = "{First Results from the JWST Early Release Science Program Q3D: Turbulent Times in the Life of a z   3 Extremely Red Quasar Revealed by NIRSpec IFU}",
      journal = {\apjl},
         year = 2022,
        month = nov,
       volume = {940},
       number = {1},
          eid = {L7},
        pages = {L7},
          doi = {10.3847/2041-8213/ac98c3},
archivePrefix = {arXiv},
       eprint = {2210.10074},
 primaryClass = {astro-ph.GA},
       adsurl = {https://ui.adsabs.harvard.edu/abs/2022ApJ...940L...7W}
}

@ARTICLE{Deconto-Machado22,
       author = {{Deconto-Machado}, A. and {Riffel}, R.~A. and {Ilha}, G.~S. and {Rembold}, S.~B. and {Storchi-Bergmann}, T. and {Riffel}, R. and {Schimoia}, J.~S. and {Schneider}, D.~P. and {Bizyaev}, D. and {Feng}, S. and {Wylezalek}, D. and {da Costa}, L.~N. and {do Nascimento}, J.~C. and {Maia}, M.~A.~G.},
        title = "{Ionised gas kinematics in MaNGA AGN. Extents of the narrow-line and kinematically disturbed regions}",
      journal = {\aap},
         year = 2022,
        month = mar,
       volume = {659},
          eid = {A131},
        pages = {A131},
          doi = {10.1051/0004-6361/202140613},
archivePrefix = {arXiv},
       eprint = {2201.01690},
 primaryClass = {astro-ph.GA},
       adsurl = {https://ui.adsabs.harvard.edu/abs/2022A&A...659A.131D}
}

@ARTICLE{Singha22,
       author = {{Singha}, M. and {Husemann}, B. and {Urrutia}, T. and {O'Dea}, C.~P. and {Scharw{\"a}chter}, J. and {Gaspari}, M. and {Combes}, F. and {Nevin}, R. and {Terrazas}, B.~A. and {P{\'e}rez-Torres}, M. and {Rose}, T. and {Davis}, T.~A. and {Tremblay}, G.~R. and {Neumann}, J. and {Smirnova-Pinchukova}, I. and {Baum}, S.~A.},
        title = "{The Close AGN Reference Survey (CARS). Locating the [O III] wing component in luminous local Type 1 AGN}",
      journal = {\aap},
         year = 2022,
        month = mar,
       volume = {659},
          eid = {A123},
        pages = {A123},
          doi = {10.1051/0004-6361/202040122},
archivePrefix = {arXiv},
       eprint = {2111.10418},
 primaryClass = {astro-ph.GA},
       adsurl = {https://ui.adsabs.harvard.edu/abs/2022A&A...659A.123S}
}

@ARTICLE{Oh22,
       author = {{Oh}, Sree and {Colless}, Matthew and {D'Eugenio}, Francesco and {Croom}, Scott M. and {Cortese}, Luca and {Groves}, Brent and {Kewley}, Lisa J. and {van de Sande}, Jesse and {Zovaro}, Henry and {Varidel}, Mathew R. and {Barsanti}, Stefania and {Bland-Hawthorn}, Joss and {Brough}, Sarah and {Bryant}, Julia J. and {Casura}, Sarah and {Lawrence}, Jon S. and {Lorente}, Nuria P.~F. and {Medling}, Anne M. and {Owers}, Matt S. and {Yi}, Sukyoung K.},
        title = "{The SAMI Galaxy Survey: the difference between ionized gas and stellar velocity dispersions}",
      journal = {\mnras},
         year = 2022,
        month = may,
       volume = {512},
       number = {2},
        pages = {1765-1780},
          doi = {10.1093/mnras/stac509},
archivePrefix = {arXiv},
       eprint = {2202.10469},
 primaryClass = {astro-ph.GA},
       adsurl = {https://ui.adsabs.harvard.edu/abs/2022MNRAS.512.1765O}
}

@ARTICLE{Riffel21,
       author = {{Riffel}, R.~A.},
        title = "{Powerful multiphase outflows in the central region of Cygnus A}",
      journal = {\mnras},
         year = 2021,
        month = sep,
       volume = {506},
       number = {2},
        pages = {2950-2962},
          doi = {10.1093/mnras/stab1877},
archivePrefix = {arXiv},
       eprint = {2106.15279},
 primaryClass = {astro-ph.GA},
       adsurl = {https://ui.adsabs.harvard.edu/abs/2021MNRAS.506.2950R}
}

@ARTICLE{Meena21,
       author = {{Meena}, Beena and {Crenshaw}, D. Michael and {Schmitt}, Henrique R. and {Revalski}, Mitchell and {Fischer}, Travis C. and {Polack}, Garrett E. and {Kraemer}, Steven B. and {Dashtamirova}, Dzhuliya},
        title = "{Radiative Driving of the AGN Outflows in the Narrow-line Seyfert 1 Galaxy NGC 4051}",
      journal = {\apj},
         year = 2021,
        month = jul,
       volume = {916},
       number = {1},
          eid = {31},
        pages = {31},
          doi = {10.3847/1538-4357/ac0246},
archivePrefix = {arXiv},
       eprint = {2103.12081},
 primaryClass = {astro-ph.GA},
       adsurl = {https://ui.adsabs.harvard.edu/abs/2021ApJ...916...31M}
}

@ARTICLE{Vayner21a,
       author = {{Vayner}, Andrey and {Zakamska}, Nadia L. and {Riffel}, Rogemar A. and {Alexandroff}, Rachael and {Cosens}, Maren and {Hamann}, Fred and {Perrotta}, Serena and {Rupke}, David S.~N. and {Bergmann}, Thaisa Storchi and {Veilleux}, Sylvain and {Walth}, Greg and {Wright}, Shelley and {Wylezalek}, Dominika},
        title = "{Powerful winds in high-redshift obscured and red quasars}",
      journal = {\mnras},
         year = 2021,
        month = jul,
       volume = {504},
       number = {3},
        pages = {4445-4459},
          doi = {10.1093/mnras/stab1176},
archivePrefix = {arXiv},
       eprint = {2101.04688},
 primaryClass = {astro-ph.GA},
       adsurl = {https://ui.adsabs.harvard.edu/abs/2021MNRAS.504.4445V}
}

@ARTICLE{Ruschel-Dutra21,
       author = {{Ruschel-Dutra}, D. and {Storchi-Bergmann}, T. and {Schnorr-M{\"u}ller}, A. and {Riffel}, R.~A. and {Dall'Agnol de Oliveira}, B. and {Lena}, D. and {Robinson}, A. and {Nagar}, N. and {Elvis}, M.},
        title = "{AGNIFS survey of local AGN: GMOS-IFU data and outflows in 30 sources}",
      journal = {\mnras},
         year = 2021,
        month = oct,
       volume = {507},
       number = {1},
        pages = {74-89},
          doi = {10.1093/mnras/stab2058},
archivePrefix = {arXiv},
       eprint = {2107.07635},
 primaryClass = {astro-ph.GA},
       adsurl = {https://ui.adsabs.harvard.edu/abs/2021MNRAS.507...74R}
}

@ARTICLE{Wylezalek20,
       author = {{Wylezalek}, Dominika and {Flores}, Anthony M. and {Zakamska}, Nadia L. and {Greene}, Jenny E. and {Riffel}, Rogemar A.},
        title = "{Ionized gas outflow signatures in SDSS-IV MaNGA active galactic nuclei}",
      journal = {\mnras},
         year = 2020,
        month = mar,
       volume = {492},
       number = {4},
        pages = {4680-4696},
          doi = {10.1093/mnras/staa062},
archivePrefix = {arXiv},
       eprint = {1911.10212},
 primaryClass = {astro-ph.GA},
       adsurl = {https://ui.adsabs.harvard.edu/abs/2020MNRAS.492.4680W}
}

@ARTICLE{Veilleux20,
       author = {{Veilleux}, Sylvain and {Maiolino}, Roberto and {Bolatto}, Alberto D. and {Aalto}, Susanne},
        title = "{Cool outflows in galaxies and their implications}",
      journal = {\aapr},
         year = 2020,
        month = apr,
       volume = {28},
       number = {1},
          eid = {2},
        pages = {2},
          doi = {10.1007/s00159-019-0121-9},
archivePrefix = {arXiv},
       eprint = {2002.07765},
 primaryClass = {astro-ph.GA},
       adsurl = {https://ui.adsabs.harvard.edu/abs/2020A&ARv..28....2V}
}

@ARTICLE{Durre19,
       author = {{Durr{\'e}}, Mark and {Mould}, Jeremy},
        title = "{The AGN Ionization Cones of NGC 5728. II. Kinematics}",
      journal = {\apj},
         year = 2019,
        month = jan,
       volume = {870},
       number = {1},
          eid = {37},
        pages = {37},
          doi = {10.3847/1538-4357/aaf000},
archivePrefix = {arXiv},
       eprint = {1811.04513},
 primaryClass = {astro-ph.GA},
       adsurl = {https://ui.adsabs.harvard.edu/abs/2019ApJ...870...37D}
}

@ARTICLE{Mingozzi19,
       author = {{Mingozzi}, M. and {Cresci}, G. and {Venturi}, G. and {Marconi}, A. and {Mannucci}, F. and {Perna}, M. and {Belfiore}, F. and {Carniani}, S. and {Balmaverde}, B. and {Brusa}, M. and {Cicone}, C. and {Feruglio}, C. and {Gallazzi}, A. and {Mainieri}, V. and {Maiolino}, R. and {Nagao}, T. and {Nardini}, E. and {Sani}, E. and {Tozzi}, P. and {Zibetti}, S.},
        title = "{The MAGNUM survey: different gas properties in the outflowing and disc components in nearby active galaxies with MUSE}",
      journal = {\aap},
         year = 2019,
        month = feb,
       volume = {622},
          eid = {A146},
        pages = {A146},
          doi = {10.1051/0004-6361/201834372},
archivePrefix = {arXiv},
       eprint = {1811.07935},
 primaryClass = {astro-ph.GA},
       adsurl = {https://ui.adsabs.harvard.edu/abs/2019A&A...622A.146M}
}

@ARTICLE{Fluetsch19,
       author = {{Fluetsch}, A. and {Maiolino}, R. and {Carniani}, S. and {Marconi}, A. and {Cicone}, C. and {Bourne}, M.~A. and {Costa}, T. and {Fabian}, A.~C. and {Ishibashi}, W. and {Venturi}, G.},
        title = "{Cold molecular outflows in the local Universe and their feedback effect on galaxies}",
      journal = {\mnras},
         year = 2019,
        month = mar,
       volume = {483},
       number = {4},
        pages = {4586-4614},
          doi = {10.1093/mnras/sty3449},
archivePrefix = {arXiv},
       eprint = {1805.05352},
 primaryClass = {astro-ph.GA},
       adsurl = {https://ui.adsabs.harvard.edu/abs/2019MNRAS.483.4586F}
}

@ARTICLE{Nevin18,
       author = {{Nevin}, R. and {Comerford}, J.~M. and {M{\"u}ller-S{\'a}nchez}, F. and {Barrows}, R. and {Cooper}, M.~C.},
        title = "{The origin of double-peaked narrow lines in active galactic nuclei - III. Feedback from biconical AGN outflows}",
      journal = {\mnras},
         year = 2018,
        month = jan,
       volume = {473},
       number = {2},
        pages = {2160-2187},
          doi = {10.1093/mnras/stx2433},
archivePrefix = {arXiv},
       eprint = {1710.00828},
 primaryClass = {astro-ph.GA},
       adsurl = {https://ui.adsabs.harvard.edu/abs/2018MNRAS.473.2160N}
}

@ARTICLE{Baron18,
       author = {{Baron}, Dalya and {Netzer}, Hagai and {Prochaska}, J. Xavier and {Cai}, Zheng and {Cantalupo}, Sebastiano and {Martin}, D. Christopher and {Matuszewski}, Mateusz and {Moore}, Anna M. and {Morrissey}, Patrick and {Neill}, James D.},
        title = "{Direct evidence of AGN feedback: a post-starburst galaxy stripped of its gas by AGN-driven winds}",
      journal = {\mnras},
         year = 2018,
        month = nov,
       volume = {480},
       number = {3},
        pages = {3993-4016},
          doi = {10.1093/mnras/sty2113},
archivePrefix = {arXiv},
       eprint = {1804.03150},
 primaryClass = {astro-ph.GA},
       adsurl = {https://ui.adsabs.harvard.edu/abs/2018MNRAS.480.3993B}
}

@ARTICLE{Venturi18,
       author = {{Venturi}, Giacomo and {Nardini}, Emanuele and {Marconi}, Alessandro and {Carniani}, Stefano and {Mingozzi}, Matilde and {Cresci}, Giovanni and {Mannucci}, Filippo and {Risaliti}, Guido and {Maiolino}, Roberto and {Balmaverde}, Barbara and {Bongiorno}, Angela and {Brusa}, Marcella and {Capetti}, Alessandro and {Cicone}, Claudia and {Ciroi}, Stefano and {Feruglio}, Chiara and {Fiore}, Fabrizio and {Gallazzi}, Anna and {La Franca}, Fabio and {Mainieri}, Vincenzo and {Matsuoka}, Kenta and {Nagao}, Tohru and {Perna}, Michele and {Piconcelli}, Enrico and {Sani}, Eleonora and {Tozzi}, Paolo and {Zibetti}, Stefano},
        title = "{MAGNUM survey: A MUSE-Chandra resolved view on ionized outflows and photoionization in the Seyfert galaxy NGC1365}",
      journal = {\aap},
         year = 2018,
        month = nov,
       volume = {619},
          eid = {A74},
        pages = {A74},
          doi = {10.1051/0004-6361/201833668},
archivePrefix = {arXiv},
       eprint = {1809.01206},
 primaryClass = {astro-ph.GA},
       adsurl = {https://ui.adsabs.harvard.edu/abs/2018A&A...619A..74V}
}

@ARTICLE{Fischer18,
       author = {{Fischer}, Travis C. and {Kraemer}, S.~B. and {Schmitt}, H.~R. and {Longo Micchi}, L.~F. and {Crenshaw}, D.~M. and {Revalski}, M. and {Vestergaard}, M. and {Elvis}, M. and {Gaskell}, C.~M. and {Hamann}, F. and {Ho}, L.~C. and {Hutchings}, J. and {Mushotzky}, R. and {Netzer}, H. and {Storchi-Bergmann}, T. and {Straughn}, A. and {Turner}, T.~J. and {Ward}, M.~J.},
        title = "{Hubble Space Telescope Observations of Extended [O III]{\ensuremath{\lambda}} 5007 Emission in Nearby QSO2s: New Constraints on AGN Host Galaxy Interaction}",
      journal = {\apj},
         year = 2018,
        month = apr,
       volume = {856},
       number = {2},
          eid = {102},
        pages = {102},
          doi = {10.3847/1538-4357/aab03e},
archivePrefix = {arXiv},
       eprint = {1802.06184},
 primaryClass = {astro-ph.GA},
       adsurl = {https://ui.adsabs.harvard.edu/abs/2018ApJ...856..102F}
}

@ARTICLE{Weinberger17,
       author = {{Weinberger}, Rainer and {Springel}, Volker and {Hernquist}, Lars and {Pillepich}, Annalisa and {Marinacci}, Federico and {Pakmor}, R{\"u}diger and {Nelson}, Dylan and {Genel}, Shy and {Vogelsberger}, Mark and {Naiman}, Jill and {Torrey}, Paul},
        title = "{Simulating galaxy formation with black hole driven thermal and kinetic feedback}",
      journal = {\mnras},
         year = 2017,
        month = mar,
       volume = {465},
       number = {3},
        pages = {3291-3308},
          doi = {10.1093/mnras/stw2944},
archivePrefix = {arXiv},
       eprint = {1607.03486},
 primaryClass = {astro-ph.GA},
       adsurl = {https://ui.adsabs.harvard.edu/abs/2017MNRAS.465.3291W}
}

@ARTICLE{Venturi17,
       author = {{Venturi}, Giacomo and {Marconi}, Alessandro and {Mingozzi}, Matilde and {Carniani}, Stefano and {Cresci}, Giovanni and {Risaliti}, Guido and {Mannucci}, Filippo},
        title = "{Ionized gas outflows from the MAGNUM survey: NGC 1365 and NGC 4945}",
      journal = {Frontiers in Astronomy and Space Sciences},
         year = 2017,
        month = nov,
       volume = {4},
          eid = {46},
        pages = {46},
          doi = {10.3389/fspas.2017.00046},
archivePrefix = {arXiv},
       eprint = {1801.05448},
 primaryClass = {astro-ph.GA},
       adsurl = {https://ui.adsabs.harvard.edu/abs/2017FrASS...4...46V}
}

@ARTICLE{Harrison17,
       author = {{Harrison}, C.~M.},
        title = "{Impact of supermassive black hole growth on star formation}",
      journal = {Nature Astronomy},
         year = 2017,
        month = jul,
       volume = {1},
          eid = {0165},
        pages = {0165},
          doi = {10.1038/s41550-017-0165},
archivePrefix = {arXiv},
       eprint = {1703.06889},
 primaryClass = {astro-ph.GA},
       adsurl = {https://ui.adsabs.harvard.edu/abs/2017NatAs...1E.165H}
}

@ARTICLE{Fiore17,
       author = {{Fiore}, F. and {Feruglio}, C. and {Shankar}, F. and {Bischetti}, M. and {Bongiorno}, A. and {Brusa}, M. and {Carniani}, S. and {Cicone}, C. and {Duras}, F. and {Lamastra}, A. and {Mainieri}, V. and {Marconi}, A. and {Menci}, N. and {Maiolino}, R. and {Piconcelli}, E. and {Vietri}, G. and {Zappacosta}, L.},
        title = "{AGN wind scaling relations and the co-evolution of black holes and galaxies}",
      journal = {\aap},
         year = 2017,
        month = may,
       volume = {601},
          eid = {A143},
        pages = {A143},
          doi = {10.1051/0004-6361/201629478},
archivePrefix = {arXiv},
       eprint = {1702.04507},
 primaryClass = {astro-ph.GA},
       adsurl = {https://ui.adsabs.harvard.edu/abs/2017A&A...601A.143F}
}

@ARTICLE{Villar-Martin16,
       author = {{Villar-Mart{\'\i}n}, M. and {Arribas}, S. and {Emonts}, B. and {Humphrey}, A. and {Tadhunter}, C. and {Bessiere}, P. and {Cabrera Lavers}, A. and {Ramos Almeida}, C.},
        title = "{Ionized outflows in luminous type 2 AGNs at z < 0.6: no evidence for significant impact on the host galaxies}",
      journal = {\mnras},
         year = 2016,
        month = jul,
       volume = {460},
       number = {1},
        pages = {130-162},
          doi = {10.1093/mnras/stw901},
archivePrefix = {arXiv},
       eprint = {1604.04577},
 primaryClass = {astro-ph.GA},
       adsurl = {https://ui.adsabs.harvard.edu/abs/2016MNRAS.460..130V}
}

@ARTICLE{Muller-Sanchez16,
       author = {{M{\"u}ller-S{\'a}nchez}, F. and {Comerford}, J. and {Stern}, D. and {Harrison}, F.~A.},
        title = "{The Nature of Active Galactic Nuclei with Velocity Offset Emission Lines}",
      journal = {\apj},
         year = 2016,
        month = oct,
       volume = {830},
       number = {1},
          eid = {50},
        pages = {50},
          doi = {10.3847/0004-637X/830/1/50},
archivePrefix = {arXiv},
       eprint = {1606.07446},
 primaryClass = {astro-ph.GA},
       adsurl = {https://ui.adsabs.harvard.edu/abs/2016ApJ...830...50M}
}

@ARTICLE{Husemann16,
       author = {{Husemann}, B. and {Scharw{\"a}chter}, J. and {Bennert}, V.~N. and {Mainieri}, V. and {Woo}, J. -H. and {Kakkad}, D.},
        title = "{Large-scale outflows in luminous QSOs revisited. The impact of beam smearing on AGN feedback efficiencies}",
      journal = {\aap},
         year = 2016,
        month = oct,
       volume = {594},
          eid = {A44},
        pages = {A44},
          doi = {10.1051/0004-6361/201527992},
archivePrefix = {arXiv},
       eprint = {1512.05595},
 primaryClass = {astro-ph.GA},
       adsurl = {https://ui.adsabs.harvard.edu/abs/2016A&A...594A..44H}
}

@ARTICLE{Carniani15,
       author = {{Carniani}, S. and {Marconi}, A. and {Maiolino}, R. and {Balmaverde}, B. and {Brusa}, M. and {Cano-D{\'\i}az}, M. and {Cicone}, C. and {Comastri}, A. and {Cresci}, G. and {Fiore}, F. and {Feruglio}, C. and {La Franca}, F. and {Mainieri}, V. and {Mannucci}, F. and {Nagao}, T. and {Netzer}, H. and {Piconcelli}, E. and {Risaliti}, G. and {Schneider}, R. and {Shemmer}, O.},
        title = "{Ionised outflows in z \raisebox{-0.5ex}\textasciitilde 2.4 quasar host galaxies}",
      journal = {\aap},
         year = 2015,
        month = aug,
       volume = {580},
          eid = {A102},
        pages = {A102},
          doi = {10.1051/0004-6361/201526557},
archivePrefix = {arXiv},
       eprint = {1506.03096},
 primaryClass = {astro-ph.GA},
       adsurl = {https://ui.adsabs.harvard.edu/abs/2015A&A...580A.102C}
}

@ARTICLE{Harrison14,
       author = {{Harrison}, C.~M. and {Alexander}, D.~M. and {Mullaney}, J.~R. and {Swinbank}, A.~M.},
        title = "{Kiloparsec-scale outflows are prevalent among luminous AGN: outflows and feedback in the context of the overall AGN population}",
      journal = {\mnras},
         year = 2014,
        month = jul,
       volume = {441},
       number = {4},
        pages = {3306-3347},
          doi = {10.1093/mnras/stu515},
archivePrefix = {arXiv},
       eprint = {1403.3086},
 primaryClass = {astro-ph.GA},
       adsurl = {https://ui.adsabs.harvard.edu/abs/2014MNRAS.441.3306H}
}

@ARTICLE{Marin14,
       author = {{Marin}, F.},
        title = "{A compendium of AGN inclinations with corresponding UV/optical continuum polarization measurements}",
      journal = {\mnras},
         year = 2014,
        month = jun,
       volume = {441},
       number = {1},
        pages = {551-564},
          doi = {10.1093/mnras/stu593},
archivePrefix = {arXiv},
       eprint = {1404.2417},
 primaryClass = {astro-ph.GA},
       adsurl = {https://ui.adsabs.harvard.edu/abs/2014MNRAS.441..551M}
}

@ARTICLE{Husemann14,
       author = {{Husemann}, B. and {Jahnke}, K. and {S{\'a}nchez}, S.~F. and {Wisotzki}, L. and {Nugroho}, D. and {Kupko}, D. and {Schramm}, M.},
        title = "{Integral field spectroscopy of nearby QSOs - I. ENLR size-luminosity relation, ongoing star formation and resolved gas-phase metallicities}",
      journal = {\mnras},
         year = 2014,
        month = sep,
       volume = {443},
       number = {1},
        pages = {755-783},
          doi = {10.1093/mnras/stu1167},
archivePrefix = {arXiv},
       eprint = {1406.4131},
 primaryClass = {astro-ph.GA},
       adsurl = {https://ui.adsabs.harvard.edu/abs/2014MNRAS.443..755H}
}

@ARTICLE{Mullaney13,
       author = {{Mullaney}, J.~R. and {Alexander}, D.~M. and {Fine}, S. and {Goulding}, A.~D. and {Harrison}, C.~M. and {Hickox}, R.~C.},
        title = "{Narrow-line region gas kinematics of 24 264 optically selected AGN: the radio connection}",
      journal = {\mnras},
         year = 2013,
        month = jul,
       volume = {433},
       number = {1},
        pages = {622-638},
          doi = {10.1093/mnras/stt751},
archivePrefix = {arXiv},
       eprint = {1305.0263},
 primaryClass = {astro-ph.CO},
       adsurl = {https://ui.adsabs.harvard.edu/abs/2013MNRAS.433..622M}
}

@ARTICLE{Liu13a,
       author = {{Liu}, Guilin and {Zakamska}, Nadia L. and {Greene}, Jenny E. and {Nesvadba}, Nicole P.~H. and {Liu}, Xin},
        title = "{Observations of feedback from radio-quiet quasars - I. Extents and morphologies of ionized gas nebulae}",
      journal = {\mnras},
         year = 2013,
        month = apr,
       volume = {430},
       number = {3},
        pages = {2327-2345},
          doi = {10.1093/mnras/stt051},
archivePrefix = {arXiv},
       eprint = {1301.1677},
 primaryClass = {astro-ph.CO},
       adsurl = {https://ui.adsabs.harvard.edu/abs/2013MNRAS.430.2327L}
}

@ARTICLE{Liu13b,
       author = {{Liu}, Guilin and {Zakamska}, Nadia L. and {Greene}, Jenny E. and {Nesvadba}, Nicole P.~H. and {Liu}, Xin},
        title = "{Observations of feedback from radio-quiet quasars - II. Kinematics of ionized gas nebulae}",
      journal = {\mnras},
         year = 2013,
        month = dec,
       volume = {436},
       number = {3},
        pages = {2576-2597},
          doi = {10.1093/mnras/stt1755},
archivePrefix = {arXiv},
       eprint = {1305.6922},
 primaryClass = {astro-ph.CO},
       adsurl = {https://ui.adsabs.harvard.edu/abs/2013MNRAS.436.2576L}
}

@ARTICLE{RodriguezZaurin13,
       author = {{Rodr{\'\i}guez Zaur{\'\i}n}, J. and {Tadhunter}, C.~N. and {Rose}, M. and {Holt}, J.},
        title = "{The importance of warm, AGN-driven outflows in the nuclear regions of nearby ULIRGs}",
      journal = {\mnras},
         year = 2013,
        month = jun,
       volume = {432},
       number = {1},
        pages = {138-166},
          doi = {10.1093/mnras/stt423},
archivePrefix = {arXiv},
       eprint = {1303.1400},
 primaryClass = {astro-ph.CO},
       adsurl = {https://ui.adsabs.harvard.edu/abs/2013MNRAS.432..138R}
}

@ARTICLE{Fischer13,
       author = {{Fischer}, T.~C. and {Crenshaw}, D.~M. and {Kraemer}, S.~B. and {Schmitt}, H.~R.},
        title = "{Determining Inclinations of Active Galactic Nuclei via their Narrow-line Region Kinematics. I. Observational Results}",
      journal = {\apjs},
         year = 2013,
        month = nov,
       volume = {209},
       number = {1},
          eid = {1},
        pages = {1},
          doi = {10.1088/0067-0049/209/1/1},
archivePrefix = {arXiv},
       eprint = {1308.4129},
 primaryClass = {astro-ph.CO},
       adsurl = {https://ui.adsabs.harvard.edu/abs/2013ApJS..209....1F}
}

@ARTICLE{Rupke13,
       author = {{Rupke}, David S.~N. and {Veilleux}, Sylvain},
        title = "{The Multiphase Structure and Power Sources of Galactic Winds in Major Mergers}",
      journal = {\apj},
         year = 2013,
        month = may,
       volume = {768},
       number = {1},
          eid = {75},
        pages = {75},
          doi = {10.1088/0004-637X/768/1/75},
archivePrefix = {arXiv},
       eprint = {1303.6866},
 primaryClass = {astro-ph.CO},
       adsurl = {https://ui.adsabs.harvard.edu/abs/2013ApJ...768...75R}
}

@ARTICLE{Husemann13,
       author = {{Husemann}, B. and {Jahnke}, K. and {S{\'a}nchez}, S.~F. and {Barrado}, D. and {Bekerait{\.{e}}}, S. and {Bomans}, D.~J. and {Castillo-Morales}, A. and {Catal{\'a}n-Torrecilla}, C. and {Cid Fernandes}, R. and {Falc{\'o}n-Barroso}, J. and {Garc{\'\i}a-Benito}, R. and {Gonz{\'a}lez Delgado}, R.~M. and {Iglesias-P{\'a}ramo}, J. and {Johnson}, B.~D. and {Kupko}, D. and {L{\'o}pez-Fernandez}, R. and {Lyubenova}, M. and {Marino}, R.~A. and {Mast}, D. and {Miskolczi}, A. and {Monreal-Ibero}, A. and {Gil de Paz}, A. and {P{\'e}rez}, E. and {P{\'e}rez}, I. and {Rosales-Ortega}, F.~F. and {Ruiz-Lara}, T. and {Schilling}, U. and {van de Ven}, G. and {Walcher}, J. and {Alves}, J. and {de Amorim}, A.~L. and {Backsmann}, N. and {Barrera-Ballesteros}, J.~K. and {Bland-Hawthorn}, J. and {Cortijo}, C. and {Dettmar}, R. -J. and {Demleitner}, M. and {D{\'\i}az}, A.~I. and {Enke}, H. and {Florido}, E. and {Flores}, H. and {Galbany}, L. and {Gallazzi}, A. and {Garc{\'\i}a-Lorenzo}, B. and {Gomes}, J.~M. and {Gruel}, N. and {Haines}, T. and {Holmes}, L. and {Jungwiert}, B. and {Kalinova}, V. and {Kehrig}, C. and {Kennicutt}, R.~C. and {Klar}, J. and {Lehnert}, M.~D. and {L{\'o}pez-S{\'a}nchez}, {\'A}. R. and {de Lorenzo-C{\'a}ceres}, A. and {M{\'a}rmol-Queralt{\'o}}, E. and {M{\'a}rquez}, I. and {Mendez-Abreu}, J. and {Moll{\'a}}, M. and {del Olmo}, A. and {Meidt}, S.~E. and {Papaderos}, P. and {Puschnig}, J. and {Quirrenbach}, A. and {Roth}, M.~M. and {S{\'a}nchez-Bl{\'a}zquez}, P. and {Spekkens}, K. and {Singh}, R. and {Stanishev}, V. and {Trager}, S.~C. and {Vilchez}, J.~M. and {Wild}, V. and {Wisotzki}, L. and {Zibetti}, S. and {Ziegler}, B.},
        title = "{CALIFA, the Calar Alto Legacy Integral Field Area survey. II. First public data release}",
      journal = {\aap},
         year = 2013,
        month = jan,
       volume = {549},
          eid = {A87},
        pages = {A87},
          doi = {10.1051/0004-6361/201220582},
archivePrefix = {arXiv},
       eprint = {1210.8150},
 primaryClass = {astro-ph.CO},
       adsurl = {https://ui.adsabs.harvard.edu/abs/2013A&A...549A..87H}
}

@ARTICLE{Maiolino12,
       author = {{Maiolino}, R. and {Gallerani}, S. and {Neri}, R. and {Cicone}, C. and {Ferrara}, A. and {Genzel}, R. and {Lutz}, D. and {Sturm}, E. and {Tacconi}, L.~J. and {Walter}, F. and {Feruglio}, C. and {Fiore}, F. and {Piconcelli}, E.},
        title = "{Evidence of strong quasar feedback in the early Universe}",
      journal = {\mnras},
         year = 2012,
        month = sep,
       volume = {425},
       number = {1},
        pages = {L66-L70},
          doi = {10.1111/j.1745-3933.2012.01303.x},
archivePrefix = {arXiv},
       eprint = {1204.2904},
 primaryClass = {astro-ph.CO},
       adsurl = {https://ui.adsabs.harvard.edu/abs/2012MNRAS.425L..66M}
}

@ARTICLE{Fabian12,
       author = {{Fabian}, A.~C.},
        title = "{Observational Evidence of Active Galactic Nuclei Feedback}",
      journal = {\araa},
         year = 2012,
        month = sep,
       volume = {50},
        pages = {455-489},
          doi = {10.1146/annurev-astro-081811-125521},
archivePrefix = {arXiv},
       eprint = {1204.4114},
 primaryClass = {astro-ph.CO},
       adsurl = {https://ui.adsabs.harvard.edu/abs/2012ARA&A..50..455F}
}

@ARTICLE{Zubovas12,
       author = {{Zubovas}, Kastytis and {King}, Andrew},
        title = "{Clearing Out a Galaxy}",
      journal = {\apjl},
         year = 2012,
        month = feb,
       volume = {745},
       number = {2},
          eid = {L34},
        pages = {L34},
          doi = {10.1088/2041-8205/745/2/L34},
archivePrefix = {arXiv},
       eprint = {1201.0866},
 primaryClass = {astro-ph.GA},
       adsurl = {https://ui.adsabs.harvard.edu/abs/2012ApJ...745L..34Z}
}

@ARTICLE{Fischer11,
       author = {{Fischer}, T.~C. and {Crenshaw}, D.~M. and {Kraemer}, S.~B. and {Schmitt}, H.~R. and {Mushotsky}, R.~F. and {Dunn}, J.~P.},
        title = "{Hubble Space Telescope Observations of the Double-peaked Emission Lines in the Seyfert Galaxy Markarian 78: Mass Outflows from a Single Active Galactic Nucleus}",
      journal = {\apj},
         year = 2011,
        month = feb,
       volume = {727},
       number = {2},
          eid = {71},
        pages = {71},
          doi = {10.1088/0004-637X/727/2/71},
archivePrefix = {arXiv},
       eprint = {1011.4213},
 primaryClass = {astro-ph.CO},
       adsurl = {https://ui.adsabs.harvard.edu/abs/2011ApJ...727...71F}
}

@ARTICLE{Muller-Sanchez11,
       author = {{M{\"u}ller-S{\'a}nchez}, F. and {Prieto}, M.~A. and {Hicks}, E.~K.~S. and {Vives-Arias}, H. and {Davies}, R.~I. and {Malkan}, M. and {Tacconi}, L.~J. and {Genzel}, R.},
        title = "{Outflows from Active Galactic Nuclei: Kinematics of the Narrow-line and Coronal-line Regions in Seyfert Galaxies}",
      journal = {\apj},
         year = 2011,
        month = oct,
       volume = {739},
       number = {2},
          eid = {69},
        pages = {69},
          doi = {10.1088/0004-637X/739/2/69},
archivePrefix = {arXiv},
       eprint = {1107.3140},
 primaryClass = {astro-ph.CO},
       adsurl = {https://ui.adsabs.harvard.edu/abs/2011ApJ...739...69M}
}

@ARTICLE{Fischer10,
       author = {{Fischer}, T.~C. and {Crenshaw}, D.~M. and {Kraemer}, S.~B. and {Schmitt}, H.~R. and {Trippe}, M.~L.},
        title = "{Modeling the Outflow in the Narrow-line Region of Markarian 573: Biconical Illumination of a Gaseous Disk}",
      journal = {\aj},
         year = 2010,
        month = aug,
       volume = {140},
       number = {2},
        pages = {577-583},
          doi = {10.1088/0004-6256/140/2/577},
archivePrefix = {arXiv},
       eprint = {1006.1875},
 primaryClass = {astro-ph.CO},
       adsurl = {https://ui.adsabs.harvard.edu/abs/2010AJ....140..577F}
}

@ARTICLE{Crenshaw10a,
       author = {{Crenshaw}, D.~M. and {Kraemer}, S.~B. and {Schmitt}, H.~R. and {Jaff{\'e}}, Y.~L. and {Deo}, R.~P. and {Collins}, N.~R. and {Fischer}, T.~C.},
        title = "{The Geometry of Mass Outflows and Fueling Flows in the Seyfert 2 Galaxy MRK 3}",
      journal = {\aj},
         year = 2010,
        month = mar,
       volume = {139},
       number = {3},
        pages = {871-877},
          doi = {10.1088/0004-6256/139/3/871},
archivePrefix = {arXiv},
       eprint = {0912.2420},
 primaryClass = {astro-ph.CO},
       adsurl = {https://ui.adsabs.harvard.edu/abs/2010AJ....139..871C}
}

@ARTICLE{Crenshaw10b,
       author = {{Crenshaw}, D.~M. and {Schmitt}, H.~R. and {Kraemer}, S.~B. and {Mushotzky}, R.~F. and {Dunn}, J.~P.},
        title = "{Radial Velocity Offsets Due to Mass Outflows and Extinction in Active Galactic Nuclei}",
      journal = {\apj},
         year = 2010,
        month = jan,
       volume = {708},
       number = {1},
        pages = {419-426},
          doi = {10.1088/0004-637X/708/1/419},
archivePrefix = {arXiv},
       eprint = {0911.0675},
 primaryClass = {astro-ph.CO},
       adsurl = {https://ui.adsabs.harvard.edu/abs/2010ApJ...708..419C}
}

@ARTICLE{Hopkins08,
       author = {{Hopkins}, Philip F. and {Hernquist}, Lars and {Cox}, Thomas J. and {Kere{\v{s}}}, Du{\v{s}}an},
        title = "{A Cosmological Framework for the Co-Evolution of Quasars, Supermassive Black Holes, and Elliptical Galaxies. I. Galaxy Mergers and Quasar Activity}",
      journal = {\apjs},
         year = 2008,
        month = apr,
       volume = {175},
       number = {2},
        pages = {356-389},
          doi = {10.1086/524362},
archivePrefix = {arXiv},
       eprint = {0706.1243},
 primaryClass = {astro-ph},
       adsurl = {https://ui.adsabs.harvard.edu/abs/2008ApJS..175..356H}
}

@ARTICLE{Das06,
       author = {{Das}, V. and {Crenshaw}, D.~M. and {Kraemer}, S.~B. and {Deo}, R.~P.},
        title = "{Kinematics of the Narrow-Line Region in the Seyfert 2 Galaxy NGC 1068: Dynamical Effects of the Radio Jet}",
      journal = {\aj},
         year = 2006,
        month = aug,
       volume = {132},
       number = {2},
        pages = {620-632},
          doi = {10.1086/504899},
archivePrefix = {arXiv},
       eprint = {astro-ph/0603803},
 primaryClass = {astro-ph},
       adsurl = {https://ui.adsabs.harvard.edu/abs/2006AJ....132..620D}
}

@ARTICLE{DiMatteo05,
       author = {{Di Matteo}, Tiziana and {Springel}, Volker and {Hernquist}, Lars},
        title = "{Energy input from quasars regulates the growth and activity of black holes and their host galaxies}",
      journal = {\nat},
         year = 2005,
        month = feb,
       volume = {433},
       number = {7026},
        pages = {604-607},
          doi = {10.1038/nature03335},
archivePrefix = {arXiv},
       eprint = {astro-ph/0502199},
 primaryClass = {astro-ph},
       adsurl = {https://ui.adsabs.harvard.edu/abs/2005Natur.433..604D}
}

@ARTICLE{Das05,
       author = {{Das}, V. and {Crenshaw}, D.~M. and {Hutchings}, J.~B. and {Deo}, R.~P. and {Kraemer}, S.~B. and {Gull}, T.~R. and {Kaiser}, M.~E. and {Nelson}, C.~H. and {Weistrop}, D.},
        title = "{Mapping the Kinematics of the Narrow-Line Region in the Seyfert Galaxy NGC 4151}",
      journal = {\aj},
         year = 2005,
        month = sep,
       volume = {130},
       number = {3},
        pages = {945-956},
          doi = {10.1086/432255},
archivePrefix = {arXiv},
       eprint = {astro-ph/0505103},
 primaryClass = {astro-ph},
       adsurl = {https://ui.adsabs.harvard.edu/abs/2005AJ....130..945D}
}

@ARTICLE{Greene05,
       author = {{Greene}, Jenny E. and {Ho}, Luis C.},
        title = "{A Comparison of Stellar and Gaseous Kinematics in the Nuclei of Active Galaxies}",
      journal = {\apj},
         year = 2005,
        month = jul,
       volume = {627},
       number = {2},
        pages = {721-732},
          doi = {10.1086/430590},
archivePrefix = {arXiv},
       eprint = {astro-ph/0503675},
 primaryClass = {astro-ph},
       adsurl = {https://ui.adsabs.harvard.edu/abs/2005ApJ...627..721G}
}

@ARTICLE{Boroson05,
       author = {{Boroson}, Todd},
        title = "{Blueshifted [O III] Emission: Indications of a Dynamic Narrow-Line Region}",
      journal = {\aj},
         year = 2005,
        month = aug,
       volume = {130},
       number = {2},
        pages = {381-386},
          doi = {10.1086/431722},
archivePrefix = {arXiv},
       eprint = {astro-ph/0505127},
 primaryClass = {astro-ph},
       adsurl = {https://ui.adsabs.harvard.edu/abs/2005AJ....130..381B}
}

@ARTICLE{Veilleux01,
       author = {{Veilleux}, S. and {Shopbell}, P.~L. and {Miller}, S.~T.},
        title = "{The Biconical Outflow in the Seyfert Galaxy NGC 2992}",
      journal = {\aj},
         year = 2001,
        month = jan,
       volume = {121},
       number = {1},
        pages = {198-209},
          doi = {10.1086/318046},
archivePrefix = {arXiv},
       eprint = {astro-ph/0010134},
 primaryClass = {astro-ph},
       adsurl = {https://ui.adsabs.harvard.edu/abs/2001AJ....121..198V}
}

@ARTICLE{Crenshaw00a,
       author = {{Crenshaw}, D. Michael and {Kraemer}, Steven B.},
        title = "{Resolved Spectroscopy of the Narrow-Line Region in NGC 1068: Kinematics of the Ionized Gas}",
      journal = {\apjl},
         year = 2000,
        month = apr,
       volume = {532},
       number = {2},
        pages = {L101-L104},
          doi = {10.1086/312581},
archivePrefix = {arXiv},
       eprint = {astro-ph/0002438},
 primaryClass = {astro-ph},
       adsurl = {https://ui.adsabs.harvard.edu/abs/2000ApJ...532L.101C}
}

@ARTICLE{Crenshaw00b,
       author = {{Crenshaw}, D.~M. and {Kraemer}, S.~B. and {Hutchings}, J.~B. and {Bradley}, L.~D., II and {Gull}, T.~R. and {Kaiser}, M.~E. and {Nelson}, C.~H. and {Ruiz}, J.~R. and {Weistrop}, D.},
        title = "{A Kinematic Model for the Narrow-Line Region in NGC 4151}",
      journal = {\aj},
         year = 2000,
        month = oct,
       volume = {120},
       number = {4},
        pages = {1731-1738},
          doi = {10.1086/301574},
archivePrefix = {arXiv},
       eprint = {astro-ph/0007017},
 primaryClass = {astro-ph},
       adsurl = {https://ui.adsabs.harvard.edu/abs/2000AJ....120.1731C}
}

@ARTICLE{Silk98,
       author = {{Silk}, Joseph and {Rees}, Martin J.},
        title = "{Quasars and galaxy formation}",
      journal = {\aap},
         year = 1998,
        month = mar,
       volume = {331},
        pages = {L1-L4},
archivePrefix = {arXiv},
       eprint = {astro-ph/9801013},
 primaryClass = {astro-ph},
       adsurl = {https://ui.adsabs.harvard.edu/abs/1998A&A...331L...1S}
}

@ARTICLE{Nelson96,
       author = {{Nelson}, Charles H. and {Whittle}, Mark},
        title = "{Stellar and Gaseous Kinematics of Seyfert Galaxies. II. The Role of the Bulge}",
      journal = {\apj},
         year = 1996,
        month = jul,
       volume = {465},
        pages = {96},
          doi = {10.1086/177405},
       adsurl = {https://ui.adsabs.harvard.edu/abs/1996ApJ...465...96N}
}

@ARTICLE{Vogt13,
       author = {{Vogt}, Fr{\'e}d{\'e}ric P.~A. and {Dopita}, Michael A. and {Kewley}, Lisa J.},
        title = "{Galaxy Interactions in Compact Groups. I. The Galactic Winds of HCG16}",
      journal = {\apj},
         year = 2013,
        month = may,
       volume = {768},
       number = {2},
          eid = {151},
        pages = {151},
          doi = {10.1088/0004-637X/768/2/151},
archivePrefix = {arXiv},
       eprint = {1303.0290},
 primaryClass = {astro-ph.CO},
       adsurl = {https://ui.adsabs.harvard.edu/abs/2013ApJ...768..151V}
}

@ARTICLE{Torres-Flores11,
       author = {{Torres-Flores}, S. and {Epinat}, B. and {Amram}, P. and {Plana}, H. and {Mendes de Oliveira}, C.},
        title = "{GHASP: an H{\ensuremath{\alpha}} kinematic survey of spiral and irregular galaxies - IX. The near-infrared, stellar and baryonic Tully-Fisher relations}",
      journal = {\mnras},
         year = 2011,
        month = sep,
       volume = {416},
       number = {3},
        pages = {1936-1948},
          doi = {10.1111/j.1365-2966.2011.19169.x},
archivePrefix = {arXiv},
       eprint = {1106.0505},
 primaryClass = {astro-ph.CO},
       adsurl = {https://ui.adsabs.harvard.edu/abs/2011MNRAS.416.1936T}
}

@ARTICLE{Bouche15,
       author = {{Bouch{\'e}}, N. and {Carfantan}, H. and {Schroetter}, I. and {Michel-Dansac}, L. and {Contini}, T.},
        title = "{GalPak$^{3D}$: A Bayesian Parametric Tool for Extracting Morphokinematics of Galaxies from 3D Data}",
      journal = {\aj},
         year = 2015,
        month = sep,
       volume = {150},
       number = {3},
          eid = {92},
        pages = {92},
          doi = {10.1088/0004-6256/150/3/92},
archivePrefix = {arXiv},
       eprint = {1501.06586},
 primaryClass = {astro-ph.IM},
       adsurl = {https://ui.adsabs.harvard.edu/abs/2015AJ....150...92B}
}

@ARTICLE{Dutton11,
       author = {{Dutton}, Aaron A. and {van den Bosch}, Frank C. and {Faber}, Sandra M. and {Simard}, Luc and {Kassin}, Susan A. and {Koo}, David C. and {Bundy}, Kevin and {Huang}, Jiasheng and {Weiner}, Benjamin J. and {Cooper}, Michael C. and {Newman}, Jeffrey A. and {Mozena}, Mark and {Koekemoer}, Anton M.},
        title = "{On the evolution of the velocity-mass-size relations of disc-dominated galaxies over the past 10 billion years}",
      journal = {\mnras},
         year = 2011,
        month = jan,
       volume = {410},
       number = {3},
        pages = {1660-1676},
          doi = {10.1111/j.1365-2966.2010.17555.x},
archivePrefix = {arXiv},
       eprint = {1006.3558},
 primaryClass = {astro-ph.GA},
       adsurl = {https://ui.adsabs.harvard.edu/abs/2011MNRAS.410.1660D}
}

@ARTICLE{Amorisco10,
       author = {{Amorisco}, N.~C. and {Bertin}, G.},
        title = "{Self-consistent nonspherical isothermal halos embedding zero-thickness disks}",
      journal = {\aap},
         year = 2010,
        month = sep,
       volume = {519},
          eid = {A47},
        pages = {A47},
          doi = {10.1051/0004-6361/201014387},
archivePrefix = {arXiv},
       eprint = {1005.3154},
 primaryClass = {astro-ph.CO},
       adsurl = {https://ui.adsabs.harvard.edu/abs/2010A&A...519A..47A}
}

@ARTICLE{Astropy22,
       author = {{Astropy Collaboration} and {Price-Whelan}, Adrian M. and {Lim}, Pey Lian and {Earl}, Nicholas and {Starkman}, Nathaniel and {Bradley}, Larry and {Shupe}, David L. and {Patil}, Aarya A. and {Corrales}, Lia and {Brasseur}, C.~E. and {N{\"o}the}, Maximilian and {Donath}, Axel and {Tollerud}, Erik and {Morris}, Brett M. and {Ginsburg}, Adam and {Vaher}, Eero and {Weaver}, Benjamin A. and {Tocknell}, James and {Jamieson}, William and {van Kerkwijk}, Marten H. and {Robitaille}, Thomas P. and {Merry}, Bruce and {Bachetti}, Matteo and {G{\"u}nther}, H. Moritz and {Aldcroft}, Thomas L. and {Alvarado-Montes}, Jaime A. and {Archibald}, Anne M. and {B{\'o}di}, Attila and {Bapat}, Shreyas and {Barentsen}, Geert and {Baz{\'a}n}, Juanjo and {Biswas}, Manish and {Boquien}, M{\'e}d{\'e}ric and {Burke}, D.~J. and {Cara}, Daria and {Cara}, Mihai and {Conroy}, Kyle E. and {Conseil}, Simon and {Craig}, Matthew W. and {Cross}, Robert M. and {Cruz}, Kelle L. and {D'Eugenio}, Francesco and {Dencheva}, Nadia and {Devillepoix}, Hadrien A.~R. and {Dietrich}, J{\"o}rg P. and {Eigenbrot}, Arthur Davis and {Erben}, Thomas and {Ferreira}, Leonardo and {Foreman-Mackey}, Daniel and {Fox}, Ryan and {Freij}, Nabil and {Garg}, Suyog and {Geda}, Robel and {Glattly}, Lauren and {Gondhalekar}, Yash and {Gordon}, Karl D. and {Grant}, David and {Greenfield}, Perry and {Groener}, Austen M. and {Guest}, Steve and {Gurovich}, Sebastian and {Handberg}, Rasmus and {Hart}, Akeem and {Hatfield-Dodds}, Zac and {Homeier}, Derek and {Hosseinzadeh}, Griffin and {Jenness}, Tim and {Jones}, Craig K. and {Joseph}, Prajwel and {Kalmbach}, J. Bryce and {Karamehmetoglu}, Emir and {Ka{\l}uszy{\'n}ski}, Miko{\l}aj and {Kelley}, Michael S.~P. and {Kern}, Nicholas and {Kerzendorf}, Wolfgang E. and {Koch}, Eric W. and {Kulumani}, Shankar and {Lee}, Antony and {Ly}, Chun and {Ma}, Zhiyuan and {MacBride}, Conor and {Maljaars}, Jakob M. and {Muna}, Demitri and {Murphy}, N.~A. and {Norman}, Henrik and {O'Steen}, Richard and {Oman}, Kyle A. and {Pacifici}, Camilla and {Pascual}, Sergio and {Pascual-Granado}, J. and {Patil}, Rohit R. and {Perren}, Gabriel I. and {Pickering}, Timothy E. and {Rastogi}, Tanuj and {Roulston}, Benjamin R. and {Ryan}, Daniel F. and {Rykoff}, Eli S. and {Sabater}, Jose and {Sakurikar}, Parikshit and {Salgado}, Jes{\'u}s and {Sanghi}, Aniket and {Saunders}, Nicholas and {Savchenko}, Volodymyr and {Schwardt}, Ludwig and {Seifert-Eckert}, Michael and {Shih}, Albert Y. and {Jain}, Anany Shrey and {Shukla}, Gyanendra and {Sick}, Jonathan and {Simpson}, Chris and {Singanamalla}, Sudheesh and {Singer}, Leo P. and {Singhal}, Jaladh and {Sinha}, Manodeep and {Sip{\H{o}}cz}, Brigitta M. and {Spitler}, Lee R. and {Stansby}, David and {Streicher}, Ole and {{\v{S}}umak}, Jani and {Swinbank}, John D. and {Taranu}, Dan S. and {Tewary}, Nikita and {Tremblay}, Grant R. and {de Val-Borro}, Miguel and {Van Kooten}, Samuel J. and {Vasovi{\'c}}, Zlatan and {Verma}, Shresth and {de Miranda Cardoso}, Jos{\'e} Vin{\'\i}cius and {Williams}, Peter K.~G. and {Wilson}, Tom J. and {Winkel}, Benjamin and {Wood-Vasey}, W.~M. and {Xue}, Rui and {Yoachim}, Peter and {Zhang}, Chen and {Zonca}, Andrea and {Astropy Project Contributors}},
        title = "{The Astropy Project: Sustaining and Growing a Community-oriented Open-source Project and the Latest Major Release (v5.0) of the Core Package}",
      journal = {\apj},
         year = 2022,
        month = aug,
       volume = {935},
       number = {2},
          eid = {167},
        pages = {167},
          doi = {10.3847/1538-4357/ac7c74},
archivePrefix = {arXiv},
       eprint = {2206.14220},
 primaryClass = {astro-ph.IM},
       adsurl = {https://ui.adsabs.harvard.edu/abs/2022ApJ...935..167A}
}

@Article{matplotlib,
  Author    = {Hunter, J. D.},
  Title     = {Matplotlib: A 2D graphics environment},
  Journal   = {Computing in Science \& Engineering},
  Volume    = {9},
  Number    = {3},
  Pages     = {90--95},
  publisher = {IEEE COMPUTER SOC},
  doi       = {10.1109/MCSE.2007.55},
  year      = 2007
}

@ARTICLE{scipy,
  author  = {Virtanen, Pauli and Gommers, Ralf and Oliphant, Travis E. and
            Haberland, Matt and Reddy, Tyler and Cournapeau, David and
            Burovski, Evgeni and Peterson, Pearu and Weckesser, Warren and
            Bright, Jonathan and {van der Walt}, St{\'e}fan J. and
            Brett, Matthew and Wilson, Joshua and Millman, K. Jarrod and
            Mayorov, Nikolay and Nelson, Andrew R. J. and Jones, Eric and
            Kern, Robert and Larson, Eric and Carey, C J and
            Polat, {\.I}lhan and Feng, Yu and Moore, Eric W. and
            {VanderPlas}, Jake and Laxalde, Denis and Perktold, Josef and
            Cimrman, Robert and Henriksen, Ian and Quintero, E. A. and
            Harris, Charles R. and Archibald, Anne M. and
            Ribeiro, Ant{\^o}nio H. and Pedregosa, Fabian and
            {van Mulbregt}, Paul and {SciPy 1.0 Contributors}},
  title   = {{{SciPy} 1.0: Fundamental Algorithms for Scientific
            Computing in Python}},
  journal = {Nature Methods},
  year    = {2020},
  volume  = {17},
  pages   = {261--272},
  adsurl  = {https://rdcu.be/b08Wh},
  doi     = {10.1038/s41592-019-0686-2},
}

@ARTICLE{numpy,
       author = {{Harris}, Charles R. and {Millman}, K. Jarrod and {van der Walt}, St{\'e}fan J. and {Gommers}, Ralf and {Virtanen}, Pauli and {Cournapeau}, David and {Wieser}, Eric and {Taylor}, Julian and {Berg}, Sebastian and {Smith}, Nathaniel J. and {Kern}, Robert and {Picus}, Matti and {Hoyer}, Stephan and {van Kerkwijk}, Marten H. and {Brett}, Matthew and {Haldane}, Allan and {del R{\'\i}o}, Jaime Fern{\'a}ndez and {Wiebe}, Mark and {Peterson}, Pearu and {G{\'e}rard-Marchant}, Pierre and {Sheppard}, Kevin and {Reddy}, Tyler and {Weckesser}, Warren and {Abbasi}, Hameer and {Gohlke}, Christoph and {Oliphant}, Travis E.},
        title = "{Array programming with NumPy}",
      journal = {\nat},
         year = 2020,
        month = sep,
       volume = {585},
       number = {7825},
        pages = {357-362},
          doi = {10.1038/s41586-020-2649-2},
archivePrefix = {arXiv},
       eprint = {2006.10256},
 primaryClass = {cs.MS},
       adsurl = {https://ui.adsabs.harvard.edu/abs/2020Natur.585..357H}
}

@ARTICLE{KLDIV,
author = {S. Kullback and R. A. Leibler},
title = {{On Information and Sufficiency}},
volume = {22},
journal = {The Annals of Mathematical Statistics},
number = {1},
publisher = {Institute of Mathematical Statistics},
pages = {79 -- 86},
year = {1951},
doi = {10.1214/aoms/1177729694},
URL = {https://doi.org/10.1214/aoms/1177729694}
}
\bibliographystyle{aasjournal}
\end{document}